\documentclass[10pt]{iopart}
\usepackage{times}
\usepackage{iopams}
\usepackage{bm,mathrsfs} 
\usepackage[caption=false]{subfig}
\usepackage{graphicx,color}
\usepackage[numbers, comma]{natbib}
\usepackage[toctitles]{titlesec}
\usepackage{balance}

\usepackage{hyperref}
\hypersetup{
    colorlinks=true,
    linkcolor=blue,
    citecolor=blue,      
    urlcolor=blue,
    breaklinks=true
}
\begin{document}

\topical[Soft hydraulics: from Newtonian to complex fluids]{Soft hydraulics: from Newtonian to complex fluid flows through compliant conduits}

\author[I.~C.~Christov]{Ivan C Christov}

\address{School of Mechanical Engineering, Purdue University, West Lafayette, IN 47907, USA}
\ead{christov@purdue.edu}
\vspace{10pt}
\begin{indented}
\item[]June 2021
\end{indented}

\begin{abstract}
Microfluidic devices manufactured from soft polymeric materials have emerged as a paradigm for cheap, disposable and easy-to-prototype fluidic platforms for integrating chemical and biological assays and analyses. The interplay between the flow forces and the inherently compliant conduits of such microfluidic devices requires careful consideration. While mechanical compliance was initially a side-effect of the manufacturing process and materials used, compliance has now become a paradigm, enabling new approaches to microrheological measurements, new modalities of micromixing, and improved sieving of micro- and nano-particles, to name a few applications. This topical review provides an introduction to the physics of these systems. Specifically, the goal of this review is to summarize the recent progress towards a mechanistic understanding of the interaction between non-Newtonian (complex) fluid flows and their deformable confining boundaries. In this context, key experimental results and relevant applications are also explored, hand-in-hand with the fundamental principles for their physics-based modeling. The key topics covered include shear-dependent viscosity of non-Newtonian fluids, hydrodynamic pressure gradients during flow, the elastic response (deformation and bulging) of soft conduits due to flow within, the effect of cross-sectional conduit geometry on the resulting fluid--structure interaction, and key dimensionless groups describing the coupled physics. Open problems and future directions in this nascent field of soft hydraulics, at the intersection of non-Newtonian fluid mechanics, soft matter physics, and microfluidics, are noted.
\end{abstract}

\vspace{2pc}

\submitto{\JPCM}

\maketitle
\tableofcontents

\ioptwocol



\section{Introduction}
\label{sec:intro}

\begin{figure}[h]
	\centering
	\includegraphics[width=\columnwidth]{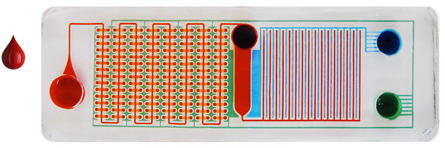}
	\caption{Schematic of a ``self-powered integrated microfluidic point-of-care low-cost enabling (SIMPLE) chip'' designed for medical diagnostics and able to detect HIV and MRSA at a cost of about \$10 per chip \cite{Allen17,Yeh17}. The entire device fits on a standard microscope slide, having dimensions about $25\times75$ mm. Blood is input in the reservoir on the left, reagents are placed in the middle reservoir, the tubing connected to the voids on the right is a battery vacuum system used to drive the flow. Microchannels are highlighted by colors. Reproduced from \cite{Yeh17} with permission from AAAS.}
	\label{fig:chip}
\end{figure}

Microfluidics, which concerns the manipulation of small (\textit{e.g.}, nanoliter) volumes of fluids at small (\textit{e.g.}, micron) scales \cite{SSA04,SQ05}, ``exploded'' around the turn of the century. The number of papers published annually grew ten-fold from 1994 to 2004 \cite[p.~7]{NW06}, with another factor of almost ten    reached by 2020, according to the Web of Science. Microfluidics has disrupted \cite{KML12} fields ranging from cellular and developmental biology \cite{EASJ06} to logical circuits \cite{PG07,KCP15} to chemical and biological warfare deterrents \cite{W06}, to name a few. The microfluidic technologies market was valued at \$18 \emph{billion} in 2020 \cite{MF_Market2}. Much of microfluidics has been enabled by polymeric gels made from \emph{polydimethylsiloxane} (PDMS) (commercially available as the SYLGARD\texttrademark\ 184 silicone elastomer). PDMS allows cheap and rapid manufacture with fine geometric control (down to the nanoscale) \cite{MW02,LOVB97} and tunable mechanical properties \cite{JMTT14}. Figure~\ref{fig:chip} shows an example PDMS-based microfluidic chip costing on the order of \$10 and designed to detect the human immunodeficiency virus (HIV) and methicillin-resistant Staphylococcus aureus (MRSA). 

\begin{figure}
	\centering
	\includegraphics[width=0.5\columnwidth]{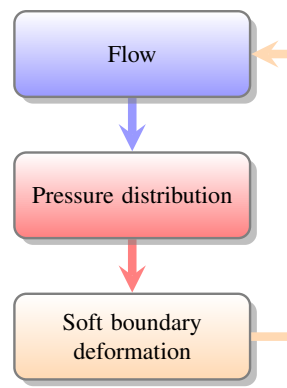}
	\caption{Schematic of the fundamental feedback mechanism of two-way-coupled fluid--structure interaction between an internal fluid flow and the elastic boundaries of the soft flow conduit; styled after a diagram of Fung \cite[figure~3.4:2]{F97}.}
	\label{fig:fsi-loop}
\end{figure}

Being made from polymeric materials (\textit{e.g.}, PDMS), channels in microfluidic devices are therefore \emph{soft} \cite{XW98} with a Young's modulus $E \approx 0.3$ to $2$ MPa \cite{LSSBC09,GEGJ06}. That is to say, these materials easily deform under an applied load. In some applications, the device's compliance can lead to blurring of high-speed optical imaging of its interior \cite{Yali20} or restrict its structural viability \cite{DSMB97,HJLK02}. In other applications, \emph{however}, flexibility is a crucial advantage to be exploited in the design of implantable and wearable electronics \cite{Xu14,YKL16}, or to emulate soft biological tissues in {organs-on-a-chip} \cite{Huhetal10,Lindetal17}. The salient physics at hand is that the fluid's pressure forces cause an elastic structure (immersed in a flow or bounding it) to deform, which in turn modifies the flow, as shown schematically in figure~\ref{fig:fsi-loop}. This is an example of a \emph{fluid--structure interaction} \cite{DS14,P14}. In some fields, another term for this phenomenon is \emph{elastohydrodynamics} \cite{G01}.

In this context, the present topical review specifically aims to address the interaction of \emph{non-Newtonian fluid flows} and their \emph{soft confining boundaries}. The discussion below is centered on the fundamental physics and seeks to enable a theoretical understanding of these coupled multi-physics phenomena. The goal is to provide the reader an understanding of key experimental results and relevant applications hand-in-hand with the fundamental principles for modeling, interpretation and design. In doing so, it is expected that the review will enable the reader to identify and pursue open problems in this nascent field of \emph{soft hydraulics}, at the intersection of non-Newtonian fluid mechanics, soft matter physics, and microfluidics.


\subsection{Scope of the review}
\label{sec:scope}

The key topics covered in this review are:
\begin{itemize}
	\item the hydraulic--electric circuit analogy (section~\ref{sec:hydrocirc});
	\item the basics of non-Newtonian (complex) fluid rheology, focusing on steady flows (section~\ref{sec:rheology});
	\item the lubrication approximation, flow in long slender conduits, and the relation between the flow rate and the hydrodynamic pressure gradient therein (sections~\ref{sec:lubrication} and \ref{sec:q_dpdz});
	\item the elastic response (deformation and bulging) of soft conduits due to flow within (section~\ref{sec:deformation}), including the effect of the flow conduit's cross-sectional geometry on the resulting fluid--structure interaction;
	\item the resulting basic laws of soft hydraulics for flow in compliant conduits (section~\ref{sec:FSI});
	\item a sampling of key applications involving non-Newtonian fluid flows in soft hydraulic conduits (section~\ref{sec:appl});
	\item and advanced topics related to the physics of these flows (section~\ref{sec:advanced}), leading into current open problems and future directions (section~\ref{sec:concl}). 
\end{itemize}

Naturally, a number of topics cannot be covered in this review. Specifically, manufacturing techniques \cite{C10_book}, design of microfluidics chips, and biomicrofluidics (see \cite[chapter~8]{C13_book} and \cite{C10_book_Ch4}) are out of scope here. However, many of these topics are covered in resources beyond the current review that are now briefly summarized.

Important overviews of the pioneering 1990s research on micro-electro-mechanical systems (MEMS) are given by Ho and Tai \cite{HT98} and Gad-el-Hak \cite{G99}. Stone \textit{et al.}~\cite{SSA04,Stone2007} provide a detailed look at the flow physics in rigid hydraulic conduits, with applications of microfluidics to the (then) emerging technology of lab-on-a-chip. Squires and Quake \cite{SQ05} take a deep-dive into (nearly) all flow physics encountered in microfluidics. Abgrall and Gu\'e \cite{AG07} give a complementary review (to \cite{SSA04,SQ05}) of the requisite micro-manufacturing techniques. The multiphysics couplings occurring in microfluidic systems, of which the fluid--solid coupling detailed in section~\ref{sec:theory} is one example, has led to the introduction of the term ``nonlinear microfluidics,'' overviews of which can be found in \cite{SDD19,Xia21}. The role of such nonlinear elastohydrodynamic effects on dynamic force measurements (with implications for, \textit{e.g.}, atomic force microscopes and surface forces apparatuses) are reviewed by Wang \textit{et al.}~\cite{WPDF17,WTF17}.

Biological and physiological implications of the coupling betwee flow and compliant boundaries (such as arteries and airways) are expounded upon by Grotberg and Jensen \cite{GJ04} and Hazel and Heil \cite{HH11} (see also \cite[chapter 8]{DS14}), building upon the research program initiated by Shapiro \cite{S77} and Pedley \cite{P80}. Lauga and Powers \cite{LP09} discuss related problems arising from the swimming of flagellated microorganisms in complex fluids. Flexible microelectronics benefitting from understanding the physics of microscale fluid--structure interactions are reviewed by Fallahi \textit{et al.}~\cite{FZPN19}. The foundational reviews on PDMS as a versatile soft polymeric material for microfluidics and the manufacture of flow conduits from it using soft lithography are given by McDonald and Whitesides \cite{MW02} and Xia and Whitesides \cite{XW98}, respectively. Recent PDMS-based microfluidics designs and applications are summarized by Raj M and Chakraborty \cite{RC20}, while the decadal review (2007--2017) by Karan \textit{et al.}~\cite{KCC18} focuses on select advances in the five categories of ``microchannels, tubes, squeeze flow, cylinder near wall and thin structures (membranes, sheets, etc.).''

\begin{figure}[hb]
	\centering	
	\includegraphics[width=\columnwidth]{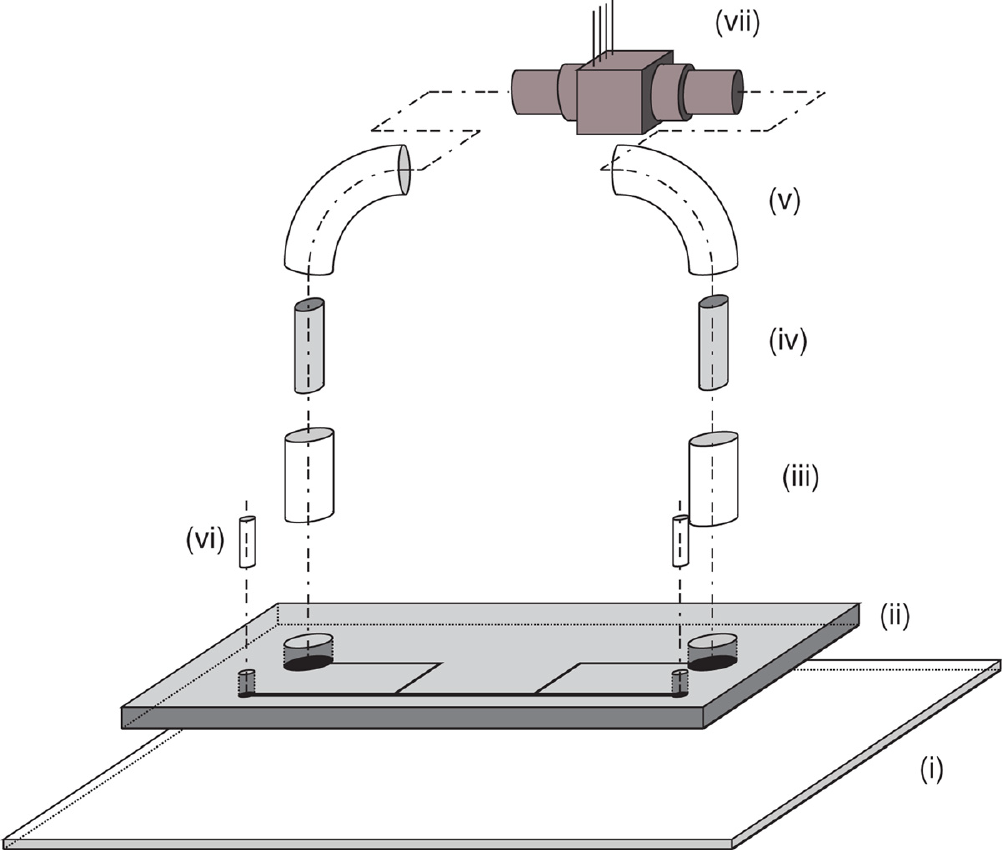}
	\caption{A schematic depicting a microfluidic chip on a glass slide (i) with a single main channel embedded in a PDMS layer (ii). A high-pressure syringe pump was connected to the channel's inlet port (vi) to maintain the flow rate. External tubing, connectors and components are needed to measure the pressure drop for a given flow rate, accounting for the flow-induced deformation. To this end, a transducer port (iii) is connected via a tubing (iv,v) to a pressure transducer (vii), which makes the pressure drop measurement. Reproduced from Cheung P, Toda-Peters K and Shen A Q 2012 \textit{Biomicrofluidics} \textbf{6} 026501 \cite{CTS12}, with permission from AIP Publishing \copyright\ 2012.}
	\label{fig:expt}
\end{figure}

Two key recent textbooks, suitable for teaching an upper undergraduate or introductory graduate level course in this field are those by Bruus \cite{B08} and Kirby \cite{K10}, building upon the earlier books by Karniadakis \textit{et al.}~\cite{KBA05}, Nguyen and Wereley \cite{NW06}, and Tabeling \cite{T05}. It should be noted, however, that these textbooks do not discuss non-Newtonian fluid flows, beyond mentioning the concept.


\section{Experimental observations: the need to understand non-Newtonian soft hydraulics}
\label{sec:expt_motiv}

Context for and the motivation to study non-Newtonian fluid flow in soft hydraulic conduits, was discussed by Anand \textit{et al.}~\cite{ADJRC19}. Specifically, recent experimental papers were reviewed to highlight the lack of broadly applicable predictive physical theory for non-Newtonian fluid flows through soft hydraulic conduits. A schematic of a typical setup of such an experiment is shown in figure~\ref{fig:expt}.

The first example is the experimental study by Raj and Sen \cite{RS16}. They performed experiments of non-Newtonian fluid flow in a rectangular microchannel with three rigid walls and a compliant top wall, which was manufactured from PDMS (Young's moduli of $E=1.362$ MPa and $E=1.184$ MPa were quoted for the two wall thicknesses used). A 0.1\% polyethylene oxyde (PEO) solution, which exhibits shear thinning (to be introduced in section~\ref{sec:rheology}) was used as the working fluid. The pressure drop along the microchannel was measured in $12$ mm increments, over its $L=30$ mm length, by a series of differential pressure sensors. As will become important in section~\ref{sec:lubrication}, the microchannel in this experiment was long and thin: having a cross-section of fixed width $w$ between $0.5$ and $2$ mm and undeformed height $h_0$ of $83$ $\mu$m. The deformation of the compliant top wall was measured using fluorescence microscopy. Raj and Sen \cite{RS16} also proposed a mathematical model for the pressure drop across the length of the microchannel, as a function of the flow rate (and the various material and geometric parameters) for Newtonian fluids. However, the model was not generalized to apply to the non-Newtonian experiments.

Kiran Raj \textit{et al.}~\cite{RCDC18} carried out experiments on non-Newtonian fluid flow in a compliant cylindrical conduit. They used 0.04\% by weight solution of Xanthan gum into deionized water as a blood-analog fluid with shear-thinning properties. A microtube of length $L=27$ mm and diameter $a\approx 500$ $\mu$m was fabricated from PDMS by pull-out soft lithography. Two PDMS mixtures were used, yielding Young's moduli of $E=2.801$ MPa and $E=0.157$ MPa. A one-way coupled theory to calculate the deformation from the known pressure drop, (as a function of the imposed flow rate) was also proposed. However, the flow regimes investigated in \cite{RCDC18} exhibited only weak fluid--structure interaction, and thus deviations from the ideal Hagen--Poiseuille law are small.

Del Giudice \textit{et al.}~\cite{DGNM16} also performed experiments with PEO solutions, which exhibit shear thinning, in square cross-section PDMS microchannels ($E\approx 1$ MPa was reported). They demonstrated that the channel's maximum height increases by $\approx 2$\% under a pressure drop $\Delta p\approx 23$ kPa for a 0.5\% PEO solution, while the increase is $\approx 12$\% under a pressure drop $\Delta p\approx 120$ kPa for a 1.6\% PEO solution. They conclude that this effect is significant and should be modeled physically.

Most recently, Nahar \textit{et al.}~\cite{NDW19} performed experiments with 1.4\% carboxymethyl cellulose (CMC) and 0.01\% polyacrylamide (PAA) aqueous solutions, which exhibit shear thinning, in silicone elastic tubes ($E=4.7$ MPa). They demonstrated the strong effect of fluid rheology by comparing to a flow of a reference Newtonian fluid (a 19\% polyethylene glycol (PEG) aqueous solution). Specifically, when a transmural pressure of 105 mbar between the inside and outside of the tube was applied, the tube's cross-sectional area decreased six times as much for both non-Newtonian fluid flows, compared to the case of the reference Newtonian flow. The observation was rationalized by noting that shear-thinning fluids have a smaller outlet pressure (drop), which is correlated to stronger compressive downstream transmural pressures. A theory for this effect was not provided.

Therefore, despite the early experimental work by Koo and Kleinstreuer \cite{KK03} noting that ``Non-Newtonian fluid effects are expected to be important for polymeric liquids and particle suspension flows,'' prior experimental measurements on non-Newtonian effects in soft hydraulic conduits (\textit{e.g.}, \cite{RS16,DGNM16,NDW19}) have not been fully rationalized by theory. Additionally, these experiments employ only shear-thinning fluids and no similar experiments with viscoelastic fluids (to be discussed in section~\ref{sec:viscoelastic}) appear to have been conducted in either rigid or compliant conduits. Therefore, a clear knowledge gap remains at the intersection of non-Newtonian fluid mechanics and soft matter physics. This research field is still in its infancy, but progress has been made in the last few years. Specifically, the fluid--structure interaction problem has been analyzed, and predictive theories are now becoming available, reducing the three-dimensional (3D) coupled problem to an ordinary differential equation (ODE) for the hydrodynamic pressure, for different types of compliant conduits (\textit{e.g.}, microchannels or microtubes). Next, the building blocks of these theories are reviewed.

\begin{figure*}
	\centering	
	\includegraphics[width=0.8\textwidth]{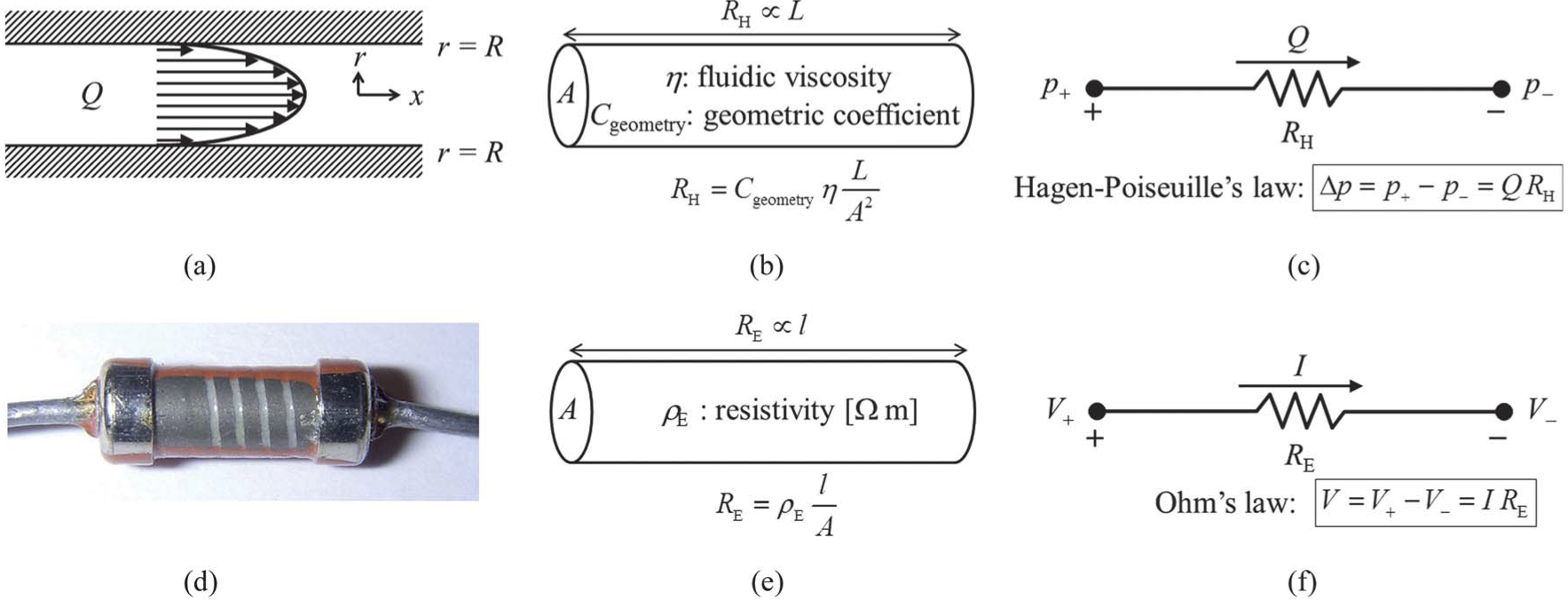}
	\caption{(a) A steady viscous Newtonian fluid flow in a channel or tube and (d) a ``partially exposed Tesla TR-212 1 k$\Omega$ carbon film resistor.'' (b,e) Analogy between hydraulic resistance and electric resistance, respectively. (c,f) The corresponding circuit notations. The Hagen--Poiseuille law (\ref{eq:dp_Rh_q}) and (\ref{eq:Pois}) is the hydraulic equivalent of Ohm's law (\ref{eq:Ohm}). In (b), $C_\mathrm{geometry}=8\pi$ for a circular flow conduit. Reproduced from \cite{Oh12} with permission from The Royal Society of Chemistry \copyright\ 2012.}
	\label{fig:circuit}
\end{figure*}

It should be noted that the problem of non-Newtonian elastohydrodynamic lubrication also comes up in tribology \cite{LME11}. However, these problems involve thin fluid films under extreme pressures and under non-isothermal conditions, in which the fluid behavior can be quite different from the microfluidic setting considered herein. Additionally, tribology problems consider complex deformations, including wall-to-wall contact, and wear (degradation of the fluid and flow conduit), which are not generally expected to occur in microchannels under normal flow conditions. One common problem between tribology and microfluidics could be roll coating flows with deformable substrates \cite{CS97,YK05}.


\section{Predictive physical theories and models}
\label{sec:theory}

To review our current understanding of the interaction of non-Newtonian (complex) fluid flows and their soft confining boundaries, in this section, it is helpful to start with the established results on Newtonian fluid flows, and build up from there.

\subsection{The hydraulic--electric circuit analogy}
\label{sec:hydrocirc}

A powerful pedagogical analogy for understanding pipe flows is the analogy between the \emph{laminar} flow of a \emph{Newtonian} fluid through a pipe and the flow of electrons in a conductor. Figure~\ref{fig:circuit} shows a schematic of this analogy. The basic laws of fluid mechanics dictate that, for a pipe of fixed cross-section, the pressure difference $\Delta p$ needed to maintain a steady volumetric flow rate $q$ through it obeys 
\begin{equation}
	\Delta p = R_h q,
	\label{eq:dp_Rh_q}
\end{equation}
where $R_h$ is the \emph{hydraulic resistance} \cite[section~4.2]{B08}. The limits of applicability of (\ref{eq:dp_Rh_q}) are explored throughout the review. Indeed, this basic law has the same form as Ohm's law, which states that the voltage difference $\Delta V$ needed to maintain a steady electrical current $I$ through a conductor with known resistance to the motion of electrons, obeys 
\begin{equation}
	\Delta V = R_e I,
	\label{eq:Ohm}
\end{equation}
where $R_e$ is the electrical resistance of the wire \cite[section~4.3]{Purcell2013}.

Thus emerges a parallel between a hydraulic and an electrical circuit: a pump provides $\Delta p$ and a battery provide $\Delta V$; the former drives the flow $q$ of a fluid, while the latter drives the electrical current (``flow'' of electrons) $I$. The resistance to conduction by a real material parallels the resistance to internal flow due to the fluid's viscous forces (friction), in particular at the bounding surfaces. Of course, the physical underpinnings of each phenomenon are entirely different, nevertheless this analogy allows for the \emph{reduced-order modeling} of fluidic systems, in particular in microfluidics \cite{Wang2008,Oh12}. Kirchhoff's currents and voltage laws at circuit junctions take on the same form for flow rates and pressures where pipes meet (due to conservation of mass and energy) \cite[section~4.7]{B08}.

\begin{figure}
	\centering	
	\includegraphics[width=\columnwidth]{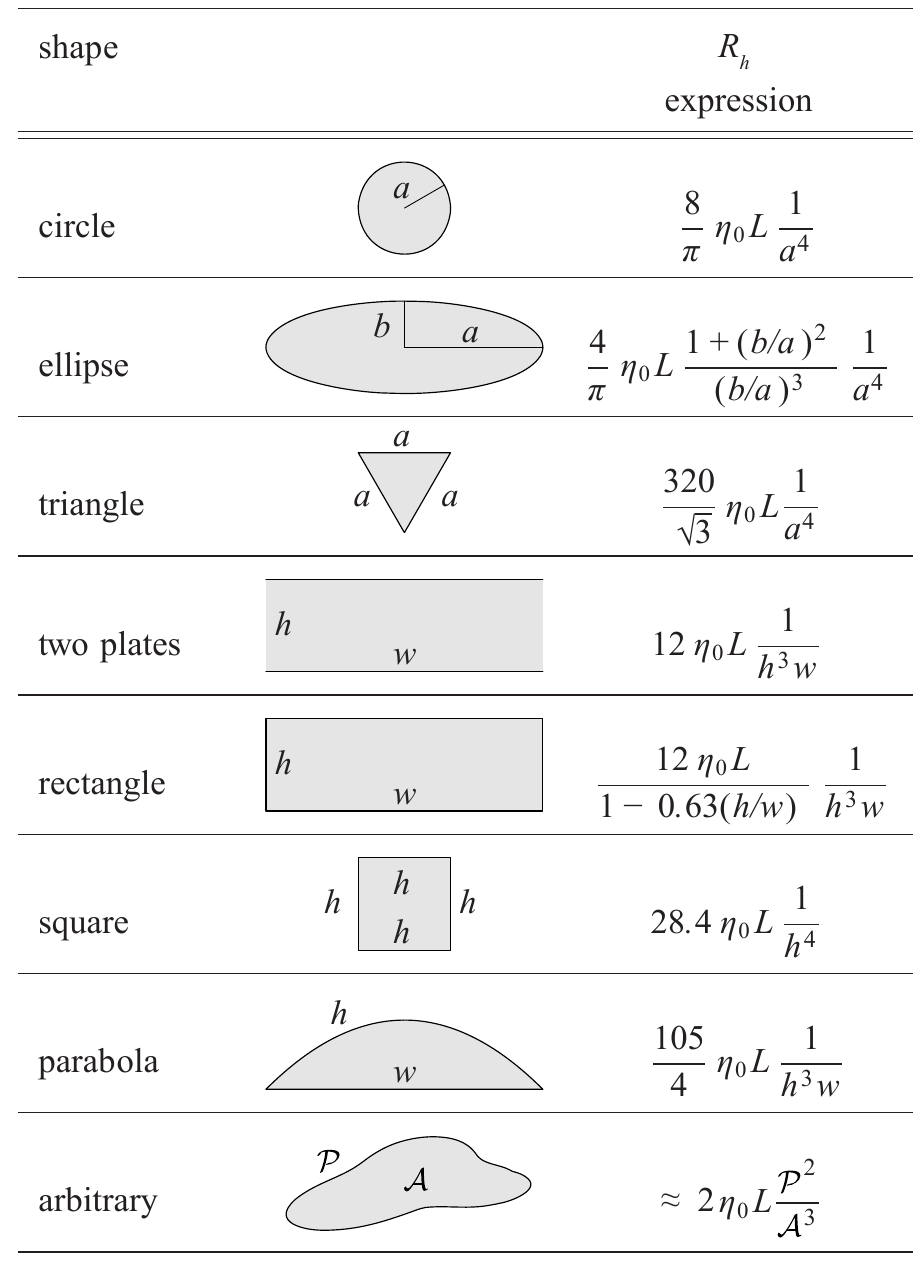}
	\caption{Analytical formulas for the hydraulic resistance $R_h$ for steady viscous flow of Newtonian fluids with \emph{constant} dynamic shear viscosity $\eta_0$ through \emph{rigid} conduits of different cross-section and fixed axial length $L$ (into the page); based on exact solutions for the fully-developed flow profile in a duct \cite[section~2-5]{HB83}. Reprinted and adapted from \cite{B08} with permission from Oxford University Press.}
	\label{fig:resistance}
\end{figure}

Now, for a \emph{rigid} pipe, the hydraulic resistance is a known function of the pipe's cross-sectional geometry. Figure~\ref{fig:resistance} shows example cross-sectional geometries of pipes, and $R_h$ calculated for each. Specifically, for the case of a rigid cylindrical pipe, (\ref{eq:dp_Rh_q}) holds with 
\begin{equation}
	R_h = \frac{8\eta_0 L}{\pi a^4},
	\label{eq:Pois}
\end{equation}
yielding the well-known \emph{Hagen--Poiseuille law} for Newtonian fluids \cite{SS93}. Further exact results are possible for a number of ``exotic'' shapes, shown in figure~\ref{fig:resistance}, by solving the basic equations of fluid mechanics exactly in unidirectional  (fully-developed) flow \cite{HB83,LSS04,W06_book,MOB05,B08}. 

But, what if the resistance were to depend on the pressure drop itself (or the flow rate)? Such behavior would correspond to a flow-responsive circuit element, \textit{i.e.}, a ``programmable'' resistor \cite{Case19}. Ajdari \cite{A04} recognized that two possible ways to generate such a nonlinear response is to employ non-Newtonian fluids and/or to allow the channel walls to deform elastically. In fact, this nonlinear behavior can be achieved in soft microfluidic devices, as demonstrated for both steady \cite{GEGJ06,HUZK09} and unsteady \cite{LESKUBL09,BULHLB09,SLHLUB09} Newtonian fluid flow. The case of a non-Newtonian working fluid has not been addressed in such detail, in part because the basic hydraulic--electric analogy, which has been so successful in understanding and designing microfluidic circuits with Newtonian fluids \cite{ZM94,A04,Oh12,B08}, requires modifications. 

In the remainder of this section, the key physics that enable a predictive theory of microscale flows of non-Newtonian fluids through compliant conduits are summarized, and it is demonstrated how to generalize (\ref{eq:dp_Rh_q}) and (\ref{eq:Pois}). Importantly, two modifications emerge: (\textit{i}) a modification of the ``hydraulic Ohm's law'' (\ref{eq:dp_Rh_q}) due to the rheology of the non-Newtonian fluid, and (\textit{ii}) a modification of the resistance $R_h$ due to the deformation of the compliant boundaries of the conduit.


\subsection{Rheological behavior of fluids}
\label{sec:rheology}

The standard textbook reference on this topic is by Bird, Armstrong and Hassager \cite{BAH87}, with other helpful textbooks by Larson \cite{L98} and Chhabra and Richardson \cite{CR08}. Owens and Phillips \cite{OP02} cover both the fundamentals of rheology and computational aspects. An exhaustive handbook entry by Nijenhuis \textit{et al.}~\cite{Netal07} describes the experimental interrogation of non-Newtonian fluids. The relevance of non-Newtonian (complex) fluids to microfludics and the basic equations of their flows are briefly discussed in encyclopedia entries by Anna \cite{A08} and Chakraborty \cite{C08}, respectively.

\begin{figure*}
	\centering
	\subfloat[][]{\includegraphics[width=0.3\textwidth]{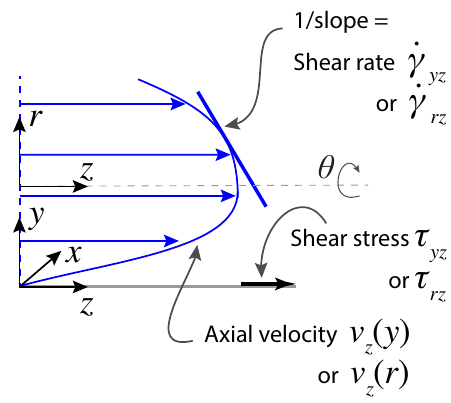}}\hfill
	\subfloat[][]{\includegraphics[height=0.275\textwidth]{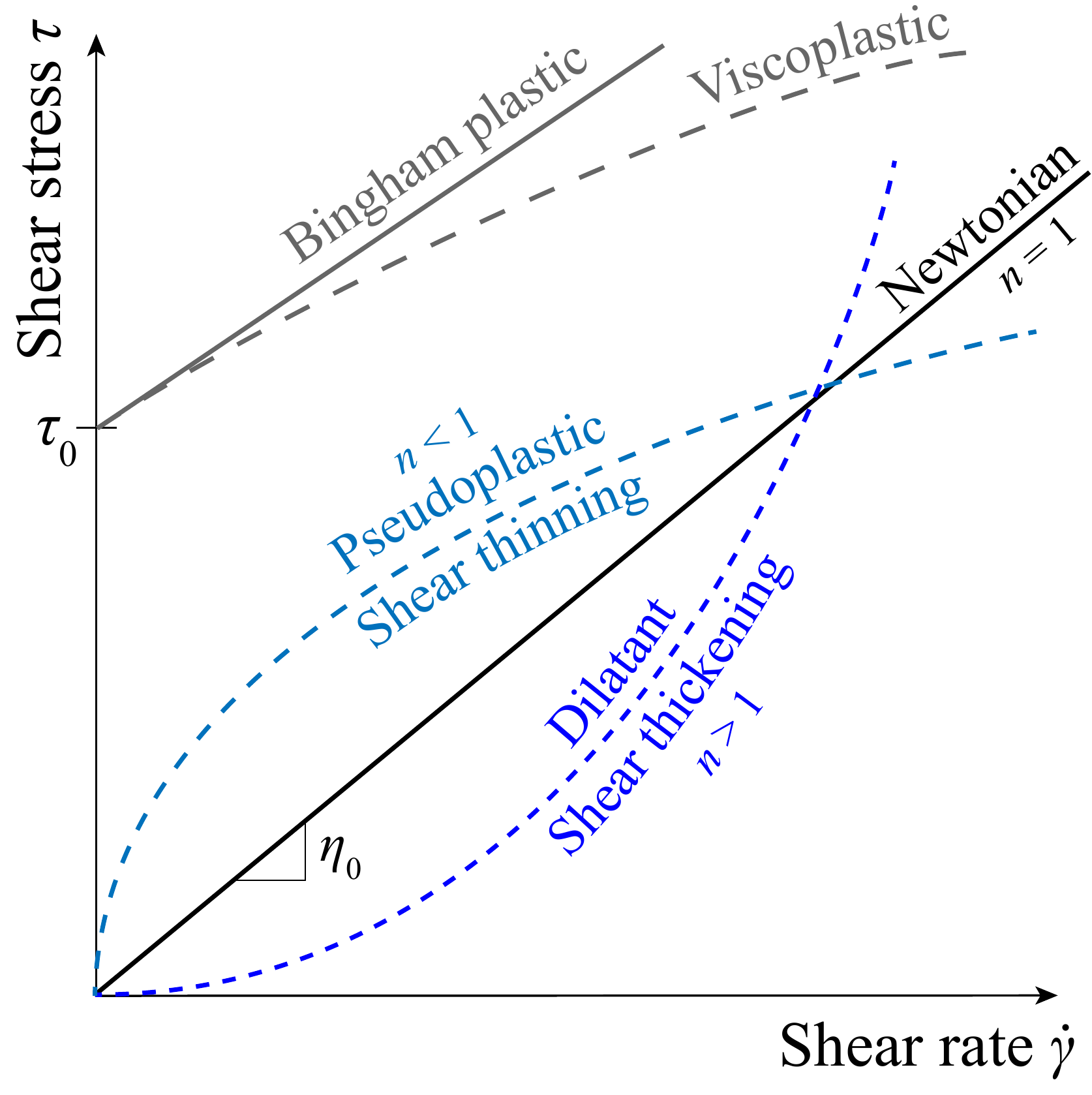}}\hfill
	\subfloat[][]{\includegraphics[width=0.35\textwidth]{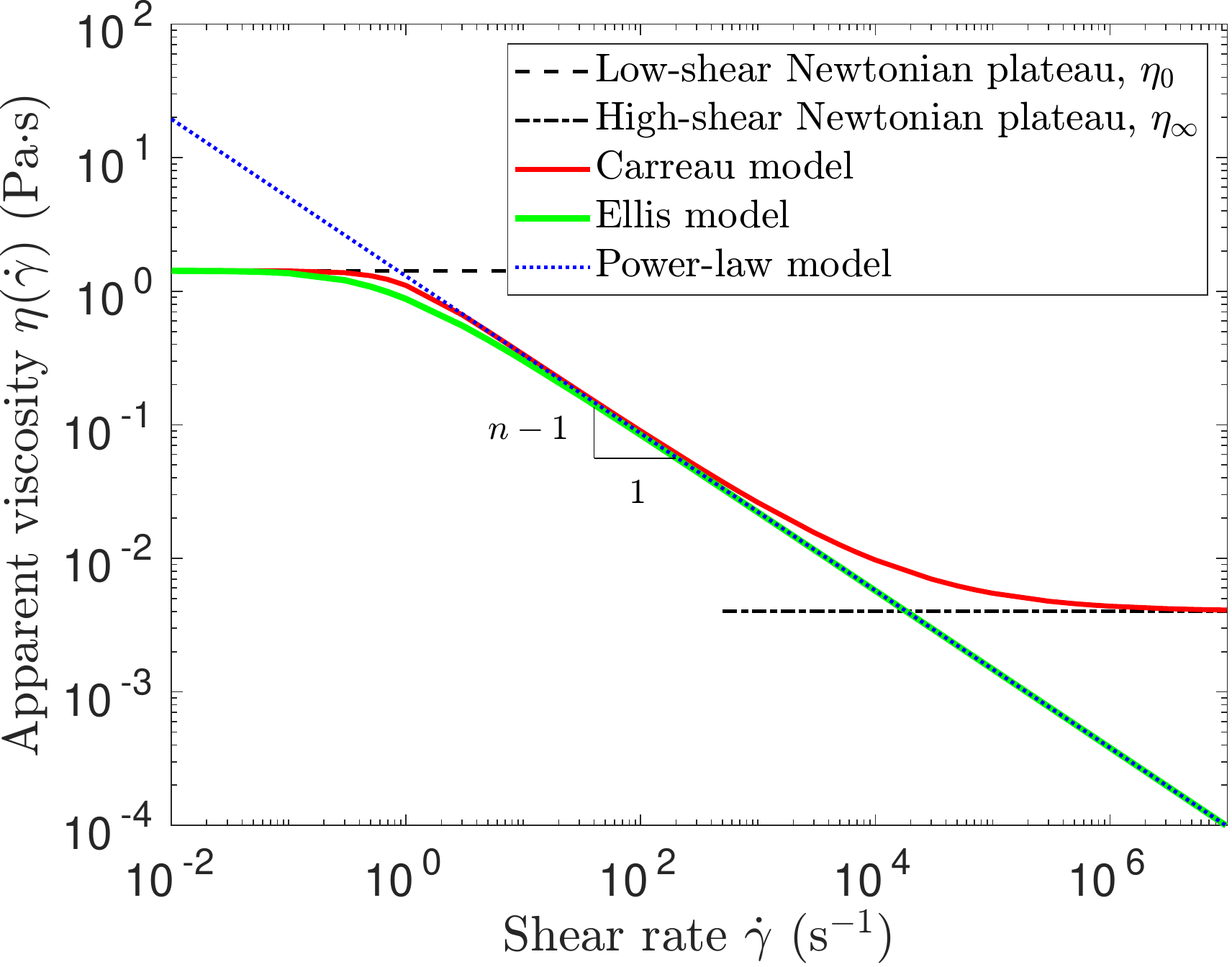}}
	\caption{(a) Sketch of a unidirectional velocity profile $v_z$, for both a Cartesian $(x,y,x)$ and a cylindrical axisymmetric $(r,\theta,z)$ coordinate system, and the attendant definition of the shear rate $\dot{\gamma}$. (b) Schematic representation of the variation of the shear stress $\tau$ with the shear rate $\dot{\gamma}$ highlighting that the effective (apparent) viscosity $\eta = \tau/\dot{\gamma} \ne const.$ for time-independent non-Newtonian fluids. (c) Quantitative plot of the three representative models for $\eta(\dot{\gamma})$ considered in the text, see table~\ref{tb:rheology} for notation, styled after a figure by Boger \cite{B77}. The rheological model parameter values used are similar to those obtained in \cite{B77} for a 0.4\% polyacrylamide solution ($\eta_0\approx1.425$ Pa$\cdot$s, $\eta_\infty\approx4\times10^{-3}$ Pa$\cdot$s, $\tau_{1/2}\approx1.211$ Pa, $n_e\approx2.43$, $n\approx0.412$, $K\approx 1.295$ Pa$\cdot$s$^n$, and $\lambda_r\approx1.177$ s). Note the logarithmic axes. The triangle indicating the slope is not to scale.}
	\label{fig:rheology}
\end{figure*}

Newton's law of viscosity states that the shear stress $\tau$ (resistance to flow and deformation) is proportional to the shear rate of strain $\dot{\gamma}$ (a measure of the deformation of fluid elements under flow) \cite{P11}. Although both $\tau$ and $\dot{\gamma}$ are, in fact, \emph{tensorial} quantities \cite{W06_book,L07,B08}, for the purposes of this subsection, they are considered to be the representative (dominant) scalar components of the respective tensors for a given flow. The proportionality \emph{constant} is the shear viscosity $\eta_0$, hence $\tau=\eta_0\dot{\gamma}$. The latter relation between shear stress and shear rate of strain is termed the \emph{constitutive equation}. Many engineering fluids (water, air, glycerol) obey Newton's law. However, in microfluidics, one deals with \emph{complex fluids}. Complex fluid are \emph{non-Newtonian}, which simply means that they do not obey Newton's law of viscosity. 

The technological focus in microfluidics has been on ``miniaturizing assays to analyze the biological, physical, and chemical properties of DNA, proteins, and biopolymers in solution, as well as suspensions of cells and bioparticles'' \cite{A08}. Nominally, when polymers, particles or cells are added to a solution, its viscosity increases. However, the stretching of initially coiled polymers (or deformation and flow-alignment of cells) under shear flow leads to a \emph{decrease} of the viscosity with shear rate (termed \emph{shear thinning}). While the shear-dependent viscosity effect may be different in extensional flows, shear flows are the most relevant class in the present context of long and thin microchannels. Further, these effects can be time-dependent (transient) as the polymers and cells relax back to equilibrium. These observations identify the critical need for understanding non-Newtonian fluid flows in microfluidics. Chip-based technologies for genomic analysis (including, but not limited, to DNA sequencing and polymerase chain reactions (PCR) detection methods) go by the name \emph{bio}microelectromechanical systems (bioMEMS) \cite{Lee06}. DNA sequencing and PCR detection have also benefited from advances in micropipetting technology, which has been impacted by new understanding of \emph{nonstandard} inkjet printers that can generate microscopic droplets of complex fluids \cite{BGB13}. 

Thus, a common complex fluid encountered in microfluidic systems is a Newtonian solvent with additives such as long-chain polymers, particles, cells or bacteria. Even a small (by percent of weight or volume) additive can drastically change the \emph{rheological} (\textit{i.e.}, flow and deformation) behavior of such a fluid. However, as noted by Chakraborty, a ``contrasting feature of the non-Newtonian constitutive behavior is a rather non-generic nature of the pertinent governing equations'' \cite{C08}. Therefore, in this subsection, several useful models for understanding the behavior of non-Newtonian fluids in microfluidics are reviewed, focusing on two key rheological behaviors of polymeric solution in flow \cite{A08}: shear-dependent viscosity and viscoelasticity.

\subsubsection{Shear-dependent viscosity at steady state}
\label{sec:eta_dotgamma}

Time-independent non-Newtonian rheological behavior is common to many complex fluids in steady shear flow. It can be accurately captured by the concept of an \emph{apparent} (or, effective) viscosity $\eta$. Generalizing Newton's law, one can write $\eta = \tau/\dot{\gamma} \ne const.$ The apparent viscosity can, in general, be a function of the shear rate $\dot{\gamma}$ (defined visually in figure~\ref{fig:rheology}(a)), namely $\eta=\eta(\dot{\gamma})$. Figure~\ref{fig:rheology}(b) shows schematically the possible ways $\tau$ might vary with $\dot{\gamma}$, highlighting that $\eta(\dot{\gamma})$ is not constant. Next, useful engineering non-Newtonian \emph{models} of shear-dependent viscosities \cite{BAH87,CR08} are summarized. The emphasis on the word `models' is to draw attention to the fact these expressions for $\eta(\dot{\gamma})$ provide reasonable (and, occasionally, excellent) agreement with experimental data, but these expressions for $\eta(\dot{\gamma})$ are not necessarily derived from first principles.

Perhaps the most common model encountered in the literature is the \emph{power-law} (also known as the Ostwald--de Waele) model:
\begin{equation}
	\eta(\dot{\gamma}) = K |\dot{\gamma}|^{n-1}.
	\label{eq:plaw}
\end{equation}
This model is not meant to be used as $\dot{\gamma} \to 0$ or $\dot{\gamma}\to \infty$, in which limits (\ref{eq:plaw}) can be singular. Depending on whether $n\gtrless1$, $\eta$ may be a concave or convex function of $\dot{\gamma}$, which corresponds to shear-thickening or shear-thinning behavior, respectively (see figure~\ref{fig:rheology}(b) for sketches of the corresponding shear stresses). Shear-thinning fluids ($n<1$) include polymeric solutions, paints, and blood. Shear-thickening fluids ($n>1$) include dense particulate suspensions and the solution of corn starch and water (sometimes referred to as ``oobleck'').

The relative simplicity of the power-law model allows a closed-form \emph{analytical} solution for the velocity profile in unidirectional flow, even when coupled to thermal and solute transport \cite{DC06}. These analytical solutions become building blocks in the theory reviewed in section~\ref{sec:lubrication}.

A better behaved model is due to Carreau:
\begin{equation}
	\eta(\dot{\gamma}) = \eta_\infty + (\eta_0-\eta_\infty) \left[ 1 + (\lambda_r\dot{\gamma})^2\right]^{(n-1)/2}.
	\label{eq:carr}
\end{equation}
Observe that, at intermediate shear rates $1\ll |\lambda_r \dot{\gamma}| < \infty$, (\ref{eq:carr}) is approximated by (\ref{eq:plaw}) (specifically, with $K=\eta_0 |\lambda_r|^{n-1}$ if $\eta_\infty=0$). Meanwhile, for $\lambda_r \dot{\gamma}\to0$ and $\lambda_r\dot{\gamma}\to\infty$, (\ref{eq:carr}) reduces to $\eta(0)=\eta_0=const.$ (a low-shear ``Newtonian plateau'') and $\eta(\infty) = \eta_\infty=const.$ (a high-shear ``Newtonian plateau''), respectively. In particular, (\ref{eq:carr}) regularizes the singularities of (\ref{eq:plaw}) for shear-thinning fluids at low shear rates. The Carreau model is meant to capture shear-thinning behavior, so typically $n$ is restricted to $0<n<1$, so $(n-1)/2<0$ in (\ref{eq:carr}). As with (\ref{eq:plaw}), (\ref{eq:carr}) reduces to the Newtonian viscosity $\eta_0$ for $n=1$.

Other useful non-Newtonian models specify the constitutive relation as shear rate in terms of shear stress. An intermediate model is that due to Ellis, which also regularizes the power-law model at low shear rates and gives the apparent viscosity in terms of the shear stress as:
\begin{equation}
	\eta(\tau) = \frac{\eta_0}{1 + (\tau/\tau_{1/2})^{n_e-1}}.
	\label{eq:ellis}
\end{equation}
Observe that, at intermediate stresses, $1\ll (\tau/\tau_{1/2})^{n_e-1} < \infty$, (\ref{eq:ellis}) is approximated by (\ref{eq:plaw}) with $n-1=(1-n_e)/n_e$ (or, $n = 1/n_e$) and $K=(\eta_0\tau_{1/2}^{n_e-1})^{1/n_e}$. Newtonian behavior is recovered as $\tau_{1/2}\to\infty$. Unlike the Carreau model, the Ellis model allows for an exact solution for the velocity profile in unidirectional flow \cite{S01,CLLDF21}. Note that (\ref{eq:ellis}) contains the effective viscosity $\eta$ on both sides (via $\tau = \eta(\dot{\gamma})\dot{\gamma}$ on the right-hand side), making this an \emph{implicit} relation for $\eta(\dot{\gamma})$, which must be solved using nonlinear root-finding.

\begin{table}
\caption{\label{tb:rheology}Constants specifying the engineering models reviewed herein for non-Newtonian fluids' rheological behavior. Typical values are quoted in the main text.}
\footnotesize
\centering
\begin{tabular}{@{\hskip 3pt}l@{\hskip 7pt}l@{\hskip 7pt}l@{\hskip 7pt}l@{\hskip 3pt}}
\br
Quantity & Notation & Units & Notes\\
\mr
Zero-shear viscosity & $\eta_0$ & Pa$\cdot$s & Newtonian viscosity,\\ &&& or $\eta(\dot{\gamma}\to0)$ as in (\ref{eq:carr}) \\[2mm]
Infinite-shear viscosity & $\eta_\infty$ & Pa$\cdot$s & $\eta(\dot{\gamma}\to\infty)$ in (\ref{eq:carr})\\[2mm]
Consistency index & $K$ & Pa$\cdot$s$^{n}$ & $K=\eta_0$ for $n=1$, see (\ref{eq:plaw}) \\[2mm]
Power-law index & $n$ & -- & $n<1$: shear thinning,\\ &&& $n>1$: shear thickening,\\ &&&see (\ref{eq:plaw})\\[2mm]
Ellis index & $n_e$ & -- & $n_e=1/n$, see (\ref{eq:ellis})\\[2mm]
Half-viscosity stress & $\tau_{1/2}$ & Pa & $\eta(\tau_{1/2}) = \eta_0/2$, see (\ref{eq:ellis}) \\[2mm]
Yield stress & $\tau_0$ & Pa & see (\ref{eq:casson})\\[2mm]
Time constant & $\lambda_r$ & s & context dependent,\\ &&&see (\ref{eq:carr}) and (\ref{eq:j-o-2})\\
\br
\end{tabular}
\end{table}

These three representative models (power-law, Carreau, and Ellis) for the shear-dependent viscosity of non-Newtonian fluids are illustrated in figure~\ref{fig:rheology}(c) and their parameters are summarized in table~\ref{tb:rheology}.

Some complex fluids exhibit a finite \emph{yield stress} at zero shear rate, understood as $\tau_0= \lim_{\dot{\gamma}\to0} \eta(\dot{\gamma})\dot{\gamma}$. In other words, these fluids do not begin to flow until $\tau_0$ is exceeded by the applied forces. Fluids with yield stress are termed \emph{viscoplastic} (see figure~\ref{fig:rheology}(b)). Viscoplastic materials do not have to be fluids; for example, meat has a finite yield stress \cite[section~1.3.2]{CR08} beyond which it deforms continuously under shear. The mechanical origin of the yield stress remains a topic of active research \cite{BFO14}. A non-Newtonian fluid model exhibiting both a yield stress and a shear-dependent viscosity is the Casson model \cite{CR08,F93,A16}:
\begin{equation}
\eta(\dot{\gamma}) = 
	\left\{\begin{array}{@{}l@{\quad}l}
     	(|\tau_0/\dot{\gamma}|^{1/2} + |\eta_0|^{1/2})^2, &|\tau|>|\tau_0|, \\[2mm]
    	\mathrm{undefined}/\mathrm{no~flow},  &|\tau|<|\tau_0|.
    \end{array}\right.
\label{eq:casson}
\end{equation}
Blood rheology is often fitted to the Casson model \cite{F93,A16}. Unidirectional flow exact solutions are possible under the Casson model~(\ref{eq:casson}). However, $\tau_0<1$ Pa for blood, a threshold easily exceeded even in microfluidic flows. Therefore, (\ref{eq:plaw}) (without a yield stress) is used in practice to capture the shear-thinning behavior of blood because the yield stress has little influence on blood flow under dynamic conditions \cite{Ch05,C08}. Beyond microflows, ``whole'' blood can even be considered to be a Newtonian fluid at sufficiently high shear rates (such as for flows in arteries) \cite[chapter~3]{F93}.

\subsubsection{Viscoelastic fluids: the relaxation time}
\label{sec:viscoelastic}

Time-dependent rheological behavior of complex fluids requires studying \emph{viscoelasticity}. Viscoelastic fluids in microfluidics were discussed in detail in \cite{SQ05,C08}. Although, in this review, the focus is on the effect of shear-dependent viscosity in steady flow, it is nevertheless instructive to introduce the basic concept of ``extra'' (polymeric) stress $\tau_\mathrm{extra}$ and its relaxation \cite[section~2.6.1]{OP02}. Now, a ``constitutive equation'' can be posited as
\begin{eqnarray}
	\tau = \eta_s \dot{\gamma} + \tau_\mathrm{extra}, \label{eq:j-o}\\
	\tau_\mathrm{extra} + \lambda_r \frac{\partial \tau_\mathrm{extra}}{\partial t} = \eta_p \dot{\gamma}, \label{eq:j-o-2}
\end{eqnarray}
where $\eta_s$ is the solvent viscosity, and $\eta_p$ is the polymeric viscosity, such that $\eta_0=\eta_s+\eta_p$ is the zero-shear viscosity of the complex fluid mixture. It should be emphasized that (\ref{eq:j-o}) and (\ref{eq:j-o-2}), being a unidirectional (scalar) description, are necessarily approximate. The \emph{relaxation time} $\lambda_r$ in (\ref{eq:j-o-2}) quantifies the exponential return to equilibrium of the extra stress due to, \textit{e.g.}, a step change in $\dot{\gamma}$. Equation~(\ref{eq:j-o}) is generally valid for a \emph{dilute} polymeric solution such that $\eta_p/\eta_s<1$.

The relaxation time $\lambda_r$ captures the time scale of the evolution of the non-Newtonian fluid's microstructure (\textit{e.g.}, the stretching of flexible polymeric chains suspended in the solvent fluid). One may identify $\lambda_r$ with $\eta_p/G$, where $G$ is a shear modulus of elasticity for the polymers \cite{BAH87}, highlighting why these fluids are called viscoelastic. At steady state, $\partial(\,\cdot\,)/\partial t=0$ and (\ref{eq:j-o})--(\ref{eq:j-o-2}) reduce to Newton's law of viscosity.

By eliminating $\tau_\mathrm{extra}$ between (\ref{eq:j-o}) and (\ref{eq:j-o-2}), the constitutive relation can be reduced to the Jeffreys model \cite[section~5.2(b)]{BAH87}:
\begin{equation}
	\tau + \lambda_r \frac{\partial \tau}{\partial t} = \eta_0\left(\dot{\gamma} + \hat{\lambda}_r\frac{\partial \dot{\gamma}}{\partial t}\right),\qquad \hat{\lambda}_r = \frac{\eta_s}{\eta_0}\lambda_r,
\label{eq:j-o-1}
\end{equation}
where $\hat{\lambda}_r$ is termed the \emph{retardation} time \cite{OP02,BAH87}. A number of relations like (\ref{eq:j-o-1}) can be derived from the general principle that the integrated time-history of the stress and that of the rate of strain are linearly related \cite{BAH87,CC16}. The tensorial generalization of the Jeffreys model is the Oldroyd-B model \cite[section~7.2]{BAH87}. This model is special in the sense that it can be justified by the molecular theory of complex fluids \cite{OP02,L98}. 

Oliveira \textit{et al.}~\cite{OAP11} reviewed viscoelastic fluid flows in microfluidics, including more general \emph{nonlinear} rheological models of this type. These nonlinear models allow for the consideration of viscoelastic effects in steady flow; recall that stress relaxation drops out of (\ref{eq:j-o})--(\ref{eq:j-o-2}) at steady state. Many nonlinear viscoelastic models exists, such as those by Phan-Thien and Tanner and by Giesekus described in textbooks \cite{OP02,L98}. The key point is that a nonlinear function(al) of $\tau_\mathrm{extra}$ and $\dot{\gamma}$ appears on the left-hand side of (\ref{eq:j-o-2}).

As mentioned in section~\ref{sec:hydrocirc}, non-Newtonian (complex) fluid rheology introduces complications in the hydraulic--electric circuit analogy, in part because the flow rate--pressure drop characteristics of steady viscoelastic flows at the microscale are not completely understood (see, \textit{e.g.}, the discussions in \cite{RM01,KSAK09}). Beyond the work of Ramos-Arzola and Bautista \cite{RAB21} using the simplified Phan-Thien-Tanner (sPTT) constitutive equation, it appears that no other recent studies have investigated steady nonlinear viscoelastic fluid flows in soft hydraulic conduits.


\subsection{Lubrication approximation}
\label{sec:lubrication}

Next, the theory of the flow within microscale conduits, used to calculate the so-called \emph{soft hydraulic resistance}, is reviewed. Flows in microchannels are \emph{laminar}, and the Reynolds number
\begin{equation}
	Re = \frac{\mathrm{flow~inertia~forces}}{\mathrm{flow~viscous~forces}} \simeq \frac{\rho \mathscr{V}_c^2}{\eta_0 \mathscr{V}_c/h_0} = \frac{\rho \mathscr{V}_c h_0}{\eta_0}
	\label{eq:reynolds}
\end{equation}
is expected to be small (at least, less than unity) \cite{SSA04}. Here, $\rho$ is the density of the fluid. Observe that $\eta_0 \mathscr{V}_c/h_0 \sim \eta_0\dot{\gamma}_c\sim\mathscr{T}_c$ is the characteristic viscous shear force per area (stress), while $\rho \mathscr{V}_c^2$ is the characteristic inertia force per area.

Additionally, microchannels are \emph{long} (height $h_0\ll$ length $L$) and \emph{shallow} (height $h_0\ll$ width $w$) leading to the so-called \emph{lubrication approximation} \cite[chapter~5]{L07}. In this case, it is more appropriate to define $\hat{Re} = (h_0/L)Re$. For example, water ($\eta_0\approx10^{-3}$ Pa$\cdot$s, $\rho\approx 1000$ kg/s \cite{B08}) flowing in a typical microchannel of height $h_0 \approx 25$ $\mu$m \cite{GEGJ06,CCSS18} at $\mathscr{V}_c = 4\times 10^{-2}$ m/s ($\Rightarrow q = \mathscr{V}_c h_0 w = 15$ $\mu$L/min) yields $Re \approx 1$. However, taking into account that the length of the microchannel is on the order of centimeters, $h_0/L\approx 10^{-2}$ \cite{GEGJ06,CCSS18}, it turns out that $\hat{Re} \approx 10^{-2}\ll1$.

\begin{table*}
\caption{\label{tb:quant}Key quantities that describe the physics of a lubrication flow of a non-Newtonian fluid in a compliant conduit. The main functional dependence for non-constant quantities is stated based on the source equation. Notation is as per figure~\ref{fig:rheology}(a).}
\footnotesize
\centering
\begin{tabular}{lllll}
\br
Quantity & Notation & Units & From & Lubrication scaling(s)\\
\mr
Shear stress & $\tau_{yz}$ or $\tau_{rz}$ & Pa & $\tau$ in section~\ref{sec:eta_dotgamma} & $\mathscr{T}_c \sim (h_0/L)\mathscr{P}_c$ or $(a/L)\mathscr{P}_c$; also $\mathscr{T}_c \sim \eta_0\dot{\gamma}_c$ \\
Pressure & $p(z)$ & Pa & (\ref{eq:RK_dpdz_NN}) & $\mathscr{P}_c$; related to $\mathscr{V}_c$ via (\ref{eq:mom})\\
Pressure gradient & $\mathrm{d}p/\mathrm{d}z$ & Pa/m & (\ref{eq:lubrication}), (\ref{eq:mom}), (\ref{eq:RK_dpdz_NN}) & $\mathscr{P}_c/L$\\
Axial velocity & $v_z(y)$ or $v_z(r)$ & m/s & (\ref{eq:mom}) & $\mathscr{V}_c$; related to $\mathscr{P}_c$ via (\ref{eq:mom})\\
Shear rate & $\dot{\gamma}_{yz}$ or $\dot{\gamma}_{rz}$ & 1/s & $\partial v_z/\partial y$ or $\partial v_z/\partial r$ & $\dot{\gamma}_c \sim \mathscr{V}_c/h_0$ or $\mathscr{V}_c/a$ \\
Volumetric flow rate & $q$ & m\textsuperscript{3}/s & (\ref{eq:q_defn}) & $h_0w\mathscr{V}_c$ or $\pi a^2 \mathscr{V}_c$; related to $\mathscr{P}_c$ via (\ref{eq:RK_dpdz_NN})\\
Wall deformation & $u_y(x,z)$ or $u_r(z)$ & m & figure~\ref{fig:sketch-geom}, (\ref{eq:uy_2D}), (\ref{eq:ur_tube}), (\ref{eq:ur_shell}), (\ref{eq:uy_3D}) & $\mathscr{U}_c$; related to $\mathscr{P}_c$ via $\beta$, see (\ref{eq:beta_tube}) or (\ref{eq:beta_chan}) \\
\br
\end{tabular}
\end{table*}

It is interesting to note that lubrication theory actually dates back to work by Osborne Reynolds \cite{R86}, which was contemporaneous with his studies on instability and flow transition \cite{R83} (see also \cite{JL07}).

Therefore, neglecting body forces, for $Re\to0$, the governing equations for the flow \cite{BAH87} are
\begin{equation}
	\underbrace{\bm{\nabla}\bm{\cdot}\bm{\tau}}_{\mathrm{viscous~forces}} = \underbrace{\bm{\nabla}p}_{\mathrm{pressure~gradient}},\quad \underbrace{\bm{\nabla}\bm{\cdot}\bm{v}=0}_{\mathrm{mass~conservation}}.
	\label{eq:stokes}
\end{equation}
However, (\ref{eq:stokes}) is still a 3D system, and $\bm{\tau}$ must be computed from the velocity field $\bm{v}$. A key simplification comes from the lubrication approximation ($\hat{Re}\ll1$), under which  (\ref{eq:stokes}) reduces to
\begin{equation}
	\bm{\nabla}_\perp\bm{\cdot}\bm{\tau} \approx \frac{\mathrm{d}p}{\mathrm{d}z} \bm{e}_z,
	\label{eq:lubrication}
\end{equation}
where $z$ is the axial (longest) direction, and $\bm{e}_z$ is the unit normal vector in the $z$-direction. In (\ref{eq:lubrication}), $\bm{\nabla}_\perp$ is the gradient in the plane perpendicular to the flow (\textit{i.e.}, the $(x,y)$ plane for a Cartesian geometry, or the $(r,\theta)$ plane in a cylindrical coordinate system, see figure~\ref{fig:rheology}(a)). 

A consequence of (\ref{eq:lubrication}) is that shear stresses (tangential fluid forces) are asymptotically smaller than the pressure (normal fluid forces) in the flow. This fact follows from a scaling analysis of (\ref{eq:lubrication}). Specifically, noting that $h_0$ and $L$ are, respectively, the cross-sectional and flow-wise length scales, $\mathscr{T}_c/h_0 \sim \mathscr{P}_c/L$, where $\mathscr{P}_c$ is the characteristic pressure scale (to be discussed in detail below). Thus, for a long and shallow microchannel, $\mathscr{T}_c \sim (h_0/L) \mathscr{P}_c$ with $h_0/L\ll1$. A more detailed discussion, also taking into account the width $w$ of the conduit, can be found in \cite{CCSS18}. Likewise, for a long and slender microtube of radius $a$, $\mathscr{T}_c \sim (a/L) \mathscr{P}_c$ with $a/L\ll1$.

In this summary of lubrication theory, the exposition of Stone \cite{Stone2017} on soft interface problems is followed. For a Newtonian fluid, $\bm{\nabla}_\perp\bm{\cdot}\bm{\tau} = \eta_0\nabla_\perp^2 \bm{v}_\|$ in (\ref{eq:lubrication}), where $\bm{v}_\|$ is the velocity parallel to the deformable channel wall, and $\nabla_\perp^2$ is the Laplacian operator in the coordinates perpendicular to the flow. In the present notation, $\bm{v}_\| = v_z \bm{e}_z$ is the axial velocity. For a non-Newtonian fluid, however, the constitutive equation (for a time-independent rheology, as in section~\ref{sec:eta_dotgamma}, neglecting viscoelasticity) is $\bm{\tau}=2\eta(\dot{\gamma})\bm{E}$, where $\bm{E}=\frac{1}{2}(\bm{\nabla}\bm{v} + \bm{\nabla}\bm{v}^\top)$ and $\dot{\gamma} = \sqrt{2\bm{E}\bm{:}\bm{E}}$ in tensorial form. In general, further scaling analysis is needed to determine the dominant components of $\bm{E}$ \cite{BBG17,AC19c}. Nevertheless, it can be shown that (\ref{eq:lubrication}) becomes
\begin{equation}
	\bm{\nabla}_\perp\bm{\cdot}\bm[\eta(\dot{\gamma})\bm{\nabla}_\perp v_z] = \frac{\mathrm{d}p}{\mathrm{d}z},
	\label{eq:mom}
\end{equation}
where the gradient operator is
\begin{equation}
\bm{\nabla}_\perp A =
	\left\{\begin{array}{@{}l@{\quad}l}
     	\frac{\partial A}{\partial x} \bm{e}_x + \frac{\partial A}{\partial y}\bm{e}_y &(\mathrm{Cartesian}), \\[3mm]
    	\frac{\partial A}{\partial r}\bm{e}_r &(\mathrm{axisymmetric}),
    \end{array}\right.
    \label{eq:grad_perp}
\end{equation}
and the divergence operator is
\begin{equation}
\bm{\nabla}_\perp\bm{\cdot}\bm{A} =
	\left\{\begin{array}{@{}l@{\quad}l}
     	\frac{\partial (\bm{A}\bm{\cdot}\bm{e}_x)}{\partial x} + \frac{\partial (\bm{A}\bm{\cdot}\bm{e}_y)}{\partial y} &(\mathrm{Cartesian}), \\[3mm]
    	\frac{1}{r}\frac{\partial(r \bm{A}\bm{\cdot}\bm{e}_r)}{\partial r} &(\mathrm{axisymmetric}).
    \end{array}\right.
    \label{eq:div_perp}
\end{equation}

Equation~(\ref{eq:mom}) describes an (almost) unidirectional flow profile $v_z$ being driven by an axial pressure gradient $\mathrm{d}p/\mathrm{d}z$, where $p$ is independent of the cross-sectional coordinates ($x$, $y$, $r$, or $\theta$). Therefore, at the leading order in the conduit slenderness ($h_0/L$), conservation of mass is automatically satisfied. Strictly speaking a unidirectional flow profile $v_z$ cannot depend on $z$ because the convective acceleration must vanishes identically: $\bm{v}\cdot\bm{\nabla}\bm{v}=\bm{0}$ \cite[chapter~3]{L07}. However, under the lubrication approximation the (almost) unidirectional flow profile $v_z$ is allowed to vary with $z$ implicitly through the deformation of the conduit's cross-section. This ``slow variation'' \cite{VD87} is introduced by the coupling (provided by $p$) of flow and deformation, as justified rigorously via perturbation expansions in \cite{CCSS18,WC19,EG14,AC19c}.

The key quantities that one needs to determine from the physics of the problem, under the lubrication approximation, are summarized in table~\ref{tb:quant}.


\subsection{Flow rate--pressure gradient relations}
\label{sec:q_dpdz}

In solving hydraulics problems, one seeks to relate the volumetric flow rate $q$ through the conduit to the driving forces represented by the hydrodynamic pressure gradient $\mathrm{d}p/\mathrm{d}z$. To this end, the flow rate is evaluated using its definition \cite{B08} as the integral of the velocity over a cross-sectional area $\mathcal{A}$ (possibly deformed): 
\begin{equation}
	q = \int\!\!\int_\mathcal{A} \bm{v}\bm{\cdot}\bm{n} \,\mathrm{d}A = \int\!\!\int_\mathcal{A} v_z \,\mathrm{d}A,
	\label{eq:q_defn}
\end{equation}
for flows primarily in the $z$ directions, and cross-sections perpendicular to $\bm{n}=\bm{e}_z$. For example, for a cross-section defined in Cartesian coordinates $\mathrm{d}A=\mathrm{d}y\,\mathrm{d}x$. 
From (\ref{eq:q_defn}), it is convenient to define the cross-sectionally-averaged axial velocity:
\begin{equation}
	\langle v_z \rangle = \frac{1}{\mathcal{A}} \int\!\!\int_\mathcal{A} v_z \,\mathrm{d}A  = \frac{q}{\mathcal{A}}.
	\label{eq:avg_vz}
\end{equation}
When (\ref{eq:mom}) can be solved analytically for $v_z$, $q$ can be evaluated from (\ref{eq:q_defn}). 

Based on lubrication theory (as in section~\ref{sec:lubrication}), as early as 1972, Rubinow and Keller \cite{RK72} hypothesized that, for a \emph{Newtonian fluid}, the result of performing the integration in (\ref{eq:q_defn}) would take the form
\begin{equation}
	-\frac{\mathrm{d}p}{\mathrm{d}z} \mathfrak{G}(p) =  q.
	\label{eq:RK_dpdz}
\end{equation}
Here, $\mathfrak{G}$ is determined by the local cross-sectional geometry of the flow conduit. The possible expansion of the conduit's boundaries by the hydrodynamic pressure is accounted for by the dependence of $\mathfrak{G}$ on $p$ (and only $p$ because shear stresses are negligible, as discussed in section~\ref{sec:lubrication}). For a non-Newtonian fluid, the shear-dependent viscosity necessitates that $\mathfrak{G}$ also depend on the pressure gradient, thus (\ref{eq:RK_dpdz}) must be generalized to
\begin{equation}
	-\frac{\mathrm{d}p}{\mathrm{d}z} \mathfrak{G}\left(p,\frac{\mathrm{d}p}{\mathrm{d}z}\right) =  q.
	\label{eq:RK_dpdz_NN}
\end{equation}

In both cases, $\mathfrak{G}$ must be determined by a \emph{detailed analysis} of the \emph{coupled} flow (see sections~\ref{sec:flow_tube} and \ref{sec:flow_channel}) and deformation (see section~\ref{sec:deformation}) problems. A relationship such as (\ref{eq:RK_dpdz_NN}) has also been interpreted as a \emph{generalized Darcy law} for flow in a deformable porous medium \cite{IMP08,PRBM20,CLLDF21}, for which $k_h=\eta_0\mathfrak{G}/\mathcal{A}$ would be a soft hydraulic permeability.

For the special case of Newtonian viscous flow in a rigid conduit, (\ref{eq:RK_dpdz}) and (\ref{eq:RK_dpdz_NN}) both reduce to
\begin{equation}
	-\frac{\mathrm{d}p}{\mathrm{d}z} \frac{L}{R_h} = q.
	\label{eq:ridig_newt_dpdz}
\end{equation}
It should now be clear how $R_h$ (recall figure~\ref{fig:resistance}) comes about from the cross-sectional geometry. 
As $q=const.$ in steady flow, (\ref{eq:ridig_newt_dpdz}) requires that $\mathrm{d}p/\mathrm{d}z=const.$, in particular one can write $\mathrm{d}p/\mathrm{d}z=-\Delta p/L$ \cite{B08,W06_book}. Then, the hydraulic ``Ohm's law'' (\ref{eq:dp_Rh_q}) follows. 

In the presence of flow-induced deformation of the conduit, (\ref{eq:RK_dpdz_NN}) is, in the most general case, a \emph{nonlinear} first-order ODE. Depending on the non-Newtonian rheological model, this ODE might be \emph{separable} (\textit{i.e.}, $\mathrm{d}p/\mathrm{d}z$ can be isolated on one side of the equation), in which case the ODE can be solved analytically for $p(z)$ (sometimes only implicitly). Even if the integration must be performed numerically (which is straightforward for such an ODE), it yields an implicit algebraic relation between pressure drop and the flow rate:
\begin{equation}
	\mathfrak{F}(\Delta p,q) = 0,
	\label{eq:soft_dp_q}
\end{equation}
in lieu of (\ref{eq:dp_Rh_q}).

In summary, the most important physical consequence of the fluid--structure interaction between the flow and the compliant wall is that the cross-sectional area varies along the flow-wise direction, $z$. In particular, an \emph{increase} in cross-sectional area $\mathcal{A}$ allows a steady flow rate $q$ to be maintained with a \emph{smaller} average axial velocity $\langle v_z \rangle$ (by (\ref{eq:avg_vz})), or \textit{vice versa} (which is the flow-control mechanism used in the celebrated ``Quake valve'' \cite{UCTSQ00}). Importantly, when  (\ref{eq:soft_dp_q}) can be resolved for $\Delta p$ in terms of $q$, then it is generally expected that the resulting soft hydraulic resistance $R_h = R_h(\Delta p)$. Deriving analytical expressions for $R_h(\Delta p)$, and obtaining a ``generalized Ohm's law'' for soft resistors, has been the goal of a number of recent studies \cite{CCSS18,SC18,ADJRC19,AC19b,AC19c,WC19,AMC20}.

Next, examples are given of how to determine the relation (\ref{eq:RK_dpdz_NN}) in two common geometries, using the exactly solvable power-law and Ellis rheological models from section~\ref{sec:eta_dotgamma}.

\subsubsection{Microtubes/micropipes}
\label{sec:flow_tube}

Consider the axisymmetric cylindrical microtube/micropipe configuration depicted in figure~\ref{fig:sketch-geom}(b,c). Denote by $R$ the deformed radius, while $a$ is the undeformed radius. Equations~(\ref{eq:mom}) and (\ref{eq:q_defn}) can be solved to obtain the version of (\ref{eq:RK_dpdz_NN}) for the power-law model of shear viscosity~(\ref{eq:plaw}) \cite{CR08,AC19c}:
\begin{equation}
	-\frac{\mathrm{d}p}{\mathrm{d}z}\left[\frac{\pi R^{3+1/n}}{2^{1/n}(3+1/n)K^{1/n}}\left|\frac{\mathrm{d}p}{\mathrm{d}z}\right|^{(1/n)-1}\right]= q,
	\label{eq:vz_tube_plaw}
\end{equation}
as well as the Ellis model~(\ref{eq:ellis}) \cite{CR08}:
\begin{eqnarray}
	-\frac{\mathrm{d}p}{\mathrm{d}z}\left[ \frac{\pi R^{4}}{8\eta_0} + \frac{\pi R^{3+n_e}}{2^{n_e}(3+n_e)\eta_0\tau_{1/2}^{n_e-1}}\left|\frac{\mathrm{d}p}{\mathrm{d}z}\right|^{n_e-1}\right]= q. \nonumber\\
	\label{eq:vz_tube_ellis}
\end{eqnarray}

For $n=1$, (\ref{eq:vz_tube_plaw}) reduces to (\ref{eq:ridig_newt_dpdz}) with $R_h$ given by (\ref{eq:Pois}). For $n_e=1$, (\ref{eq:vz_tube_ellis}) also reduces to the latter Newtonian relation, but with shear viscosity $\eta_0/2$.

\subsubsection{Microchannels}
\label{sec:flow_channel}
Consider the two-dimensional (2D) configuration depicted in figure~\ref{fig:sketch-geom}(a) with width $w$ into the page. Equations~(\ref{eq:mom}) and (\ref{eq:q_defn}) can be solved to obtain the version of (\ref{eq:RK_dpdz_NN}) for the power-law model of shear viscosity~(\ref{eq:plaw}) \cite{ADJRC19}:
\begin{equation}
	-\frac{\mathrm{d}p}{\mathrm{d}{z}}\left[\frac{h^{2+1/n}w}{2^{1+1/n}(2+1/n)K^{1/n}}\left|\frac{\mathrm{d}p}{\mathrm{d}{z}}\right|^{(1/n)-1}\right] = q,
	\label{eq:vz_chan_plaw}
\end{equation}
as well as the Ellis model~(\ref{eq:ellis}) \cite{S01,CLLDF21}:
\begin{eqnarray}
	-\frac{\mathrm{d}p}{\mathrm{d}{z}}\left[\frac{h^3w}{12\eta_0} + \frac{h^{2+n_e}w}{2^{1+n_e}(2+n_e)\eta_0\tau_{1/2}^{n_e-1}}\left|\frac{\mathrm{d}p}{\mathrm{d}{z}}\right|^{n_e-1}\right] = q. \nonumber\\
	\label{eq:vz_chan_ellis}
\end{eqnarray}

For $n=1$, (\ref{eq:vz_chan_plaw}) reduces to (\ref{eq:ridig_newt_dpdz}) with $R_h$ given by the ``two plates'' expression from figure~\ref{fig:resistance}. 
For $n_e=1$, (\ref{eq:vz_chan_ellis}) also reduces to the latter Newtonian relation, but with shear viscosity $\eta_0/2$.

For a rigid hydraulic conduit, $h=h_0$ and $R=a$ are known geometric constants in (\ref{eq:vz_tube_plaw})--(\ref{eq:vz_chan_ellis}). For a soft hydraulic conduit, however, flow-induced deformation makes $h$ and $R$ functions of $p$. This relationship, which is needed to complete the theory, is reviewed next.


\begin{figure}
	\centering
	\includegraphics[width=0.8\columnwidth]{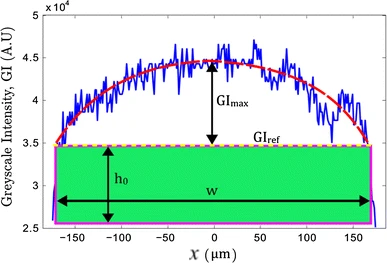}
	\caption{Example fluorescent microscopy measurement of the bulging of a microchannel's cross-section in the $(x,y)$ plane at fixed axial location $z$ (out of the page). The noisy curve is the measurement, and a normalized, filtered grayscale intensity (GI) is calibrated and mapped to a deformed channel height $h(x,z)$. 
Reprinted by permission from Springer Nature's \textit{Microfluid.\ Nanofluid.}\ \cite{RDC17} \copyright\ 2017.}
	\label{fig:bulge_expt}
\end{figure}

\subsection{Deformation--pressure relations}
\label{sec:deformation}

Flow conduits manufactured from soft polymeric materials deform due to the transmural pressure difference caused by the flow within \cite{GEGJ06,HUZK09}, bulging out when the external pressure is lower than the internal one. When dealing with complex fluids and a PDMS-based elastomer, care must be taken that the PDMS does not uptake fluid (such as a mineral oil for which it has affinity) from the channel, because such infiltration can lower the PDMS' Young's modulus by a factor of two over the course of hours \cite{Hunter20}. Figure~\ref{fig:bulge_expt} shows an example fluorescence microscopy measurement of the deformation of an initially rectangular cross-section of a microchannel due to the flow within it. 

\begin{figure}
	\centering	
	\includegraphics[width=\columnwidth]{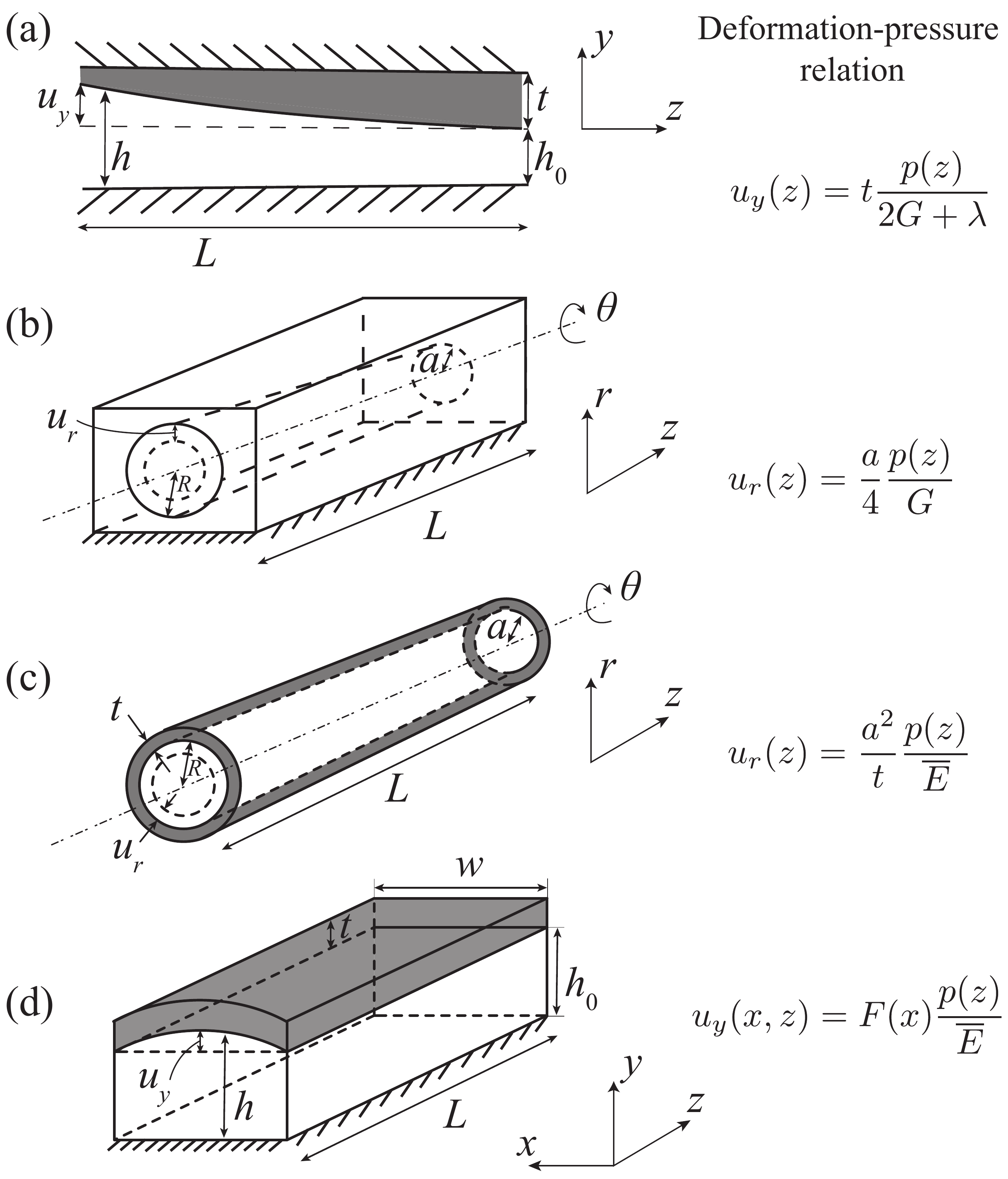}
	\caption{Building blocks for soft hydraulic resistance calculations. Four model geometries are shown schematically in which the flow-induced deformation can be analyzed mathematically, yielding a fitting-parameter-free deformation--pressure relationship.}
	\label{fig:sketch-geom}
\end{figure}

Under the theory of linear elasticity, the deformation $u_y(z)$ (or $u_r(z)$) is expected to be proportional to $p(z)$ scaled by geometric factors ($a$, $t$, $h_0$, etc.) and elasticity constants (table~\ref{tb:elasticity}) \cite{HKO09}. Conventionally, the strain ($u_y(z)/h_0$ or $u_r(z)/a$) is taken to be $\propto p(z)/E$ (see, \textit{e.g.}, \cite{GEGJ06}). The idea is that the proportionality constant can be calibrated from experiments \cite{GEGJ06,HUZK09}. However, a predictive model requires calculating the proportionality constant from the governing equations of elasticity. After initial attempts \cite{RS16,RDC17}, our current understanding has settled on a set of canonical relations, illustrated in figure~\ref{fig:sketch-geom}, between $u_y(z)/h_0$ (or $u_r(z)/a$) and $p(z)/E$, with the proportionality factor having been determined by solving a suitable elasticity problem. Importantly, this observation that the deformation at any fixed-$z$ cross-section depends only on the local hydrodynamic pressure $p(z)$, while the axial bending and tension (as well as the fluid shear stresses) are negligible, has been justified by perturbation methods \cite{CCSS18,WC19,AMC20,EG14,AC19c} within the long-and-shallow-conduit scaling that leads to the lubrication approximation.

\begin{table}
\caption{\label{tb:elasticity}Elasticity constants and their interrelations. Typical values are quoted in the main text.}
\centering
\footnotesize
\begin{tabular}{@{\hskip 3pt}l@{\hskip 9pt}l@{\hskip 9pt}l@{\hskip 9pt}l@{\hskip 3pt}}
\br
Quantity & Notation & Units & Definition\\
\mr
Young's modulus & $E$ & Pa & --\\[2mm]
Poisson ratio & $\nu$ & -- & --\\[2mm]
Plane-strain $E$ & $\overline{E}$ & Pa & $\frac{E}{(1-\nu^2)}$\\[2mm]
Shear modulus & $G$ & Pa & $\frac{E}{2(1+\nu)}$\\[2mm]
First Lam\'e parameter & $\lambda$ & Pa & $\frac{E\nu}{(1+\nu)(1-2\nu)}$\\[2mm]
Compliance & $\mathcal{C}$ & m/Pa & $\frac{t}{2G+\lambda}$, $\frac{a}{4G}$, $\frac{a^2}{t\overline{E}}$, $\frac{\alpha w}{\overline{E}}$\\[2mm]
&&& (context dependent) \\
\br
\end{tabular}
\end{table}

Thus, unlike rigid pipes (figure~\ref{fig:resistance}), calculating the hydraulic resistance of a soft conduit requires the solution of a non-trivial elasticity problem (figure~\ref{fig:sketch-geom}), \emph{in addition} to the flow problem. Depending on the conduit geometry, the balance of elastic forces can be entirely different. Example mathematical expressions for the function $\mathfrak{G}(p,\mathrm{d}p/\mathrm{d}z)$ can be ``read off'' from the equations in sections~\ref{sec:flow_tube} and \ref{sec:flow_channel}. However, these expressions depend on $p$ \emph{implicitly} through the tube radius $R$ (deformed from $a$) and channel height $h$ (deformed from $h_0$). To fully specify $\mathfrak{G}$, suitable deformation--pressure relationships must be determined. Figure~\ref{fig:sketch-geom} shows four geometries that can be analyzed by the perturbative approach:
\begin{itemize}	
	\item[(a)] 2D planar configuration with vertical deformation \cite{SM04,SM05,CC11}:
\begin{eqnarray}
	h = h(z) = h_0 + u_y(z),\\
	u_y(z) = t\frac{p(z)}{2G+\lambda}. 
	\label{eq:uy_2D}
\end{eqnarray}
    \item[(b)] 3D axisymmetric exclusion, within an infinite elastic medium, with radial deformation \cite{RDC19}:
\begin{eqnarray}
	R = R(z) = a + u_r(z),\\
	u_r(z) = \frac{a}{4} \frac{p(z)}{G}. 
	\label{eq:ur_tube}
\end{eqnarray}    
    \item[(c)] 3D axisymmetric thin shell with radial deformation \cite{EG14,AC19c}:
\begin{eqnarray}
	R = R(z) = a + u_r(z),\\
	u_r(z) = \frac{a^2}{t} \frac{p(z)}{\overline{E}}.
	\label{eq:ur_shell}
\end{eqnarray}     
	\item[(d)] 3D Cartesian configuration (thick or thin) with vertical deformation \cite{CCSS18,SC18,WC19,AMC20}:
\begin{eqnarray}
	h = h(x,z) = h_0 + u_y(x,z),\\
	u_y(x,z) = F(x) \frac{p(z)}{\overline{E}},
	\label{eq:uy_3D}
\end{eqnarray}
for some $F(x)$ determined from linear elasticity.
\end{itemize}

Note that case (d) is unconfined on top, unlike (a). However, it was shown by Wang and Christov \cite{WC19} that the vertical displacement decays exponentially with vertical distance from the fluid--solid interface for a thick elastic top wall. Further, the result in case (d) requires the use of the shallowness assumption of the microchannel ($h_0\ll w$). For cross-sections that are close to square ($h_0\sim w$) techniques based on conformal mappings \cite{RM18} could potentially be employed to obtain the deformation from the equations of linear elasticity. However, this approach has not yet been coupled to flow (via the hydrodynamic pressure $p(z)$).

It should also be emphasized that these relationships rest on the result from lubrication theory that the pressure varies only in the flow-wise direction, \textit{i.e.}, $p=p(z)$, for a long and slender conduit (as in figure~\ref{fig:sketch-geom}) \cite{B08}. The most challenging case is the 3D microchannel in figure~\ref{fig:sketch-geom}(d), for which the expression for the function $F(x)$ depends on whether the rectangular wall is thin (\textit{i.e.}, plate theory applies \cite{CCSS18,SC18}) or thick (\textit{i.e.}, the ``full'' equations of linear elasticity must be employed \cite{WC19}). Recently, it has been shown \cite{WC21} that spanwise-averaging the deformation as
\begin{equation}
	\langle u_y \rangle_x(z) = \int_{-w/2}^{+w/2} u_y(x,z) \,\mathrm{d}x = \alpha w\frac{p(z)}{\overline{E}},
	\label{eq:avg_uy}
\end{equation}
where $\alpha=\int_{-1/2}^{+1/2} F(x/w) \,\mathrm{d}(x/w)$ is dimensionless, commits only a small error in the final solution to the coupled problem. Although this result was established for Newtonian fluids, its generalization to non-Newtonian fluids is highlighted below. The constant $\alpha$ has been calculated from theory (no fitting or calibration required) \cite{WC21} as
\begin{equation}
\alpha =
	\left\{\begin{array}{@{}l@{\quad}l}
     	\frac{1}{60}\left(\frac{w}{t}\right)^3\left[1 + \frac{10(t/w)^2}{\varkappa(1-\nu)}\right] &\quad(\mathrm{plate}), \\[2mm]
    	0.542742 &\quad(\mathrm{halfspace}).
    \end{array}\right.
    \label{eq:alpha}
\end{equation}
The ``plate'' case considers thick-plates $t/w <1$ (but not necessarily restricted to $t/w\ll 1$), $\varkappa$ is a shear-correction factor taken to be unity \cite{SC18}. The ``halfspace'' case refers to the limit $t^2/w^2\to0$, in which the wall thickness drops out \cite{WC19}. 

Importantly, the span-averaged deformation--pressure relation (\ref{eq:avg_uy}) allows us to treat the 3D channel geometry in figure~\ref{fig:sketch-geom}(d) in the same way as those in figure~\ref{fig:sketch-geom}(a,b,c). Then, all the local deformation--pressure relations reviewed here \emph{resemble} a Winkler ``mattress'' foundation \cite{Dillard2018,CV20} with $p=k_w u_y$ (or $p=k_w u_r$). The \emph{effective} stiffness $k_w$ can be obtained from the expressions in figure~\ref{fig:sketch-geom}. For convenience, the \emph{compliance} $\mathcal{C}=1/k_w$ is now introduced. Here, $\mathcal{C}$ is not to be confused with the \emph{hydraulic capacitance}, which also arises from compliance, within the hydraulic--electric circuit analogy \cite[section~4.6]{B08}. Capacitance will not be discussed in this review.


\subsection{Analytical models for the coupled problem and their solution}
\label{sec:FSI}

Combing the physics of the flow reviewed in section~\ref{sec:q_dpdz} with the physics of the flow-induced deformation reviewed in section~\ref{sec:deformation}, yields the \emph{basic laws} of soft hydraulics. Specifically, these laws will take the form of nonlinear ODEs for the hydrodynamic pressure $p(z)$, from which the deformation and velocity field can be reconstituted.

\subsubsection{Microtubes/micropipes}

For the geometries in figure~\ref{fig:sketch-geom}(b,c), (\ref{eq:vz_tube_plaw}) becomes:
\begin{equation}
	-\frac{\mathrm{d}p}{\mathrm{d}z} = \frac{2(3+1/n)^nKq^n}{\pi^n [a+\mathcal{C}p(z)]^{1+3n}},
	\label{eq:dpdz_tube_plaw}
\end{equation}
where $\mathrm{d}p/\mathrm{d}z<0$ so that $|\mathrm{d}p/\mathrm{d}z| = -\mathrm{d}p/\mathrm{d}z$ for flow in the $+z$-direction. Here, $\mathcal{C} = a/(4G)$ for the cylindrical exclusion geometry (``micropipe,'' figure~\ref{fig:sketch-geom}(b)), while $\mathcal{C} = a^2/(t\overline{E})$ for the cylindrical shell (``microtube,'' figure~\ref{fig:sketch-geom}(c)). A typical cylindrical exclusion in PDMS has $\mathcal{C} \approx 0.3$ $\mu$m/kPa (experimental fit), with values up to 4 $\mu$m/kPa possible when using silicone rubbers like ``Dragon Skin'' \cite{SGOG20}.
Equation (\ref{eq:dpdz_tube_plaw})  can be solved as a separable first-order ODE to obtain \cite{AC19c}:
\begin{eqnarray}
	p(z) = \frac{a}{\mathcal{C}} \left\{\left[ 1 + \hat{n}_t\left(\frac{\mathcal{C}KLq^n}{\pi^n a^{2+3n}}\right)\left(1-\frac{z}{L}\right)\right]^{\frac{1}{2+3n}}-1\right\}, \nonumber\\
	\label{eq:P_vs_Z_Microtube}
\end{eqnarray}
where $\hat{n}_t=2(3+1/n)^n(2+3n)$ for convenience.

A non-trivial dimensionless group, which is interpreted as the \emph{fluid--structure interaction parameter}  $\beta$ \cite{CCSS18}, arises from this calculation:
\begin{equation}
\eqalign{
	\beta &= \frac{\mathcal{C}KLq^n}{\pi^n a^{2+3n}} = \frac{KL\langle v_z \rangle^n/a^{1+n}}{a/\mathcal{C}} \\ 
	      &\simeq \frac{\mathrm{hydrodynamic~forces~on~wall}}{\mathrm{elastic~resistance~of~wall}}.
}
\label{eq:beta_tube}
\end{equation}
Here, $\langle v_z \rangle = q/(\pi a^2)$. By balancing (\ref{eq:mom}) with the power-law model (\ref{eq:plaw}) for the viscosity, $\mathscr{P}_{c} = KL\langle v_z \rangle^n/a^{1+n}$ can be shown to be the characteristic pressure scale (hydrodynamic force per area) for this flow. It follows that $\mathscr{U}_c = \beta a = \mathscr{P}_c\mathcal{C}$ is a characteristic deformation scale.

\begin{figure*}
	\centering
	\subfloat[][]{\includegraphics[height=0.3\textwidth]{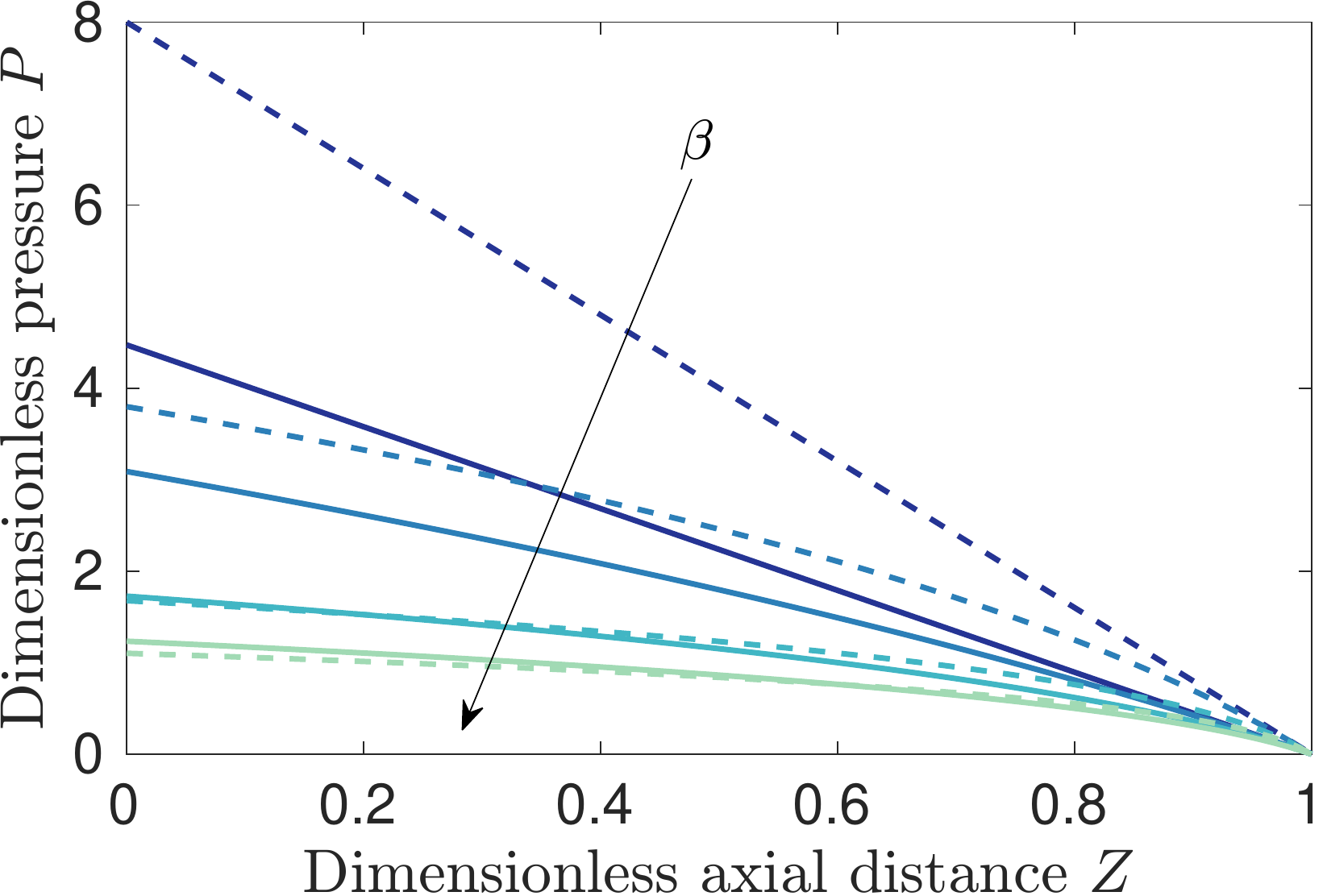}}\quad\qquad
	\subfloat[][]{\includegraphics[height=0.3\textwidth]{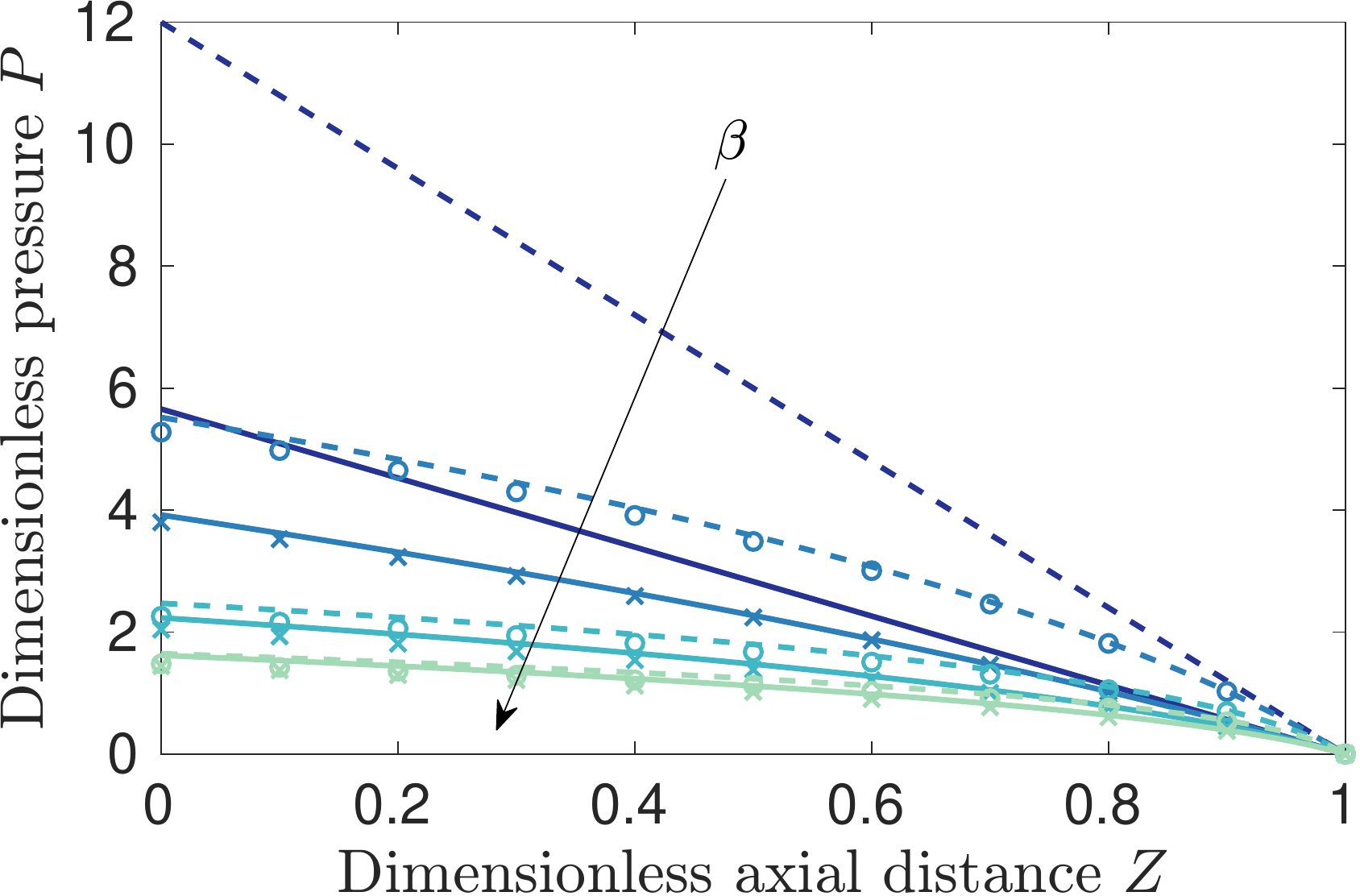}}
	\caption{(a) Dimensionless pressure $P(Z)=p(z)/\mathscr{P}_c$ variation along a compliant microtube/micropipe ($Z=z/L$) for a Newtonian fluid (\ref{eq:P_vs_Z_Microtube}) with $n=1$ (\dashed) and a shear-thinning fluid  (\ref{eq:P_vs_Z_Microtube}) with $n=0.5$ (\full). (b) Dimensionless pressure $P(Z)=p(z)/\mathscr{P}_c$ variation along a compliant microchannel ($Z=z/L$) for a Newtonian fluid (\ref{eq:P_vs_Z_Microchannel}) with $n=1$ (\dashed) and a shear-thinning fluid  (\ref{eq:P_vs_Z_Microchannel}) with $n=0.5$ (\full). Symbols show the corresponding ($\opencircle$, $n=1$) and ($\times$, $n=0.5$) pressure profiles found from numerically integrating the ``full'' microchannel theory (\ref{eq:p_ode_thick_dimensional}). In both cases, a thin plate compliant top wall, with negligible $(t/w)^2/[\varkappa(1-\nu)]$, is considered. Colors represent different values of the fluid--structure interaction parameter $\beta = 0,0.1,0.5,1$ (top to bottom, dark to light). Note that $\mathscr{P}_c$ and $\beta$ are context-dependent and defined differently for (a) and (b), as discussed in the text.}
	\label{fig:p_of_z}
\end{figure*}

\subsubsection{Microchannels: span-averaged theory}

For the geometry in figure~\ref{fig:sketch-geom}(d) and using (\ref{eq:avg_uy}), (\ref{eq:vz_chan_plaw}) becomes:
\begin{equation}
	-\frac{\mathrm{d}p}{\mathrm{d}{z}} = \frac{2^{1+n}(2+1/n)^nKq^n}{[h_0 + \mathcal{C} p(z)]^{1+2n}w^n},
	\label{eq:dpdz_chan_plaw}	
\end{equation}
where $\mathcal{C} = \alpha w/\overline{E}$ with $\alpha$ given in (\ref{eq:alpha}) for a flow conduit with Cartesian geometry. A typical microchannel with a thin deformable membrane as its top wall has $\mathcal{C} \approx 1-10$ $\mu$m/kPa (experimental fit) \cite{LLMH21}. 
Equation (\ref{eq:dpdz_chan_plaw}) can be solved as a separable first-order ODE to obtain:
\begin{eqnarray}
	p(z) = \frac{h_0}{\mathcal{C}} \left\{\left[ 1 + \hat{n}_c\left(\frac{\mathcal{C}KLq^n}{h_0^{2+2n}w^n}\right)\left(1-\frac{z}{L}\right)\right]^{\frac{1}{2+2n}}-1\right\}, \nonumber\\
	\label{eq:P_vs_Z_Microchannel}
\end{eqnarray}
where $\hat{n}_c=2^{1+n}(2+1/n)^n(2+2n)$ for convenience.
Once again, a fluid--structure interaction dimensionless parameter emerges from (\ref{eq:P_vs_Z_Microchannel}):
\begin{equation}
	\beta = \frac{\mathcal{C}KLq^n}{h_0^{2+2n}w^n} = \frac{KL\langle v_z \rangle^n/h_0^{1+n}}{\overline{E}h_0/(\alpha w)}.
	\label{eq:beta_chan}
\end{equation}
Here, $\langle v_z \rangle = q/(h_0w)$, and $\mathscr{P}_{c} = KL\langle v_z \rangle^n/h_0^{1+n}$ can be shown to the be characteristic pressure scale for this flow \cite{ADJRC19}.  It follows that $\mathscr{U}_c = \beta h_0 = \mathscr{P}_c \mathcal{C}$ is a characteristic deformation scale.

To estimate a typical value of $\beta$ for a non-Newtonian soft hydraulic system, consider the typical compliant ($\overline{E}\approx 1$ MPa, $\alpha\approx1$), long and shallow microchannel ($h_0 \approx 25$ $\mu$m, $w \approx 10h_0$, $L \approx 100h_0$) \cite{GEGJ06}. For the power-law model's parameters used in figure~\ref{fig:rheology} ($K\approx 1.25$ Pa$\cdot$s$^n$, $n\approx 0.41$) and a flow at $\langle v_z \rangle \approx 1$ m/s (for $q\approx 300$ $\mu$L/s \cite{GEGJ06}), yields $\beta \approx 0.1$. Importantly, however, the theory reviewed above does not require $\beta\ll1$, and it is applicable up to $\beta=\mathcal{O}(1)$ \cite{CCSS18,WC19}.

\subsubsection{Microchannels: ``full'' theory}

Without using the spanwise-averaging idea to reduce $u_y(x,z)$ to a function of $z$ alone, the coupled two-way fluid--structure interaction problem can still be reduced to single nonlinear ODE for $p(z)$. To understand the challenges in using the ``full'' theory of the 3D Cartesian geometry's deformation, consider (\ref{eq:q_defn}), which becomes:
\begin{equation}
    q = \int_{-w/2}^{+w/2}\!\!\int_{0}^{h_0+u_y(x,z)} v_z \,\mathrm{d}x\, \mathrm{d}y.
\label{eq:q_def_chan}    
\end{equation}
Therefore, (\ref{eq:vz_chan_plaw}) becomes 
\begin{equation}
\eqalign{
    -\frac{\mathrm{d}p}{\mathrm{d}z}\left[\frac{1}{2^{1+1/n}(2+1/n)K^{1/n}}\left|\frac{\mathrm{d}p}{\mathrm{d}{z}}\right|^{(1/n)-1}\right] \\ 
	\qquad\times\int_{-w/2}^{+w/2}[h_0+u_y(x,z)]^{2+1/n}  \,\mathrm{d}x = q.
}
\label{eq:q_int_chan}
\end{equation}
The last integral is to be evaluated using a suitable deformation--pressure relationship, such as $u_y(x,z)=F(x)p(z)/\overline{E}$. However, this integral does not always yield a closed-form expression due to the \emph{fractional power} $2+1/n$ involved. 

Anand \textit{et al}.~\cite{ADJRC19} used a generalized binomial expansion to handle the latter difficulty, and finally obtained the desired ODE for $p(z)$:
\begin{eqnarray}\label{eq:p_ode_thick_dimensional}
	-\frac{\mathrm{d}p}{\mathrm{d}z} = \frac{Kq^n}{h_0^{1+2n}w^n} {2^{1+2n}(2+1/n)^n} \pi^{-n/2}\nonumber\\
	\times \Bigg[ \sum_{k=0}^{\infty}C_{k,n} \Bigg\{\frac{1}{2^5}\frac{w}{\overline{E}h_0}\frac{w^3}{t^3}\left[1+\frac{8(t/w)^2}{\varkappa(1-\nu)}\right]p(z)\Bigg\}^k \nonumber\\
	\quad \times  \, \Gamma{(k+1)} \,{}_2\tilde{F}_1\Bigg(\frac{1}{2},-k;\frac{3}{2}+k;\frac{1}{1+\frac{8(t/w)^2}{\varkappa(1-\nu)}}\Bigg)\Bigg]^{-n},
\end{eqnarray}
where ${}_2\tilde{F}_1$ is the regularized hypergeometric function, $\Gamma$ is the gamma function, 
\begin{equation}
	C_{k,n} = \frac{\Gamma (3+1/n)}{\Gamma (k+1) \Gamma (3-k+1/n)}
\end{equation}
are binomial-like coefficients, and $\varkappa$ is again a shear correction factor taken to be unity. Once the pressure $p(z)$ is determined via (\ref{eq:p_ode_thick_dimensional}), the deformed channel shape can be found as $h(x,z) = h_0 + F(x) p(z)/\overline{E}$ (recall figure~\ref{fig:sketch-geom}). An example Python code for solving this ODE numerically is available in the supplementary material associated with \cite{ADJRC19}.

To illustrate and summarize the basic results on the pressure distribution in a soft hydraulic conduit, figure~\ref{fig:p_of_z} shows the dimensionless pressure $p(z)/\mathscr{P}_c$ variation along the compliant microchannel, \textit{i.e.}, against dimensionless axial distance $z/L$, for both Newtonian and non-Newtonian fluids and different values of the dimensionless fluid--structure interaction parameter $\beta \propto \Delta p/E$. Both the shear-thinning rheology and fluid--structure interaction decrease the pressure drop, which is evaluated from $p(0)$. For $\beta=\mathcal{O}(1)$, it is interesting to note that all the profiles saturate, showing that the strongest effect of the interaction between the non-Newtonian fluid's rheology and the compliance of the conduit is for small $\beta$, which is typical in practice. This observation has implications for the sensitivity of microfluidic rheological measurements (see section~\ref{sec:microrheol}). In figure~\ref{fig:p_of_z}(b), observe also that (\ref{eq:P_vs_Z_Microchannel}) (curves) is a good approximation to $p(z)$ found by solving the nonlinear ODE (\ref{eq:p_ode_thick_dimensional}) (symbols). 

The reference pressure distributions for Newtonian flow in a rigid conduit ($n=1$ and $\beta=0$) are (a) $p(z)/\mathscr{P}_c = 8(1-z/L)$ and (b) $p(z)/\mathscr{P}_c = 12(1-z/L)$, which can be obtained by a Taylor series expansion in $\beta\ll1$ (or a limit using L'H\^{o}pital's rule) from (\ref{eq:P_vs_Z_Microtube}) and (\ref{eq:P_vs_Z_Microchannel}), respectively.

\subsubsection{Special case: flow rate--pressure drop relations}
\label{sec:q-dp}

Whenever the ODE governing the coupled soft hydraulics problem can be solved analytically for $p(z)$, subject to gauge outlet pressure $p(L)=0$, the relationship between the pressure drop $\Delta p = p(0)-p(L)$ and the steady volumetric flow rate $q$ can be obtained, as suggested by (\ref{eq:soft_dp_q}). When (\ref{eq:soft_dp_q}) can be resolved algebraically as $\Delta p = R_h(\Delta p) q$ \emph{or} as $q = \Delta p/R_h(q)$, the tunable soft hydraulic resistance $R_h(\Delta p)$ or $R_h(q)$ can be determined.

From the two prototypical pressure distributions for non-Newtonian fluid flow in a soft hydraulic conduit, namely (\ref{eq:P_vs_Z_Microtube}) for the microtube/micropipe and (\ref{eq:P_vs_Z_Microchannel}) for the microchannel, it is trivial to evaluate $\Delta p$ in each case. However the resulting expressions cannot be algebraically resolved as $\Delta p = R_h q$, unless an expansion in $\beta\ll1$ is performed. However, these flow rate--pressure drop relations can, of course, be easily inverted numerically. In fact, an implicit relation of the form (\ref{eq:soft_dp_q}) can also be obtained under the Carreau model (\ref{eq:carr}) \cite{S15} (but note that the validity of the variational method used for validation in \cite{S15} has been questioned \cite{PC15}). Most recently, Boyko and Stone \cite{BS21} obtained explicit analytical results for small, intermediate and large Carreau numbers [$\lambda_r \Delta p h_0/(\eta_0L)$] via perturbation methods.

Similarly, the ``full'' microchannel result (\ref{eq:p_ode_thick_dimensional}) does not allow for an analytical evaluation of $\Delta p$. However, its Newtonian counterpart ($n=1$, $K=\eta_0$) does, and yields:
\begin{equation}\label{qdpthick}
	\Delta p = \underbrace{\overbrace{\frac{12\eta_0 L} {h_0^3w}}^{\mathrm{rigid~channel}}\left[\sum_{k=0}^3 \left(\frac{w}{\overline{E}h_0}\right)^k S_k(\Delta p)^k \right]^{-1}}_{\mathrm{Newtonian~soft~hydraulic~resistance,~}R_h(\Delta p)} q,
\end{equation}
the parameters $S_1$, $S_2$ and $S_3$ are \emph{known} functions of the geometry (\textit{e.g.}, wall thickness $t$) \cite{SC18,WC19}, or potentially pre-stress in the wall \cite{AMC20}. The term outside of the bracket can be compared to figure~\ref{fig:resistance} (row ``two plates''). 

Equation~(\ref{qdpthick}) is an analytical result capturing the tunable nonlinear resistance \cite{A04,Case19} of a soft hydraulic conduit. This expression is \emph{not} a perturbation series in $\Delta p$. The dependence of $S_1$, $S_2$ and $S_3$ on geometry (and even pre-stress) enables \emph{tuning} of the nonlinear resistance. Using this approach, the \emph{ultra-low} aspect ratio regime ($w/h_0 \gg 1$) was studied experimentally by Mehboudi and Yeom \cite{MY19}, highlighting the (perhaps unexpected) importance of the higher powers of $\Delta p$ in (\ref{qdpthick}), which are dominant in this regime. It would be of interest to further develop these concepts for non-Newtonian flows in compliant conduits.


\section{Maturing applications areas}
\label{sec:appl}


\begin{figure}
	\centering
	\includegraphics[width=0.45\textwidth]{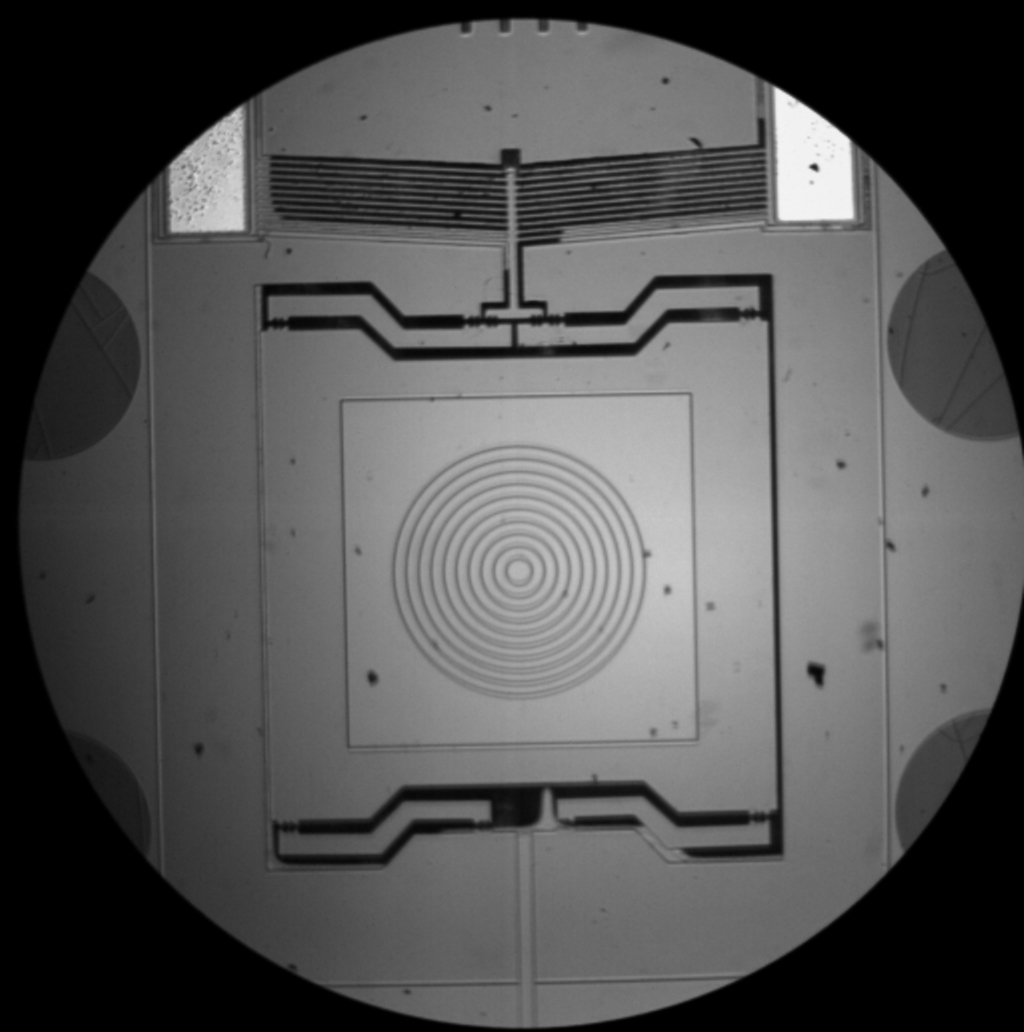}
	\caption{A MEMS-based dynamic shear microrheometer (``MEMS-$\mu$R'') \cite{CYDHM10,MR_NIST}. The chip shown is a few millimeters in width and height, with the central square region, where the fluid sample is placed, being $\approx 500$ $\mu$m. Oscillatory shear is generated by the ``chevron-like structure'' at the top of the chip. The nested beams in the chevron are actuated by resistance heating elements (contact pads to the left and right), which in turn sets the central stage into motion. The gap thickness into the page is $\approx 10$~$\mu$m. The chip can be placed in an optical microscope to detect the motion of and flow within the sample using a high-speed camera. From the latter optical measurement, the rheological parameters of the material can be estimated. The MEMS-$\mu$R uses only $5$ nL of material. Reproduced from \cite{MR_NIST}, photo credit: G.\ Christopher/NIST, by courtesy of the National Institute of Standards and Technology. All rights reserved, U.S.\ Secretary of Commerce.}
	\label{fig:murheo}
\end{figure}

\subsection{Microfluidic rheometry}
\label{sec:microrheol}

It has been proposed that rheological measurements can also be performed with miniaturized setups, involving small amounts of liquid (obviating the need for multiple dilutions) and improved optical measurement capabilities (via microscopy) \cite{A08}. Utilizing flows in microchannels gives rise to \emph{microfluidic rheometry} \cite{PM09,GWV16} also known as ``rheometry-on-a-chip'' \cite{DDGM15}. An example of such a (rigid) chip, designed and manufactured by the US National Institute of Standards and Technology (NIST), is shown in figure~\ref{fig:murheo}. Srivastava and Burns \cite{SB06} microfabricated a transient capillary
viscometer capable of measuring $n$ for the power-law non-Newtonian model (\ref{eq:plaw}) in $2$--$8$ min from 1 $\mu$L of fluid. More generally, microfluidic rheometry enables the characterization of low-viscosity fluids \cite{GAO13} and the measurement of the relaxation time of weakly-elastic fluids down to milliseconds \cite{DDGM15}, both of which are normally challenging using macroscopic rheometers. In the biopharamaceutical industry, it is desirable to perform rheological measurements not only using a small-volume samples of fluid, but also across a significant range of shear rates, and without interaction between the sample and air (such as at a free surface). A microrheometer fits the bill \cite{HSPM15}. Another benefit of miniaturization is the ability to easily generate high-frequency flows \cite{VJ20} in PDMS-based microchannels and induce \emph{acoustic streaming} \cite{R01,S12} past a cylindrical pillar in the channel. Oscillatory-flow techniques are not only useful for measuring the shear viscosity of low-viscosity liquids \cite{VJ19a}, but also for characterizing the storage and loss moduli and relaxation time of viscoelastic dilute and semi-dilute polymeric solutions \cite{VJ19b}.

In the most basic form of microrheometry, experiments that characterize the pressure drop for a given flow rate are compared to a theoretical prediction. The flow rate--pressure drop curve (recall section~\ref{sec:q-dp}) is a cornerstone of capillary viscometry, even for non-Newtonian fluids \cite[chapter~2]{CR08}. The shear-rate-dependent viscosity of non-Newtonian fluids can be taken into account, yielding correlations \cite{ME08} from which the steady shear viscosity and its rate dependence can be inferred from simultaneous measurements of $\Delta p$ and $q$. Essentially, these approaches rely on a correlation for the \emph{friction factor}, which is a dimensionless pressure drop (Darcy--Weisbach friction factor):
\begin{equation}
	f_D = \frac{\Delta p/L}{\frac{1}{2}\rho \mathscr{V}_c^2/D_h},
	\label{eq:dwf}
\end{equation}
Equations such as (\ref{eq:dwf}) are key to microfluidic system design \cite{SASM01}, much like their use for analyzing industrial pipe networks \cite[chapter~8]{P11}, including non-Newtonian ones \cite[section~3.8]{CR08}. The friction factor allows for a convenient parametrization of viscous (energy) losses in both laminar and turbulent flows \cite{P11}. In (\ref{eq:dwf}), $\mathscr{V}_c$ is  suitable velocity scale, \textit{e.g.}, $\langle v_z \rangle$; $D_h$ is a suitable \emph{hydraulic diameter}, \textit{e.g.}, $4\mathcal{A}/\mathcal{P}$, where $\mathcal{A}$ and $\mathcal{P}$ are the area and  perimeter of an axial cross-section of the flow conduit, respectively. Typically, the same $\mathscr{V}_c$ and $D_h$ are substituted for $\mathscr{V}_c$ and $h_0$ in (\ref{eq:reynolds}) to define the appropriate Reynolds number for the flow. 

If the pressure drop is not know \textit{a priori}, it is more convenient to define the friction factor as a dimensionless (local) \emph{mean} wall shear stress $\bar{\tau}_w$ (Fanning friction factor) \cite{ME08,SASM01}:
\begin{eqnarray}
	f_F &=& \frac{\bar{\tau}_w}{\frac{1}{2}\rho \mathscr{V}_c^2},\label{eq:fff}\\
	\bar{\tau}_w &=& \frac{1}{\mathcal{P}}\oint_\mathcal{P} \tau_w \, \mathrm{d}\ell = \frac{D_h}{4}\left(-\frac{\mathrm{d}p}{\mathrm{d}z}\right),\label{eq:tauw}
\end{eqnarray}
where $\tau_w$ is the shear stress evaluated at the wall. Under these definitions, $f_F = f_D/4$ \cite{SASM01}, but care must be taken to properly relate $\bar{\tau}_w$, $\Delta p$ and $\mathscr{V}_c$ in a non-Newtonian flow via the momentum equation~(\ref{eq:mom}). For example, for a power-law fluid one can define a generalized Reynolds number $Re_g$ and obtain \cite{ME08} (see also \cite[chapter~3]{CR08}):
\begin{eqnarray}
	f_F &=& \frac{16}{Re_g}, \label{eq:ff_plaw}\\
	Re_g &=& \frac{\rho \mathscr{V}_c^{2-n}D_h^n}{8^{n-1}(c_2+c_1/n)^nK}, \label{eq:Re_g}	
\end{eqnarray}
which is also valid for a Newtonian fluid ($n=1$, $K=\eta_0$). In (\ref{eq:ff_plaw}),  $c_1$ and $c_2$ are the so-called Rabinowitsch--Mooney constants depending on the flow conduit's geometry. For a circular pipe, $c_1=1/4$ and $c_2=3/4$; for a parallel plate channel, $c_1=1/2$ and $c_2=1$ \cite{ME08}. Observe that, from (\ref{eq:ff_plaw}), the \emph{Poiseuille number} $Po = f_F Re_g$ is a constant determined solely by the cross-sectional shape of the (rigid) conduit \cite{W06_book,SASM01}. 

The goal of using $Re_g$ is to eliminate ambiguities and challenges that arise due to the different shear rates experienced by fluids in microchannels of different shapes \cite{YWHFJZM19}. Then, for example, given experimental measurements of $f_F$, the power-law rheological parameters $K$ and/or $n$ can be back-calculated from (\ref{eq:ff_plaw}) via (\ref{eq:Re_g}). Clearly, an accurate friction factor theory of non-Newtonian flow in microfluidic conduits is key to microrheometry. However, beyond the (\ref{eq:ff_plaw})--(\ref{eq:Re_g}) for the power-law model, few similar parametrizations exist for other shear-dependent viscosity models. It is of interest to develop such relations because the power-law model is only applicable over some intermediate range of shear rates (recall figure~\ref{fig:rheology}(c)), failing to capture the high-shear Newtonian plateau (for  $\dot{\gamma}_c \sim 10^4-10^6$ s$^{-1}$) now being accessed using microrheometers \cite{KLK05,PMM08} (see also the discussion in \cite{BS21}).

The MEMS-based dynamic shear microrheometer shown in figure~\ref{fig:murheo} uses glass plates to confine the sample. It can be easier and cheaper to manufacture the confined channels from, \textit{e.g.}, PDMS using soft lithography. However, given that PDMS-based microchannels are compliant, an open problem in microrheology \citep{DGNM16} is whether such measurements are affected by the dependence $\propto \Delta p/E$ introduced by fluid--structure interaction (recall section~\ref{sec:deformation}). Taking this idea one step further, Shiba \textit{et al.}~\cite{SLVYW21} exploited compliance to measure the viscosity of Newtonian fluids (both liquids and gases). They used a strain gauge to quantify a PDMS microchannel's flow-induced deformation, then calibrated the strain--viscosity relationship empirically.

Recently, Wang and Christov \cite{WC21} addressed the $\Delta p/E$ dependence of the friction factor for Newtonian fluid flow through a soft hydraulic conduit. Implementing their derivation for the power-law fluid through a microchannel (a microtube can be handled analogously), one first needs to account for the flow-induced conduit deformation in the expressions for $\mathscr{V}_c$ in (\ref{eq:fff}) and $D_h$ in (\ref{eq:tauw}), replacing them by \emph{local} expressions: 
\begin{eqnarray}
	\mathscr{V}_c &=& q/[h_0(1+\mathcal{C}p/h_0)w] = \mathscr{V}_{c,h_0}/(1+\mathcal{C}p/h_0), \label{eq:Vc_d}\\
	D_h &=& D_{h_0}(1+2\mathcal{C}p/h_0). \label{eq:Dh_d}
\end{eqnarray}
In (\ref{eq:Vc_d}), $\mathscr{V}_{c,h_0} = q/(h_0w)$ is the mean axial velocity in the undeformed channel and $D_{h_0} = 4h_0w/[2(h_0+w)]$ in (\ref{eq:Dh_d}) is its hydraulic diameter. To arrive at (\ref{eq:Dh_d}), the force balance used to evaluate (\ref{eq:tauw}) had to be rederived for a compliant flow conduit \cite{WC21}.
Then, combining (\ref{eq:fff}), (\ref{eq:tauw}) and (\ref{eq:dpdz_chan_plaw}) yields
\begin{eqnarray}
	f_F &=& \frac{(1+\mathcal{C}p/h_0)^2}{\frac{1}{2}\rho \mathscr{V}_{c,h_0}^2} \frac{D_{h_0}(1+2\mathcal{C}p/h_0)}{4}\nonumber\\
	&\phantom{=}& \times \frac{2^{1+n}K(2+1/n)^nq^n}{h_0^{1+2n}(1 + \mathcal{C} p/h_0)^{1+2n}w^n} \nonumber\\
	&=& \frac{D_{h_0}^{1+n}}{h^{1+n}} \frac{2^{2n}[1+1/(2n)]^n K}{\rho \mathscr{V}_{c,h_0}^{2-n} D_{h_0}^n}\frac{(1+2\mathcal{C}p/h_0)}{(1 + \mathcal{C} p/h_0)^{2n-1}}.
\label{eq:fff_d}
\end{eqnarray}

\begin{figure*}
	\centering
	\subfloat[][]{\includegraphics[width=0.35\textwidth]{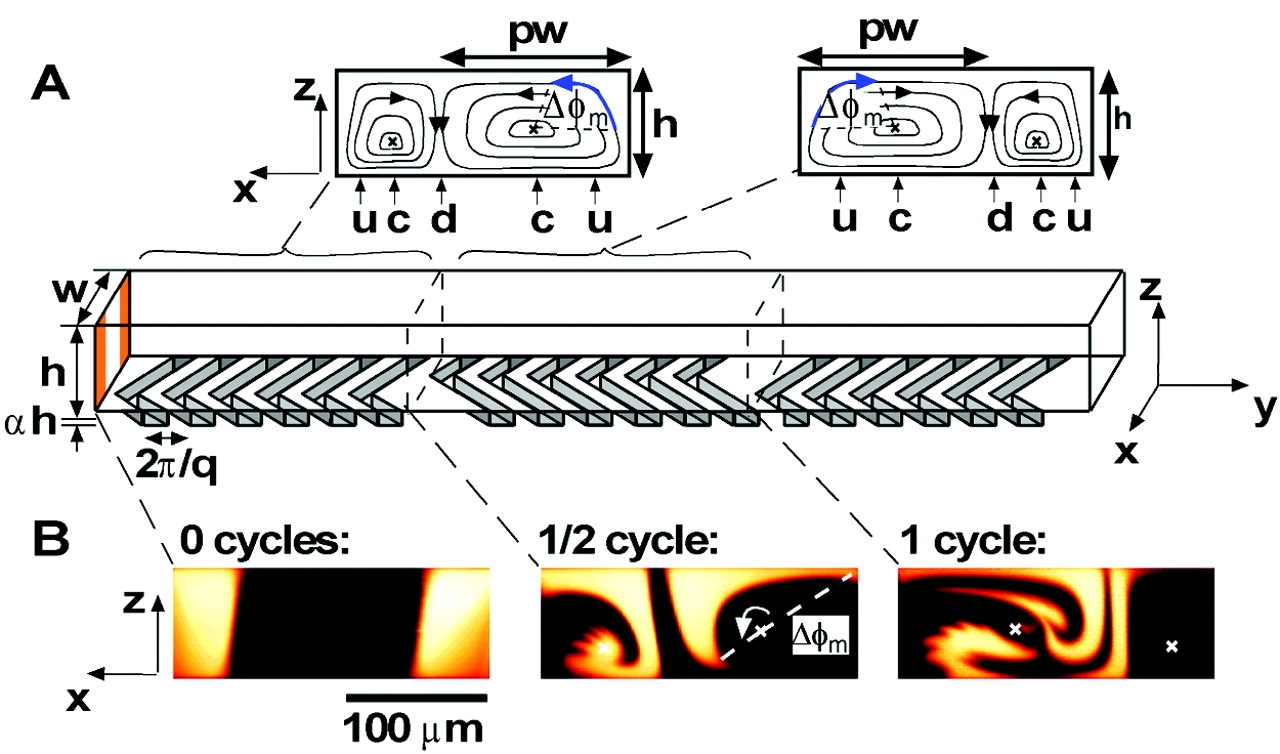}}\hfill
	\subfloat[][]{\includegraphics[width=0.64\textwidth]{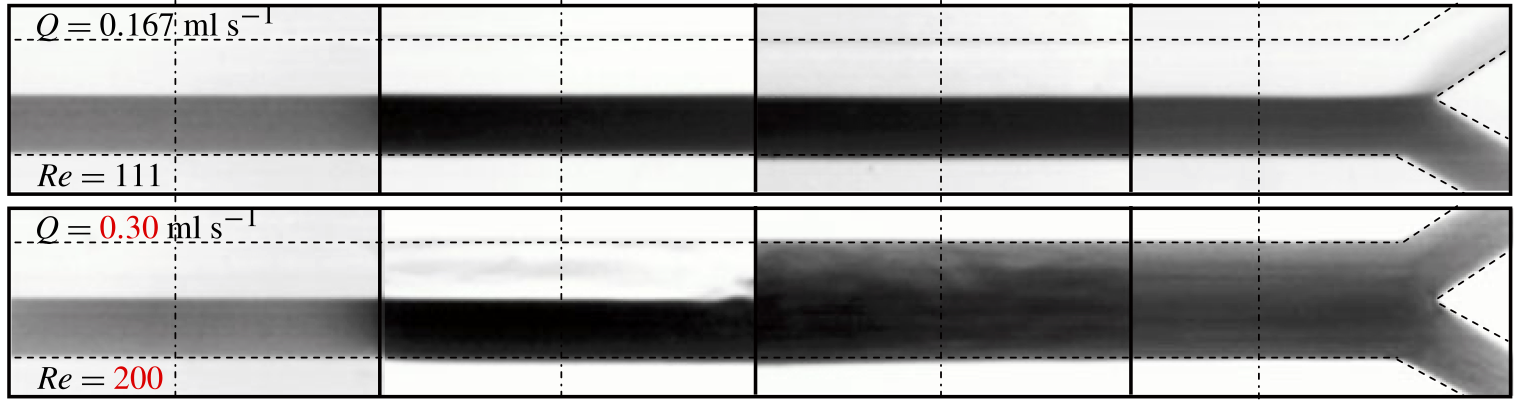}}
	\caption{(a) Staggered herringbone mixer (SHM); subpanel {A} shows the periodic pattern on the lower channel wall (rigid) and resulting secondary flows, while subpanel {B} shows conformal micrographs of the mixing of dyes in the channel at $Re\approx0$. From Stroock \textit{et al.}\ 2002 \textit{Science} \textbf{295} 647--651 \cite{sdamsw02}. Reprinted with permission from AAAS. (b) Instability leads to mixing in a soft-walled microchannel flow, without patterning the wall but, rather, due to fluid--structure interaction at an unusually low, but finite, $Re\ne0$. The channel is manufactured from a PDMS-based elastomer to yield a  low shear modulus $G\approx 10-50$ kPa. Reproduced from \cite{VK13} with permission from Cambridge University Press \copyright\ 2013. An optical visualization is shown, in the style of Reynolds' dye breakup experiment \cite{R83}, at four representative (non-continuous) segments of the long microchannel. The working fluid in (b) is water, however the same flow--structure instability is observed with dilute polymeric solutions \cite{SK17,CSD19}. In both panels, the flow is left to right.}
	\label{fig:examples}
\end{figure*}

Next, observing that $[1+1/(2n)]^n = (c_2 + c_1/n)^n$ for a microchannel and $D_{h_0}/h_0 = 2/(1+h_0/w) \simeq 2$ for $h_0\ll w$, one can introduce the definition of $Re_g$ (based on $D_{h_0}$ and denoted $Re_{g,h_0}$) from (\ref{eq:Re_g}) into (\ref{eq:fff_d}) to obtain:
\begin{equation}
	f_F = \frac{16}{Re_{g,h_0}} \frac{(1+2\mathcal{C}p/h_0)}{(1 + \mathcal{C} p/h_0)^{2n-1}}.
	\label{eq:fff_d2}
\end{equation}
Then, from (\ref{eq:Vc_d}) and (\ref{eq:Dh_d}), the generalized Reynolds number for the deformed channel is 
\begin{equation}
	Re_g = Re_{g,h_0} \frac{(1+2\mathcal{C}p/h_0)^n}{(1 + \mathcal{C} p/h_0)^{2-n}}.
	\label{eq:Re_g_d}
\end{equation}
Finally, the Poiseuille number for the flow of a non-Newtonian fluid under the power-law model in a deformable microchannel is obtained from (\ref{eq:fff_d2}) and (\ref{eq:Re_g_d}):
\begin{equation}
	Po = f_F Re_g = 16 \left(1 + \frac{\mathcal{C}p}{h_0 + \mathcal{C} p}\right)^{n+1}.
\label{eq:Po_plaw_d}
\end{equation}
Equation (\ref{eq:Po_plaw_d}) highlights that compliance alone can increase $Po$ by up to a factor of $2^{n+1}$ (for $\mathcal{C}p \gg h_0$), which is not negligible. Note that $p=p(z)$ in (\ref{eq:Po_plaw_d}) is given by (\ref{eq:P_vs_Z_Microchannel}). Due to the flow-induced deformation, the Poiseuille number now varies axially.


\subsection{Microfluidic mixing}

Fast and efficient mixing of fluids in labs-on-a-chip is of tremendous technological importance: micro total analysis systems ($\mu$TAS) \cite{RDIAM02,AIRM02}, which have enabled low-cost disposable medical diagnostics \cite{Yetisen13,SFB14} with public health implications \cite{Yager06}, rely on reagents and biological fluid samples mixing and reacting thoroughly. However, ``[t]he design and implementation of mixers in microfluidics differs considerably from that on the macro\-scale'' \cite{K13}. At the macroscale, the laminar--turbulent transition enables efficient mixing, which Osborne Reynolds \cite{R83} used to visualize the instability. Meanwhile, at the microscale, flows are dominated by viscosity and mixing is limited by diffusion. For a Newtonian fluid, G.~I.\ Taylor demonstrated, in a classic film \cite{Taylor_NCFMF}, the well-known kinematic reversibility of Stokes flow ($Re=0$) . Mixing by molecular diffusion between fluids requires channel lengths of 0.1 to 1 \emph{meters} \cite{OW04,K13}. For example, the diffusivity of hemoglobin into water is $\kappa\simeq 70$ $\mu$m$^2$/s \cite[section~9.1]{KBA05}. The diffusion time across a microchannel of height $h_0\approx25$ $\mu$m is $t_D = h_0^2/\kappa \approx 9$ s. During this time, for the typical flow speed of $\mathscr{V}_c = 4 \times 10^{-2}$ m/s quoted in section~\ref{sec:lubrication}, a fluid parcel would have traveled a distance $\mathscr{V}_c t_D \approx 0.4$ m. Relying on diffusive mixing over such lengths is not always feasible for a lab-on-a-chip processing blood.

In stark contrast to the laminar--turbulent transition, reversibility hinders mixing in microfluidics (``micromixing''), as there is no way to introduce asymmetries in laminar flow in a constant cross-section conduit (see also \cite[section~2.4]{DS14}). To mix, one must ``outsmart'' kinematic reversibility. Within the present context, one way to do so is to consider viscoelastic fluids. In the flows of such non-Newtonian fluids, \emph{purely elastic} instabilities can occur even at vanishing Reynolds number (\textit{i.e.,} in the absence of inertia) \cite{L92,Shaq96,S21}. Li \textit{et al}.~\cite{Li12} and Galindo-Rosales \textit{et al.}~\cite{GR14} reviewed the experimental characterization of these instabilities in \emph{rigid} microchannels with an outlook towards micromixing (see also \cite{OAP11}). 

In the absence of inertia or purely elastic flow instabilities, another approach to circumventing reversibility involves breaking symmetries in the equations governing Lagrangian trajectories, to induce \emph{chaotic advection} \cite{a84,o89}. Static mixer designs based on this principle have uses in the process industry \cite{Thakur03,Ghanem14}. In microfluidics, patterning the wall of a channel induces secondary flows and chaotic mixing (subpanel {A} in figure~\ref{fig:examples}(a)) \cite{sdamsw02}. However, this techniques requires more complex micromanufacturing than just soft lithography. 

A third approach is to \emph{harness} the flow--structure instabilities \emph{inherent} to soft (\textit{e.g.}, PDMS-based devices). Microchannels manufactured from polymeric materials are soft; consequently, flow--structure instabilities occur, inducing mixing without the need for patterning any channel surfaces (figure~\ref{fig:examples}(b)). First mentioned in the 1970s \cite{KS79}, the application of this discovery did not become clear until Kumaran \textit{et al.}\ provided an experimental demonstration of ``ultrafast'' micromixing (figure~\ref{fig:examples}(b)) in the 2010s \cite{VK12,VK13,KB16}. For a Reynolds number $Re=\rho q/(w\eta_0)$ as low as $\approx200$, transitional unstable flow is observed. This is in \emph{stark contrast} with the critical Reynolds number value $Re_c\approx2300$ (for rigid-pipe flow) quoted to undergraduates \cite[p.~43]{P11} (admittedly an oversimplification \cite{ESHW07} but nevertheless valid at the microscale \cite{SA04}). This new modality of mixing is akin to an \emph{active} micromixer \cite{LCWF11} in which an external force (typically via an acoustic field \cite{FY11,RWH17} but, here, via instability) agitates the fluid sample. These instabilities and ultrafast micromixing have also been demonstrated both theoretically \cite{CK07,CBK15} and experimentally \cite{SK17,CSD19} in non-Newtonian fluid flows of dilute polymeric solutions. 

Microfluidics has traditionally resided in the domain of $Re < 1$. However, the need to achieve efficient and high-throughput cell sorting has given rise to \emph{extreme microfluidics}, as Toner terms it in a 2018 Nobel Symposium lecture \cite{Toner2018}, or \emph{nonlinear microfluidics}, as Di Carlo terms it in a contemporaneous review \cite{SDD19}. Focusing particles (or cells) in microchannels requires inertia \cite{BKP08,DC09}. Current PDMS-based soft microfluidic platforms access the inertial regime up to $Re \simeq 10^2$ \cite{DiCarlo2007}, while epoxy-based ``hard'' casting allows microflows to achieve $Re\simeq 10^4$ \cite{Lim2014}. Clearly, the range of $Re$ in which flow--structure instabilities occur is now relevant to current microfluidic technologies. Such instabilities have previously been of interest due to the possibility of the diametrically opposed goal of drag reduction of high-speed flow over hard surfaces via \emph{compliant coatings} \cite{RGHM88,G96}. The complexity of these coatings has not yielded a working theory for applications~\cite[p.~367]{W06_book}, despite a significant amount of research \cite{G96,G02} since Kramer's work on dolphins' swimming efficiency \cite{K61} and Benjamin's 1960s analysis of linear instabilities \cite{B60}. On the other hand, at the microscale where flow-induced deformation is common, ultrafast mixing due to flow--structure instabilities in soft hydraulic conduits has, at least, been conclusively demonstrated \cite{VK13}.  

Kumaran \textit{et al.}~\cite{VK12,VK13,KB16} characterized the low-Reynolds-number flow--structure instability in compliant conduits. Specifically, their experiments \cite{VK13} provided support for a scaling $Re_c \sim \Sigma^{5/8}$ between the critical Reynolds number $Re_c$ at the transition and a dimensionless elastic modulus $\Sigma = \rho G D_h^2/\eta_0^2$ (albeit over just a single decade in $\Sigma$). This scaling is ``in between'' those predicted for instability of inviscid (bulk) modes ($Re_c \sim \Sigma^{1/2}$) and viscous (wall) modes ($Re_c\sim\Sigma^{3/4}$), both previously identified theoretically for flows in compliant conduits. More recent experiments \cite{NS15}, \emph{however}, do not quite support this scaling, suggesting a \emph{much stronger} dependence $Re_c \propto \Sigma^{3/2}$ near the transition point. As recently as 2019, inconsistencies have been found in previous calculations \cite{PS19}. A theoretical re-analysis of a planar Couette flow over a compliant surface of a non-Newtonian fluid under the Carreau model showed that shear-thinning has a strongly stabilizing effect \cite{TPS18}, with $Re_c \sim \Gamma^{3/(2n-1)}$, where $\Gamma = Gh_0/(V_p\eta_0)$ and $V_p$ is the velocity of the rigid plate driving the Couette flow. Meanwhile, another experimental study \cite{CSD19} motivated the scaling $Re_c \sim El^{-3/2}$, where $El = \lambda_r\eta_0/(a^2\rho)$ is an \emph{elasticity number} that characterizes the viscoelasticity of the polyacrylamide into water solution used. Therefore, at this time, it appears that a complete understanding and a predictive theory of the instability that enables efficient mixing of non-Newtonian fluids in compliant conduits at low Reynolds numbers is lacking. Nevertheless, the scalings reviewed here reveal that the critical Reynolds number for transition in a compliant conduit is \emph{highly tunable} using soft materials (such as PDMS with a low shear modulus $G$) and viscoelastic fluids (with sufficiently long relaxation time $\lambda_r$). A comprehensive  overview/perspective on the problem of stability and transition of flows compliant conduits is available in \cite{K21}.


\section{Advanced topics}
\label{sec:advanced}


\subsection{Biomimicry}

PDMS is biocompatible, which opens the possibility of both microfluidic analogues of organ functions \cite{Huhetal10,Lindetal17} and the embedding of microfluidic devices and components \textit{in vivo} \cite{W06}. As mentioned in section~\ref{sec:intro}, biological, physiological or even zoological implications of the coupling of flow and compliant boundaries are not topics that are covered in this review, and reader is referred to the reviews in \cite{S77,GJ04,LP09,HH11} (see also \cite{F93,F97} and \cite[chapter~8]{DS14}). Nevertheless, it is worth pointing out a few studies that build upon the concepts reviewed herein.

For example, Kiran Raj \textit{et al.}~\cite{RCDC18} suggested that flow of a Xanthan gum solution (a shear-thinning fluid) in a cylindrical pipe embedded in a PDMS block could mimic blood flow in a vein or artery, leading to ``biomimetic in-vitro models for lab-on-a-chip applications.'' Meanwhile, the non-Newtonian fluid flow in a slender elastic shell considered by Anand and Christov \cite{AC19c} was motivated by the mechanics of a compliant blood vessel sketched out in Fung's textbook \cite{F97} (see also \cite{A16}). More recently, Karan \textit{et al.}~\cite{Karan2020b} reconsidered this type of fluid--structure interaction in the presence of axial gradients in the elastic properties of the compliant wall. This kind of setup can be considered an \textit{in vitro} model of micro-circulation, in which the gradients in the elastic properties would be due to diseased conditions \textit{in vivo}. It would be of interest to extend the latter study to non-Newtonian fluid flows.


\subsection{Unsteady dynamics}
\label{sec:unsteady}

Perhaps the most obvious unsteady problem related to the topics reviewed herein concerns the inflation (or deflation) of a compliant flow conduit due to a suddenly imposed (or ceased) axial pressure drop. This problem has applications to microfluidic \emph{stop-flow lithography} \cite{DGPHD07}, for which it is important to accurately quantify the time scale of inflation (or deflation) of the microchannel. Dendukuri \textit{et al.}~\cite{DGPHD07} analyzed this unsteady fluid--structure interaction for the 2D microchannel geometry (figure~\ref{fig:sketch-geom}(a)). Mukherjee \textit{et al}.~\cite{MCC13} generalized the latter to electroosmotic flow (see also section~\ref{sec:electro}). Elbaz and Gat \cite{EG14} considered the case of a cylindrical shell geometry (figure~\ref{fig:sketch-geom}(c)). Meanhile Mart{\'{i}}nez-Calvo \textit{et al.}~\cite{MCSPS19} considered the 3D microchannel geometry (figure~\ref{fig:sketch-geom}(d)). These studies focused on characterizing the coupled physics (and parameters) that determine the transient's time scale, as well as the actual inflation/deflation dynamics. Elbaz \textit{et al.}~\cite{EJG18} additionally analyzed compressible (gas) flow in the 2D microchannel geometry (figure~\ref{fig:sketch-geom}(a)), highlighting a variety of self-similar behaviors set by different physical balances. Meanwhile, Anand and Christov \cite{AC20} considered compressible flow within the cylindrical shell geometry (figure~\ref{fig:sketch-geom}(c)) with viscoelastic damping in the structure. They showed that fluid--structure interaction can generate a streaming flow within the compliant conduit. 

All these works \cite{DGPHD07,MCC13,EG14,EJG18,MCSPS19,AC20} are restricted to Newtonian working fluids. The basic idea in these papers is to generalize the governing nonlinear ODE (\ref{eq:RK_dpdz_NN}) for the hydrodynamic pressure $p(z)$ to a nonlinear \emph{partial} differential equation (PDE) for the unsteady pressure $p(z,t)$. In this subsection, $t$ denotes time, rather than wall thickness. Such a model can be derived by cross-sectionally averaging the governing equations, in which case conservation of mass requires that (see, \textit{e.g.}, \cite{S77,GJ04}):
\begin{equation}
	\frac{\partial \mathcal{A}}{\partial t} + \frac{\partial q}{\partial z} = 0,
	\label{eq:unsteady_com}
\end{equation}
where now $\mathcal{A} = \mathcal{A}(z,t)$ and $q = q(z,t)$ due to unsteady fluid--structure interaction. For a soft hydraulic system, both $\mathcal{A}$ and $q$ can be calculated using the theory reviewed in section~\ref{sec:theory} and substituted into (\ref{eq:unsteady_com}). An example of the end-result of such a calculation is (\ref{eq:forced_cyl}) below. These unsteady models, which are one-dimensional (1D), depending only on the axial coordinate $z$, are more complex than the steady models considered in section~\ref{sec:theory}. Nevertheless, the unsteady 1D models can be solved (sometimes analytically) by perturbation \cite{H13} and self-similarity methods \cite{B96}. 

A more detailed 1D model, consisting of coupled PDEs for the \emph{finite}-Reynolds-number lubrication flow (\textit{i.e.}, $Re=\mathcal{O}(1)$ and $\hat{Re}\not\to0$) of a Newtonian fluid underneath an elastic membrane with inertia, bending and nonlinear stretching (von K\'arm\'an strains), was developed by Inamdar \textit{et al.}~\cite{IWC20} and solved numerically. Further, they showed that the inflated shapes of such channels are linearly stable to global perturbations. It is of interest to generalize all these unsteady models to non-Newtonian working fluids. Here, perhaps the most intriguing aspect is the possible breakdown of lubrication theory for a shear-thinning fluid under the power-law model \cite{GKATB17}, when inertia is included at the leading order in the lubrication theory (as in \cite{IWC20}).

\begin{figure}
	\centering
	\subfloat[][]{\includegraphics[width=\columnwidth]{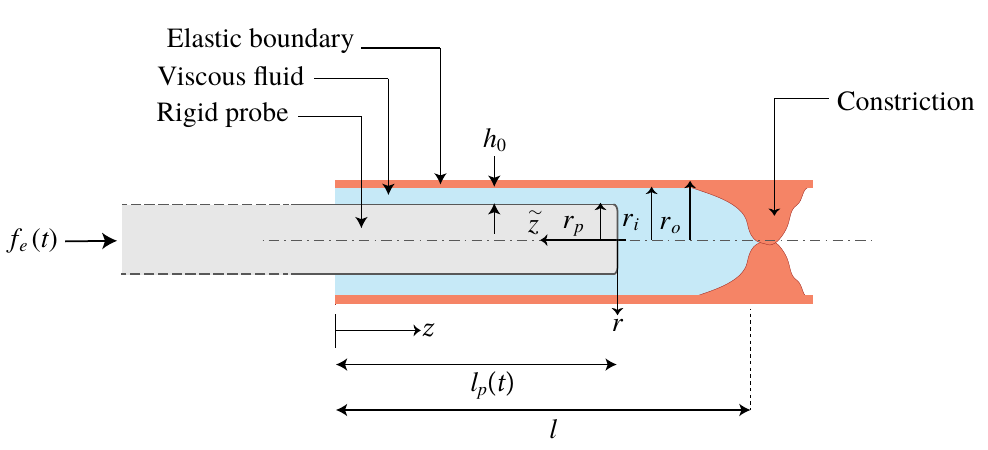}}\\
	\subfloat[][]{\includegraphics[width=0.9\columnwidth]{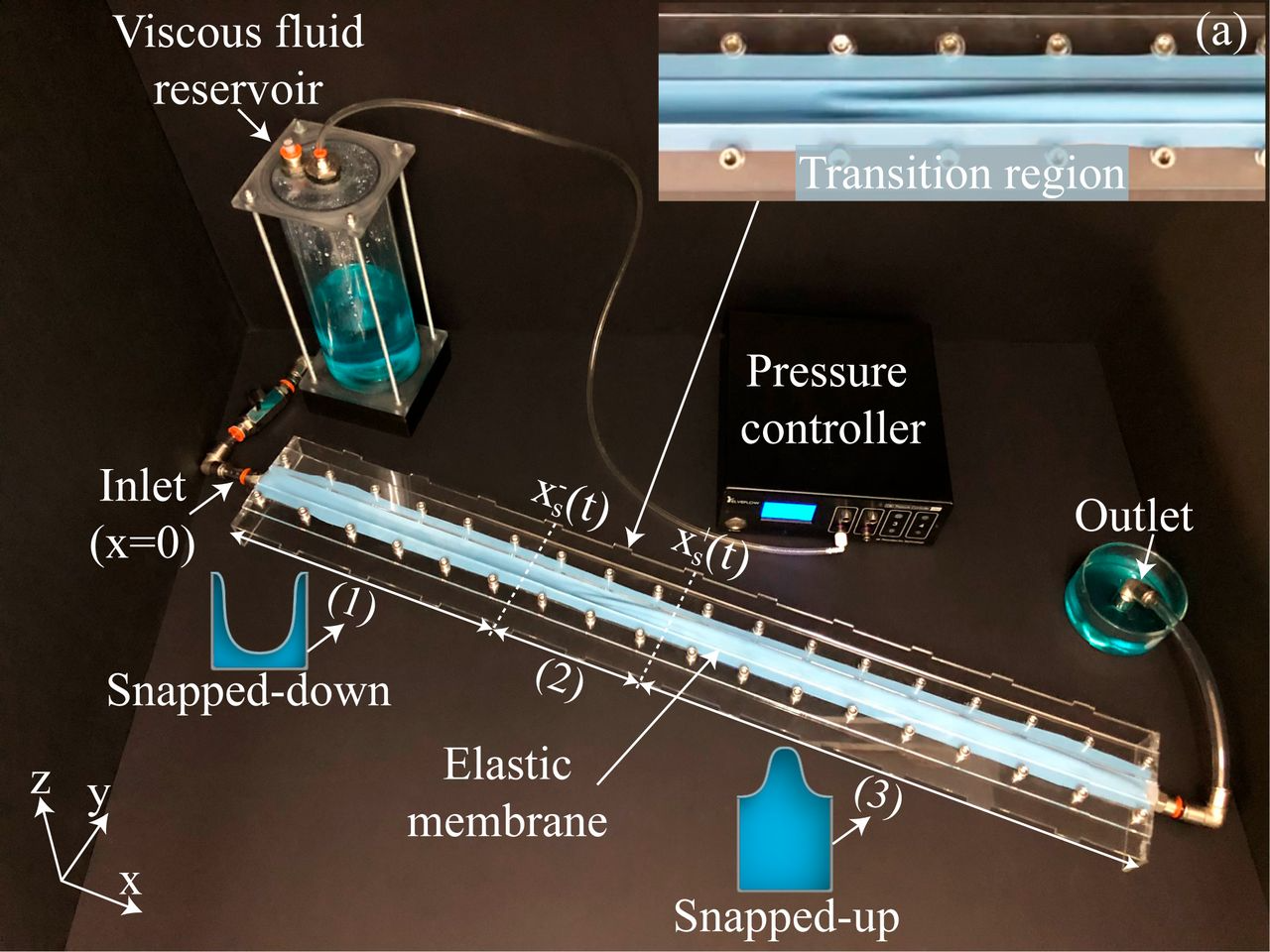}}
	\caption{(a) Schematic of a model of a minimally invasive medical procedure. A rigid cylinder is forced by $f_e(t)$ into a fluid-filled elastic tube. The fluid--structure interaction between the flow generated around the cylinder and the attendant elastic deformation sets the intrusion dynamics $l_p(t)$. Reproduced from \cite{VEG19}  with permission from Cambridge University Press \copyright\ 2019. (b) Experimental setup used to demonstrate multistability and underactuated fluidic control of an elastic membrane by a low-Reynolds-number viscous channel flow beneath. The working fluid is Newtonian: glycerol with $\rho = 1260$ kg/m$^3$ and  $\eta_0 = 1.412$ Pa$\cdot$s. A reservoir controls the inlet pressure at $x=0$. A transition region with two fronts, $x_s^\pm(t)$, slowly moves to the right along the channel and actuates the shape. Reproduced from \cite{PSMG20} with permission.}
	\label{fig:forced_cyl}
\end{figure}

Unsteady fluid--structure interactions also arise during the forced motion of a cylindrical probe into a fluid-filled elastic tube. The latter is a model system for a number of minimally invasive medical procedures such as endoscopies and laparoscopies \cite{VEG19}. As shown in figure~\ref{fig:forced_cyl}(a), the motion of the rigid cylinder induces a flow in the gap ($h_0=r_i-r_p$) between itself and the soft wall. The hydrodynamic pressure of this flow can deform the compliant outer wall. The basic equation governing the unsteady dynamics of the insertion or retraction of the rigid cylinder, in the case of a Newtonian fluid, is \cite{VEG19}:
\begin{equation}
	\frac{\partial u_r}{\partial t} - \frac{1}{12\eta_0}\frac{\partial}{\partial z}\left[(h_0+u_r)^3\frac{\partial p}{\partial z}\right] 
	+ \frac{1}{2}\frac{\partial l_p}{\partial t}\frac{\partial u_r}{\partial z} = 0,
	\label{eq:forced_cyl}
\end{equation}
subject to two integral constraints enforcing global conservation of mass and conservation of momentum on the inserted rod. The quantities of interest to be determined in this problem are the radial wall deformation $u_r(z,t)$ (or the pressure distribution $p(z,t)$) and the protrusion motion $l_p(t)$, given different types of forcing $f_e(t)$. Like the results reviewed in section~\ref{sec:theory}, (\ref{eq:forced_cyl}) is also based on the lubrication theory of the flow coupled to the equations of linear elasticity for the deformation of the outer wall. The deformation--pressure relation in this case is $u_r(z,t) = \frac{r_i^2}{r_o-r_i} \frac{p(z,t)}{E}$ \cite{EG16} (compare to (\ref{eq:ur_shell})), which can be used to eliminate $u_r$ and write (\ref{eq:forced_cyl}) as a PDE for $p$.  

In another variation on the unsteady dynamics, one considers \emph{viscous peeling} of the front (\textit{e.g}., at $z=l_p(t)$ in figure~\ref{fig:forced_cyl}(a)). Salem \textit{et al.}~\cite{SGOG20} successfully leveraged the viscous peeling mechanism for the design of actuators for applications in soft robotics \cite{Poly17}. Importantly, these are multistable systems that lead to ``underactuated fluidic control,'' which enables patterns to be \emph{continuously switched} via a single control parameter (\textit{e.g.}, the inlet pressure of the channel) \cite{PSMG20}. Peretz \textit{et al.}~\cite{PSMG20} demonstrated underactuated control of the multistable shape of soft hydraulic conduits. Their experimental setup is shown in figure~\ref{fig:forced_cyl}(b), including a typical pattern of the elastic-walled channel.

To understand the effect of non-Newtonian rheology on unsteady soft hydraulics, Boyko \textit{et al.}~\cite{BBG17} analyzed the flow in a slender elastic tube under the power-law model. As discussed in section~\ref{sec:theory}, the coupling of the nonlinear pressure gradient--flow rate relation for a non-Newtonian fluid with a deformation--pressure relation for the flow-induced deformation can lead to an intractable model. Therefore, a key simplification made in \cite{BBG17} is that the tube's cross-sectional area does not change significantly, which is akin to linearizing the $(h_0+u_r)^3$ term in (\ref{eq:forced_cyl}). Then, an unsteady nonlinear PDE governs the pressure evolution in the tube:
\begin{equation}\label{eq:p_laplace}
\eqalign{
 	\frac{\partial p}{\partial t} = \left(\frac{1}{3+1/n}\right)\left[\frac{E (r_o-r_i) r_i}{5-4\nu}\right]\frac{\partial}{\partial z}\left[\frac{1}{\tilde{\eta}(\dot{\gamma})}\frac{\partial p}{\partial z}\right],\\
	\tilde{\eta}(\dot{\gamma}) = \left(\frac{K^{1/n}r_i^{1-1/n}}{2^{1-1/n}}\right)\left|\frac{\partial p}{\partial z}\right|^{1-1/n},
}
\end{equation}
where $r_i=a$ is the fluid domain's (constant) radius, and $r_o-r_i$ is the thickness of the tube, keeping with the notation from figure~\ref{fig:forced_cyl}(a). In (\ref{eq:p_laplace}), the effective viscosity $\tilde{\eta}$ has an ``inverse role'' compared to the momentum equation (\ref{eq:mom}). Using similarity methods, (\ref{eq:p_laplace}) was solved for a number of unsteady problems, such as an instantaneous injection of mass at the tube's inlet, sudden change of the inlet pressure, and for the post-transient regime following an oscillatory inlet pressure \cite{BBG17}. Observe that (\ref{eq:p_laplace}) at steady state ($\partial p/\partial t=0$) reduces to (\ref{eq:vz_tube_plaw}) (subject to properly defining the flow rate to match), but not to (\ref{eq:dpdz_tube_plaw}) (due to the linearization of the $(h_0+u_r)^3$ term).

Similar unsteady problems involving viscous non-Newtonian flows in compliant conduits arise when analyzing the relaxation of an elastic fracture filled with a complex fluid \cite{CCLDF19,CLLDF21}. These types of fluid--structure interactions involving complex fluids are common in \emph{hydraulic fracturing} \cite{BDRM16}. In \cite{CCLDF19,CLLDF21}, as a simplification, the deformation is considered to be uniform in the axial direction, so that the channel (fracture) height is just $h(t)$. Then, from (\ref{eq:unsteady_com}), a suitable PDE for the non-Newtonian rheology can be derived. The fracture (of length $L$) is considered to have some resistance to opening/proclivity to closing (quantified by a Winkler-like effective stiffness $k_w=E/l$, where $l$ is the fracture spacing in the transverse direction). This coupling between flow and elasticity is captured by a global force balance between the hydrodynamic pressure applied on the fracture walls, its elastic properties, and any overload pressure $p_0$: $\int_0^L p(z,t) \,\mathrm{d}z = (E/l)h(t)L + p_0L$ \cite{Dana18}.

From the results on unsteady soft hydraulics reviewed in this subsection, it is evident that the same basic equations that govern soft hydraulics in microfluidics also allow one to analyze biomedical (minimally invasive surgery) and even geophysical (hydraulic fracturing) problems. The reason for this generality is that the physics (and resultant models) reviewed in section~\ref{sec:theory} hold across vastly different scales, as long as the basic assumptions (low-Reynolds-number flow, small aspect ratio geometry, lubrication approximation, linearly elastic deformation, etc.) are satisfied.


\subsection{Electrohydrodynamics}
\label{sec:electro}

Electrohydrodynamics, also often referred to as \emph{electrokinetics}, concerns the flow of electrically conducting fluids. Electric fields cause transport of charged ions within electrolyte solutions. The motion of the ions can ``drag'' the surrounding fluid, which is perhaps the most striking electrohydrodynamic phenomenon of relevance to microfluidics. The bulk motion (flow), relative to a charged surface, of an electrically conducting fluid due to an imposed electric field is called \emph{electroosmosis}. \textit{Vice versa}, a pressure-driven flow of an electrically conducting fluid can lead to the build up of an electrokinetic potential (and, consequently, an electric current in the flow-wise direction), which is called a \emph{streaming potential}. The standard textbook on this topic is by Probstein \cite{P94}, while the monograph by Li \cite{L04} specifically focuses on electrokinetics in the microfluidics context. Further coverage is available in \cite{SSA04,NW06,B08,C10_book,C13_book}. 

Electroosmotic flows are generally ``weak'' and difficult to realize at the macroscale. Stone \textit{et al.}~\cite{SSA04} estimate that an electric field strength on the order of few kV/cm is needed to achieve flow speeds on the order of mm/s. A high-voltage power supply would be necessary to generate electric fields of this strength in a channel of length of a few cm. Nevertheless, the flow speeds are in the range relevant to microfluidics. A general introduction to electrokinetic actuation of microscale flows is provided by Chakraborty and Chakraborty \cite[section~1.4.5]{C10_book_Ch1}, while Ghosal \cite{C10_book_Ch2} covers the mathematical modeling of charge transport during electroosmotic flows. Alizadeh \textit{et al.}~\cite{AWWD21} provide an updated tutorial review on electroosmotic flows in micro- and nanofluidic applications, including a historical overview and discussion of applications of confined flows through micro- and nanoporous media. Electrohydrodynamics is a broad field, and only a few key results related to non-Newtonian flows and flows in compliant conduits are summarized here.

Chakraborty and Chakraborty \cite{CC10,CC11} appear to have been the first to consider the effect of a compliant boundary on the electroosmotic flow of a Newtonian fluid in a narrow confinement, motivated by the phenomenological deformation--pressure relation of Steinberger \textit{et al.}~\cite{SCKC08}. Specifically, by analyzing the classical ``slider bearing'' steady flow problem from lubrication theory, they showed \cite{CC11} that the load bearing capacity can increase by a factor of up to $\approx 1.5$ under an applied electric field. However, the electrokinetic augmentation effect weakens for highly compliant substrates. Das \textit{et al}.~\cite{C13_book_Ch3} summarize a number of these theoretical results on coupled microscale problems involving lubrication flows of electrically conducting fluids in compliant conduits, or coupled to heat and mass transfer, or in the presence of capillarity.

More recently, the actuation of thin elastic membranes by nonuniform electroosmotic flow was demonstrated by Rubin \textit{et al.}~\cite{RTGB17} and Boyko \textit{et al.}~\cite{BEGGB19}, both theoretically and experimentally. These studies highlight that electrohydrodynamics is an effective flow control and actuation mechanism for soft hydraulic systems, with a high degree of tunability, which allows for nontrivial pre-set wall deformations to be achieved. On the other hand, a computational study by de Rutte \textit{et al.}~\cite{Rutte16} on electroosmotic flow in a compliant microchannel concluded that a narrow and wide flow conduit can collapse over a range of imposed electric field strengths, and the effect is ``exacerbated for soft materials such as PDMS.''  Boyko \textit{et al.}~\cite{BEGB20,BIBG20} expounded on this idea, showing that ``above a certain electric field threshold, negative gauge pressure induced by electro-osmotic flow causes the collapse of its elastic wall.'' They experimentally demonstrated this novel type of fluid--structure instability and showed that an electroosmotic fluid--structure interaction parameter $\beta \propto B/(k_w L h_0^3/\eta_0)$ (recall section~\ref{sec:FSI}) controls the (in)stability in a 2D Cartesian configuration as in figure~\ref{fig:sketch-geom}(a). Here, $B$ has units of m$^3$/s and quantifies the strength of the electroosmotic flow, while $k_w$ is a Winkler-like stiffness quantifying the elastic resistance to deformation of the compliant wall. The channel deformation $h=h(t)$ is taken to be uniform in the flow-wise direction, as in the hydraulic fracture example reviewed in section~\ref{sec:unsteady}. 

To highlight the novel features of these electroosmotic flows in soft hydraulic conduits, it is instructive to consider the unsteady lubrication model from \cite{BIBG20}. In this case, (\ref{eq:forced_cyl}) becomes
\begin{eqnarray}
	\frac{\partial h}{\partial t} - \frac{1}{12\eta_0}\frac{\partial}{\partial z}\left(h^3 \frac{\partial p}{\partial z}\right) = -\frac{1}{2}\frac{\partial}{\partial z}(h u_\mathrm{EOF}), \label{eq:eof_com}\\
	u_\mathrm{EOF} = -\frac{\varepsilon \zeta(z)}{\eta_0} \frac{E_0}{h(z,t)} \left\{\begin{array}{@{}l@{\quad}l}
     	h_0, &\mathrm{const.~current},\\[2mm]
     	\frac{L}{\int_0^{L}h^{-1}\mathrm{d}z}, &\mathrm{const.~voltage}.
    \end{array}\right.\nonumber\\ \label{eq:eof_u}
\end{eqnarray}
Here, $\varepsilon$ is the fluid's permittivity, $\zeta$ is electrokinetic potential along the channel wall, and $E_0$ is the strength of the imposed electric field. Note that the electric field $E$ is nonuniform in this narrow confinement, and it is given by the product of the second and third terms on the right-hand side of (\ref{eq:eof_u}). In model from \cite{BIBG20}, the pressure $p(z,t)$ in (\ref{eq:eof_com}) is \emph{not} obtained from a local deformation--pressure relation, as in section~\ref{sec:deformation}. Instead $p(z,t)$ is related to $\partial^4 u_y/\partial z^4$, $\partial^2 u_y/\partial z^2$, $\rho g h$, and $\varepsilon E^2$ to account for axial bending, axial tension, gravity, and Maxwell stresses induced by the electric field, respectively (see also \cite{HM04}). The nontrivial coupling of flow and deformation through the electrohydrodynamics, represented by the right-hand side of (\ref{eq:eof_com}), drives the unsteady nonlinear dynamics of the system (including instability).

As discussed in section~\ref{sec:rheology}, complex fluids are commonly encountered in microfluidics. Just as their response to shear differs from Newtonian fluids, so does their response to an electric fields. Zhao and Yang \cite{ZY13} surveyed non-Newtonian effects that might arise in conjunction with electrokinetics, generally concluding that electrokinetic phenomena are enhanced for shear-thinning fluids. Das and Chakraborty \cite{DC06}, followed by Zhao \textit{et al.}~\cite{ZZMY08,ZY11}, obtained approximate and exact solutions for the electroosmotic flow profile in a 2D channel under the power-law model. Solutions for electroosmotic flow of viscoelastic fluids between rigid parallel plates have also been found, as mentioned in \cite[p.~99]{ZY13}. From these results, it would be possible to obtain a generalized flow rate--pressure gradient--electric field relation to incorporate as a building block of a predictive physical theory, as in section~\ref{sec:theory}. Meanwhile, Zimmerman \textit{et al.}~\cite{ZRC06} considered a T-junction microchannel and numerically solved the governing equations for the electrokinetic flow of a Carreau fluid through it. They demonstrated the potential of this setup for microrheometry by showing that a one-to-one mapping exists between the parameters in (\ref{eq:carr}) and the end-wall pressure in the junction. The current outlook \cite{S19} in the field of non-Newtonian electrokinetics is towards enabling interactions with the constituents of blood in microchannels. The ultimate goal is new methods for blood plasma separation and isolation of circulatory tumor cells in the microcirculation \cite{S19}. 

Overall, it appears that prior studies have not considered the combined effect of non-Newtonian rheology, deformation of the compliant flow conduit, and electrokinetics all together, leaving this problem open. Furthermore, beyond electrokinetics, non-hydrodynamic effects, such as van der Waals and solvation forces \cite{KCC20,KCC21} as well as finite-size effects and steric interactions \cite{NCC17}, arise at nanometer level confinements. These effects can also be incorporated into the physical theories and mechanistic models reviewed herein, assuming the continuum hypothesis still holds.


\subsection{Non-invasive measurements, particles, and sieving}

\begin{figure}
	\centering
	\includegraphics[width=0.45\textwidth]{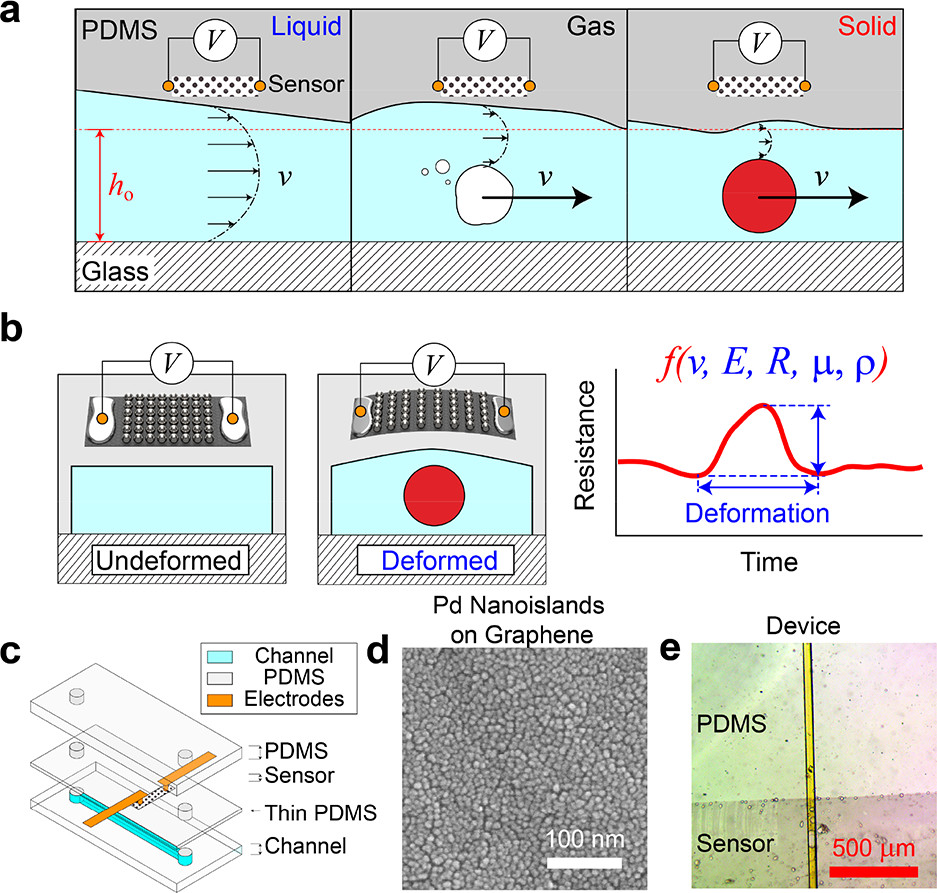}
	\caption{(a) Schematic of how a clear fluid (left), a bubbly fluid (center) and a particle-laden fluid (right) flow induces deformation of a soft channel wall made from PDMS. (b) The deformation is measured from a voltage change generated by a piezoresistive thin film embedded into the soft wall. The voltage depends on the elastic wall's properties ($E$, $\nu$), fluid properties ($\mu=\eta_0$, $\rho$) and particle/bubble size ($R$). (c) Schematic of how the microchip with a deformable channel and embedded sensing electrodes is manufactured. (d) Scanning electron microscope (SEM) image of the piezoresistive thin films, which is made from ``palladium nano-islands on graphene.'' (e) Bright-field microscope image of a portion of the device. The fluid channel is visible as the vertical line in the center. Reprinted with permission from Dhong \textit{et al.}\ 2018 \textit{Nano Lett.}\ \textbf{18} 5306--5311 \cite{Dhong18}. Copyright (2018) American Chemical Society.}
	\label{fig:optics-free}
\end{figure}

Dhong \textit{et al.}~\cite{Dhong18} proposed (and realized experimentally) a novel optics-free, non-contact technique for measuring the motion of liquids, bubbles and particles in microchannels. An earlier technique by Ozsun \textit{et al.}~\cite{OYE13} was based on ``constitutive pressure--deformation'' curves, measured ahead of time, so that the axial pressure (and deformation) profiles could be inferred from a single pressure drop measurement across a microchannel. Similarly, Anand \textit{et al.}~\cite{AMC20} demonstrated a ``hydrodynamic bulge test'' \textit{in silico} for measuring the material properties of thin elastic films sealing a microchannel, from the pressure drop across the length of the channel. Once again, the key innovation of these methods is to harness the flow-induced deformation of a PDMS microchannel. In \cite{Dhong18}, the bulging of the conduit due to either the hydrodynamic pressure during flow, or the presence of a large particle or bubble in the channel, is transferred into a voltage by a metal/graphene strain sensor, as shown in figure~\ref{fig:optics-free}. Therefore, no optical information is needed to perform this measurement. A related idea was discussed by Niu \textit{et al.}~\cite{NNBR17}, who interrogated the change in impedance for the flow of an electrolyte solution in a compliant flow conduit. They showed that the flow-induced deformation can be characterized via the measured decrease in the resistance (between two electrodes along the channel) of the flowing solution.

In a similar vein, using the approach outlined in section~\ref{sec:theory}, Chen \textit{et al}.~\cite{Chen19} proposed  (and realized experimentally) a \emph{nanosieve}. Separation of particles and cells by size or by functionalization is a common operation in labs-on-a-chip \cite{DC09}. Using the elastic deformation of a microchannel as a mechanism to control the degree of confinement in the channel, they proposed a formula for the distance that a particle of a given size can travel before becoming ``stuck'' and, therefore, sieved out of the flow. Further, the efficiency of filtration was estimated using the theory of the flow-induced deformation. Even with significant conduit deformation at high flow rates, the efficiency was found to be as high as 80\% and verified experimentally. Beyond \cite{Chen19}, the topic of objects (particles, bubbles, cells, spheres or cylinders) near compliant surfaces (originally named \emph{soft lubrication} \cite{SM04,SM05,EPKVS21}) is too broad to cover here. Some aspects are reviewed by Karan \textit{et al}.~\cite{KCC18}.

Although the working fluids in these examples are Newtonian, the proposed experimental methods are aimed at the rheologically complex biofluids encountered in lab-on-a-chip applications. Therefore, the predictive physical theories reviewed in section~\ref{sec:theory} are expected to provide insights when applied to these problems. Importantly, the viscoelasticity of the suspending fluid can have a profound effect on particle motion and migration in microfluidic flows \cite{DGM17,ZP20}, which would have to be addressed in the context of studies like \cite{Dhong18,Chen19}.

\begin{figure*}
	\centering
	\subfloat[][]{\includegraphics[height=0.26\textwidth]{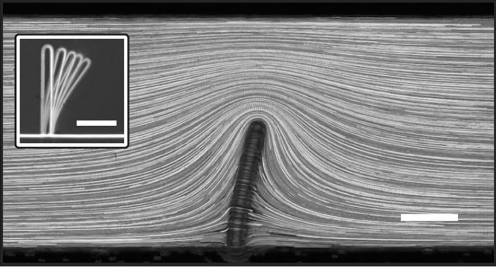}}\qquad
	\subfloat[][]{\includegraphics[height=0.26\textwidth]{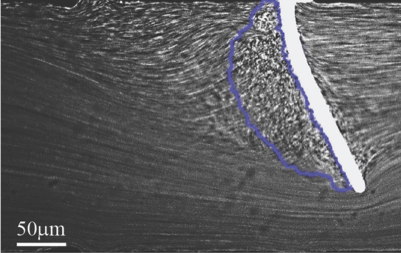}}
	\caption{(a) Bending of a soft PDMS beam ($E=64\pm22$ kPa) due to low-Reynolds-number Newtonian fluid (100\% PEGDA-575) flow (left to right, visualized by pathlines). The beam height is $h_b=226$ $\mu$m in a channel that is $h_0=400$ $\mu$m tall. The flow is confined in the transverse direction across a depth of $d_c=45$ $\mu$m. The inset shows the deformation for flow rates $q=0, 3, 8, 15$ and $30$ $\mu$L/min, while the main image is for $3$ $\mu$L/min. Scale bars represent $100$ $\mu$m. Reproduced from \cite{Wexler13} with permission from Cambridge University Press \copyright\ 2013.
	(b) Bending of a soft PDMS beam ($E=2\pm0.3$ MPa) due to low-Reynolds-number non-Newtonian polymeric fluid (Flopaam 3630 mixed with deionized water at a concentration of 0.02\% by weight) flow (left to right, visualized by streaklines) at finite Weissenberg number. Dimensions are similar to (a). Dark field imaging reveals flow fluctuations and vortex shedding in the circled region, despite the negligible flow inertia. Reproduced from \cite{DMSLR20} with permission from The Royal Society of Chemistry \copyright\ 2020.}
	\label{fig:fsi_beam}
\end{figure*}

\subsection{Rigid walls, soft obstacles}

Viscous flow past a deformable (soft) beam is a classical problem in fluid--structure interactions (see, \textit{e.g.}, the numerous examples in \cite{DS14}). In fact, a version of this problem (in which a channel flow is parallel to the beam) is a benchmark for numerical methods for fluid--structure interaction (the so-called Turek--Hron problem \cite{TH06}). In a rigid channel, an anchored series of elastic fibers (``hairs'') can deform due to flow, thus increase the flow area much like a compliant wall (as in section~\ref{sec:theory}). However, when the fibers are cantilevered at an angle towards the wall, they were shown to decrease the flow area, producing a ``a rectification nonlinearity'' \cite{ACLH17}, which can be harnessed for designing diodes and generating pumping in microfluidics.

In this context, Wexler \textit{et al.}~\cite{Wexler13} performed experiments (figure~\ref{fig:fsi_beam}(a)) and modeling of this problem, demonstrating that the tip deflection $u_\mathrm{tip}$ of the beam due to Newtonian fluid flow obeys
\begin{equation}
\eqalign{
u_\mathrm{tip} \simeq \frac{6\eta_0 (d_c+d) h_0^4q}{EId_c^3}\nonumber\\
\quad\times
	\left\{\begin{array}{@{}l@{\quad}l}
     	0.15(h_b/h_0)^5, & h_b\ll h_0, \\
    	\frac{1}{2\pi}\log(1-h_b/h_0)^{-1}-0.176, & h_b\to h_0,
    \end{array}\right.
}
\label{eq:beam_deflect}
\end{equation}
where $I=dw^3/12$ is the moment of inertia of the beam, $h_b$ is the deformed length of the beam, $h_0$ is the height of the channel, $d_c$ is the transverse confinement dimension, and the beam has rectangular cross-section $d\times w$.

For the Newtonian case shown in figure~\ref{fig:fsi_beam}(a), no instabilities or unsteady effects are observed. However, if the working fluid is a polymeric solution (with stress relaxation time $\lambda_r$, as in section~\ref{sec:viscoelastic}), then purely elastic instabilities can occur even at vanishing Reynolds number (in the absence of inertia) \cite{L92,Shaq96,S21}. The dimensionless number quantifying this non-Newtonian effect is the Weissenberg number \cite{Poole12}:
\begin{equation}
	Wi = \frac{\mathrm{flow~elastic~forces}}{\mathrm{flow~viscous~forces}} \simeq \lambda_r \dot{\gamma} = \frac{\lambda_r q}{d_c^2(h_0-h_b)}
\end{equation}
for the flow in figure~\ref{fig:fsi_beam}. 
Dey \textit{et al.}~\cite{DMSLR20,DMSR20b} demonstrated that at finite $Wi=\mathcal{O}(1)$, these \emph{viscoelastic} fluid--structure interactions lead to instabilities, vortex shedding off the beam, and generally unsteady flows (figure~\ref{fig:fsi_beam}(b)). The physical mechanism behind this sequence of events  is distinctly different from vortex shedding past a beam due to inertia (finite Reynolds number) in Newtonian fluid flows. Understanding the physics of these viscoelastic microscale fluid--structure interactions past soft obstacles, and if and when they lead to predictable steady flows, is relevant for flow rate sensors and actuators for non-Newtonian fluids in microfluidics. As the problem is now unsteady and the flow is complex, a theoretical result such as (\ref{eq:beam_deflect}) is not available. Nevertheless, the frequency and amplitude of the beam's oscillations were measured as functions of $Wi$ for different types of polymeric liquids \cite{DMSLR20,DMSR20b}.


\section{Conclusions and outlook}
\label{sec:concl}

From labs-on-a-chip that analyze biological fluid samples (such as blood and saliva) to miniaturized devices that mimic the behavior of organs, microfluidics requires the manipulation of not only Newtonian but also non-Newtonian (complex) fluids. Cheap and facile manufacturing of the flow conduits in these devices from soft polymeric materials, such as PDMS, leads to flow-induced deformation of the passages. On the one hand, this is a side effect to be analyzed and avoided. On the other hand, compliance opens up new avenues for flexible microdevices, pumping, valving, mixing and so on. At the intersection of non-Newtonian fluid mechanics and soft matter physics, flows of complex fluids through compliant conduits remains an active topic of research in this emerging field of \emph{soft hydraulics}. Experiments have provided a number of observations, as well as new puzzles, that merit careful study. As theory catches up, predictive physical models are emerging to rationalize experiments. Much of the ``conventional wisdom'' on flows in compliant conduits has come from understanding the physics of Newtonian fluid flows. The focus has been on this case because it allows for significant progress in the mathematical analysis of challenging problems involving the coupled 3D equations of fluid mechanics and elasticity. Thus, non-Newtonian fluid flows through soft hydraulic conduits is a topic at the foreground of ongoing research, as summarized in this review.

From the practical point of view, this review surveyed the key physics from the recent literature, and provided a ``recipe'' for building predictive analytical models, as outlined in the stated scope (section~\ref{sec:scope}). First, the relationship between volumetric flow rate and axial pressure gradient is needed for a given flow (section~\ref{sec:q_dpdz}). This relationship is challenging (or impossible) to obtain in closed analytical form for most non-Newtonian fluids with shear-dependent viscosity. Notable exceptions are the power-law and Ellis models of the effective viscosity. Then, the channel height or tube radius in the pressure gradient--flow rate relation is to be replaced with an expression that accounts for the pressure-induced deformation of the height/radius of the conduit (section~\ref{sec:deformation}), suitably derived from the theory of elasticity. To solve the coupled problem, and obtain an analytical model, it is necessary to be able to ``separate'' (and integrate) the resulting ODE for the pressure (section~\ref{sec:FSI}). This manipulation is also not always possible analytically. However, at this point, even a straightforward numerical integration can be used to obtain the pressure distribution in the soft hydraulic conduit, from which all other quantities follow. It should be emphasized that these new theoretical developments require quantitative validation against novel, precise experiments and across a range of flow conditions (\textit{i.e.}, shear rates) for which the Carreau rheological model is best suited \cite{ML18}.

The discussion in this review focused mainly on non-Newtonian fluids with a shear-dependent viscosity at steady state. It is worth noting that there are few conclusive experimental or theoretical results on the flow rate--pressure drop characteristics of steady viscoelastic flows at the microscale (low Reynolds number), much less through non-uniform or deformable conduits (see, \textit{e.g.}, the discussions in \cite{RM01,KSAK09}). Notably, Boyko and Stone \cite{BS21b} recently showed how to use the reciprocal theorem for Stokes flow to calculate the leading-order non-Newtonian correction to the pressure drop--flow rate relation, using only the Newtonian base flow solution in a two-dimensional slowly-varying conduit of given shape. Indeed, the relevance of these non-Newtonian fluids flows to microfluidics is established \cite{A08}. However, the physical mechanism for why viscoelasticity (due to, \textit{e.g.}, just 0.01\% of a high-molecular-weight polymer added to an aqueous solution) leads to anisotropic flow resistance remains to be understood theoretically \cite{GQ04}. Nevertheless, Groisman \textit{et al.}~\cite{GEQ03} were able to exploit this viscoelastic fluid flow effect to design circuit elements, such as rectifiers and diodes, for fluidic control.

While certain open problems and suggested generalizations of the quoted results were noted along the way, it is now important to identify a number of fundamental topics that have not been addressed by research and experiments on soft hydraulics. Complex fluids (especially biological fluids) might exhibit wall-shear-stress dependent slip \cite{GZL03}, for which some exact solutions generalizing those in section~\ref{sec:q_dpdz} are known \cite{FNP12}. (For Newtonian fluid flow in a compliant microchannel with a patterned bottom solid wall that engenders linear Navier slip, Karan \textit{et al.}~\cite{KCCWC21} recently developed a model along the lines of section~\ref{sec:FSI}.) Next, although the deformation of PDMS-based microfluidics channels is well described by the theory of linear elasticity, at large enough pressures, geometric nonlinearities (such as stretching) can arise \cite{RS16,SC18}. Buckling is also possible, which can be harnessed to design a ``passive fuse'' \cite{GMV17}. In these cases, the present state of the art, is to employ a deformation--pressure \emph{correlation} for stretching of thin shells (\textit{e.g.}, \cite[pp.~29--33]{S11}) for the deformation--pressure relationship (section~\ref{sec:deformation}). A more general approach is to derive such a relationship from nonlinear beam \cite{IWC20} and shell \cite{WHJW10} theories, as commonly done in the literature on \emph{collapsible tubes}, which undergo large elastic deformation \cite{S77,GJ04,HH11}. 

\balance

Under large loads, PDMS can also exhibit material nonlinearity (hyperelastic response) \cite{N11}, which can also be handled within the framework reviewed herein \cite{AC19b}. Viscoelastic behavior of the working fluid was discussed in this review. However, the compliant conduit's walls can themselves exhibit viscoelastic response \cite{HS12} (although typical PDMS mixtures have a weak viscoelastic response \cite{LOLLCZ09}). Furthermore, \emph{poroelasticity} of the conduit might also be important if dealing with hydrogel materials \cite{HS12}, within which biomimetic ``hierarchically branched vascular networks'' of microchannels with porous walls resemble tissues \cite{Huang11}. Auton and MacMinn \cite{AM17,AM18} considered the flow-induced deformation a of poroelastic cylinder saturated with a Newtonian fluid. It would be of interest to generalize the results reviewed herein (focused on elasticity of the conduit walls only) to such hyper-, visco-, and poroelastic response of the walls due to the flow of a non-Newtonian fluid within \cite{AENY19}.

In conclusion, it is fitting to recall the last sentence of the 2004 review by Stone \textit{et al.}~\cite{SSA04}: ``These [microfluidic] systems should provide challenges in fluid dynamics, transport, and engineering design for generations of readers.'' Indeed, as the latter review matures to two decades, the next generation of readers has emerged. Undoubtedly, the challenges of non-Newtonian fluid dynamics, elastic deformation of soft materials, and engineering design at the microscale will continue to puzzle and delight them.


\ack

ICC would like to thank V Anand, O A Basaran, M Bercovici, E Boyko, J Chakraborty, S Chakraborty, V Cognet, A D Gat, T C Inamdar, G Juarez, P Karan, S C Muchandimath, J D J Rathinaraj, P Sanaei, T C Shidhore, H A Stone, D Vella, G Vishwanathan, X Wang, S T Wereley for fruitful discussions, collaborations, and insights on fluid--structure interactions at the microscale. The genesis of the ideas discussed in this review was also enabled the by research activities supported  by the US National Science Foundation under grant No.\ CBET-1705637 and the Scheme for Promotion of Academic and Research Collaboration (SPARC), a Government of India Initiative, under Project Code SPARC/2018-2019/P947/SL.

\section*{Data availability statement}
The data that support the findings of this study are available upon reasonable request from the authors.

\section*{ORCID iDs}
Ivan C.\ Christov \includegraphics[width=.75em]{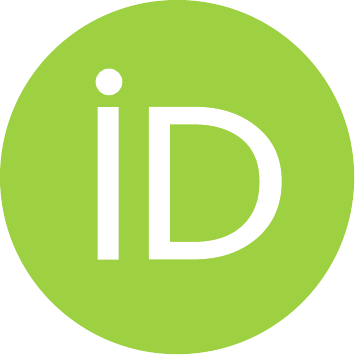}~\href{https://orcid.org/0000-0001-8531-0531}{https://orcid.org/0000-0001-8531-0531}

\cleardoublepage

\phantomsection
\addcontentsline{toc}{section}{References}

\bibliographystyle{iopart-num-mod}
\bibliography{jpcm-review-refs}

\providecommand{\newblock}{}
\begin{thebibliography}{100}
\expandafter\ifx\csname url\endcsname\relax
  \def\url#1{{\tt #1}}\fi
\expandafter\ifx\csname urlprefix\endcsname\relax\def\urlprefix{URL }\fi
\providecommand{\eprint}[2][]{\url{#2}}

\bibitem{SSA04}
Stone H~A, Stroock A~D and Ajdari A 2004 {\em Annu. Rev. Fluid Mech.\/}
  \href{http://dx.doi.org/10.1146/annurev.fluid.36.050802.122124}{{\bf 36}
  381--411}

\bibitem{SQ05}
Squires T~M and Quake S~R 2005 {\em Rev. Mod. Phys.\/}
  \href{http://dx.doi.org/10.1103/RevModPhys.77.977}{{\bf 77} 977--1026}

\bibitem{NW06}
Nguyen N~T and Wereley S~T 2006 {\em Fundamentals and Applications of
  Microfluidics\/} 2nd ed Integrated Microsystems Series (Norwood, MA: Artech
  House)

\bibitem{KML12}
Kim Y, Messner W~C and LeDuc P~R 2012 {\em Disruptive Sci. Tech.\/}
  \href{http://dx.doi.org/10.1089/dst.2012.0003}{{\bf 1} 41--53}

\bibitem{EASJ06}
El-Ali J, Sorger P~K and Jensen K~F 2006 {\em Nature\/}
  \href{http://dx.doi.org/10.1038/nature05063}{{\bf 442} 403--411}

\bibitem{PG07}
Prakash M and Gershenfeld N 2007 {\em Science\/}
  \href{http://dx.doi.org/10.1126/science.1136907}{{\bf 315} 832--835}

\bibitem{KCP15}
Katsikis G, Cybulski J~S and Prakash M 2015 {\em Nat. Phys.\/}
  \href{http://dx.doi.org/10.1038/nphys3341}{{\bf 11} 588--596}

\bibitem{W06}
Whitesides G~M 2006 {\em Nature\/}
  \href{http://dx.doi.org/10.1038/nature05058}{{\bf 442} 368--373}

\bibitem{MF_Market2}
{Grand View Research} 2021 Microfluidics market size, share \& trends analysis
  report Tech. Rep. GVR-1-68038-056-9
  \urlprefix\url{https://www.grandviewresearch.com/industry-analysis/microfluidics-market}

\bibitem{MW02}
McDonald J~C and Whitesides G~M 2002 {\em Acc. Chem. Res.\/}
  \href{http://dx.doi.org/10.1021/ar010110q}{{\bf 35} 491--499}

\bibitem{LOVB97}
L{\"otters} J~C, Olthuis W, Veltink P~H and Bergveld P 1997 {\em J. Micromech.
  Microeng.\/} \href{http://dx.doi.org/10.1088/0960-1317/7/3/017}{{\bf 7}
  145--147}

\bibitem{JMTT14}
Johnston I~D, McCluskey D~K, Tan C~K~L and Tracey M~C 2014 {\em J. Micromech.
  Microeng.\/} \href{http://dx.doi.org/10.1088/0960-1317/24/3/035017}{{\bf 24}
  35017}

\bibitem{Allen17}
Allen M 2017 Microfluidic chip can detect {HIV} and {MRSA} Physics World
  \urlprefix\url{https://physicsworld.com/a/microfluidic-chip-can-detect-hiv-and-mrsa/}

\bibitem{Yeh17}
Yeh E~C, Fu C~C, Hu L, Thakur R, Feng J and Lee L~P 2017 {\em Sci. Adv.\/}
  \href{http://dx.doi.org/10.1126/sciadv.1501645}{{\bf 3} e1501645}

\bibitem{XW98}
Xia Y and Whitesides G~M 1998 {\em Annu. Rev. Mater. Sci.\/}
  \href{http://dx.doi.org/10.1146/annurev.matsci.28.1.153}{{\bf 28} 153--184}

\bibitem{LSSBC09}
Liu M, Sun J, Sun Y, Bock C and Chen Q 2009 {\em J. Micromech. Microeng.\/}
  \href{http://dx.doi.org/10.1088/0960-1317/19/3/035028}{{\bf 19} 035028}

\bibitem{GEGJ06}
Gervais T, El-Ali J, G\"unther A and Jensen K~F 2006 {\em Lab Chip\/}
  \href{http://dx.doi.org/10.1039/b513524a}{{\bf 6} 500--507}

\bibitem{Yali20}
Yalikun Y, Ota N, Guo B, Tang T, Zhou Y, Lei C, Kobayashi H, Hosokawa Y, Li M,
  Enrique~Mu{\~n}oz H, Di~Carlo D, Goda K and Tanaka Y 2020 {\em Cytometry Part
  A\/} \href{http://dx.doi.org/10.1002/cyto.a.23944}{{\bf 97} 909--920}

\bibitem{DSMB97}
Delamarche E, Schmid H, Michel B and Biebuyck H 1997 {\em Adv. Mat.\/}
  \href{http://dx.doi.org/10.1002/adma.19970090914}{{\bf 9} 741--746}

\bibitem{HJLK02}
Hui C~Y, Jagota A, Lin Y~Y and Kramer E~J 2002 {\em Langmuir\/}
  \href{http://dx.doi.org/10.1021/la0113567}{{\bf 18} 1394--1407}

\bibitem{Xu14}
Xu S, Zhang Y, Jia L, Mathewson K~E, Jang K~I, Kim J, Fu H, Huang X, Chava P,
  Wang R, Bhole S, Wang L, Na Y~J, Guan Y, Flavin M, Han Z, Huang Y and Rogers
  J~A 2014 {\em Science\/}
  \href{http://dx.doi.org/10.1126/science.1250169}{{\bf 344} 70--74}

\bibitem{YKL16}
Yeo J~C, {Kenrya} and Lim C~T 2016 {\em Lab Chip\/}
  \href{http://dx.doi.org/10.1039/c6lc00926c}{{\bf 16} 4082--4090}

\bibitem{Huhetal10}
Huh D, Matthews B~D, Mammoto A, Montoya-Zavala M, Hsin H~Y and Ingber D~E 2010
  {\em Science\/} \href{http://dx.doi.org/10.1126/science.1188302}{{\bf 328}
  1662--1668}

\bibitem{Lindetal17}
Lind J~U, Busbee T~A, Valentine A~D, Pasqualini F~S, Yuan H, Yadid M, Park S~J,
  Kotikian A, Nesmith A~P, Campbell P~H, Vlassak J~J, Lewis J~A and Parker K~K
  2017 {\em Nat. Mat.\/} \href{http://dx.doi.org/10.1038/nmat4782}{{\bf 16}
  303--308}

\bibitem{DS14}
Duprat C and Stone H~A 2014 {\em Fluid-Structure Interactions in
  Low-{Reynolds}-Number Flows\/} (Cambridge, UK: Royal Society of Chemistry)

\bibitem{P14}
Pa\"{i}doussis M~P 2014 {\em Fluid-Structure Interactions: Slender Structures
  and Axial Flow\/} 2nd ed (San Diego, CA: Academic Press)

\bibitem{G01}
Gohar R 2001 {\em Elastohydrodynamics\/} 2nd ed (London: Imperial College
  Press)

\bibitem{F97}
Fung Y~C 1997 {\em Biomechanics: Circulation\/} 2nd ed (New York: Springer)

\bibitem{C10_book}
Chakraborty S (ed) 2010 {\em Microfluidics and Microfabrication\/} (New York:
  Springer Science+Business Media)

\bibitem{C13_book}
Chakraborty S (ed) 2013 {\em Microfluidics and Microscale Transport
  Processes\/} IIT Kharagpur Research Monograph Series (Boca Raton, FL: CRC
  Press)

\bibitem{C10_book_Ch4}
Das T and Chakraborty S 2010 Bio-microfluidics: Overview {\em Microfluidics and
  Microfabrication\/} ed Chakraborty S (New York: Springer Science+Business
  Media) \href{http://dx.doi.org/10.1007/978-1-4419-1543-6_4}{chap~4, pp
  131--179}

\bibitem{HT98}
Ho C~M and Tai Y~C 1998 {\em Annu. Rev. Fluid Mech.\/}
  \href{http://dx.doi.org/10.1146/annurev.fluid.30.1.579}{{\bf 30} 579--612}

\bibitem{G99}
Gad-el Hak M 1999 {\em ASME J. Fluids Eng.\/}
  \href{http://dx.doi.org/10.1115/1.2822013}{{\bf 121} 5--33}

\bibitem{Stone2007}
Stone H~A 2007 Introduction to fluid dynamics for microfluidic flows {\em CMOS
  Biotechnology\/} Series on Integrated Circuits and Systems ed Lee H,
  Westervelt R~M and Ham D (New York, NY: Springer Science+Business Media, LLC)
  \href{http://dx.doi.org/10.1007/978-0-387-68913-5{\_}2}{pp 5--30}

\bibitem{AG07}
Abgrall P and Gu{\'e} A~M 2007 {\em J. Micromech. Microeng.\/}
  \href{http://dx.doi.org/10.1088/0960-1317/17/5/R01}{{\bf 17} R15}

\bibitem{SDD19}
Stoecklein D and {Di Carlo} D 2019 {\em Anal. Chem.\/}
  \href{http://dx.doi.org/10.1021/acs.analchem.8b05042}{{\bf 91} 296--314}

\bibitem{Xia21}
Xia H~M, Wu J~W, Zheng J~J, Zhang J and Wang Z~P 2021 {\em Lab Chip\/}
  \href{http://dx.doi.org/10.1039/D0LC01120G}{{\bf 21} 1241--1268}

\bibitem{WPDF17}
Wang Y, Pilkington G~A, Dhong C and Frechette J 2017 {\em Curr. Opin. Colloid
  Interface Sci.\/} \href{http://dx.doi.org/10.1016/j.cocis.2016.09.009}{{\bf
  27} 43--49}

\bibitem{WTF17}
Wang Y, Tan M~R and Frechette J 2017 {\em Soft Matter\/}
  \href{http://dx.doi.org/10.1039/C7SM01061C}{{\bf 13} 6718--6729}

\bibitem{GJ04}
Grotberg J~B and Jensen O~E 2004 {\em Annu. Rev. Fluid Mech.\/}
  \href{http://dx.doi.org/10.1146/annurev.fluid.36.050802.121918}{{\bf 36}
  121--147}

\bibitem{HH11}
Heil M and Hazel A~L 2011 {\em Annu. Rev. Fluid Mech.\/}
  \href{http://dx.doi.org/10.1146/annurev-fluid-122109-160703}{{\bf 43}
  141--162}

\bibitem{S77}
Shapiro A~H 1977 {\em ASME J. Biomech. Eng.\/}
  \href{http://dx.doi.org/10.1115/1.3426281}{{\bf 99} 126--147}

\bibitem{P80}
Pedley T~J 1980 {\em The Fluid Mechanics of Large Blood Vessels\/} (Cambridge:
  Cambridge University Press)

\bibitem{LP09}
Lauga E and Powers T~R 2009 {\em Rep. Prog. Phys.\/}
  \href{http://dx.doi.org/10.1088/0034-4885/72/9/096601}{{\bf 72} 096601}

\bibitem{FZPN19}
Fallahi H, Zhang J, Phan H~P and Nguyen N~T 2019 {\em Micromachines\/}
  \href{http://dx.doi.org/10.3390/mi10120830}{{\bf 10} 830}

\bibitem{RC20}
Raj~M K and Chakraborty S 2020 {\em J. Appl. Polym. Sci.\/}
  \href{http://dx.doi.org/10.1002/app.48958}{{\bf 137} 48958}

\bibitem{KCC18}
Karan P, Chakraborty J and Chakraborty S 2018 {\em J. Indian Inst. Sci.\/}
  \href{http://dx.doi.org/10.1007/s41745-018-0073-5}{{\bf 98} 159--183}

\bibitem{CTS12}
Cheung P, Toda-Peters K and Shen A~Q 2012 {\em Biomicrofluidics\/}
  \href{http://dx.doi.org/10.1063/1.4720394}{{\bf 6} 026501}

\bibitem{B08}
Bruus H 2008 {\em Theoretical Microfluidics\/} (Oxford: Oxford University
  Press)

\bibitem{K10}
Kirby B~J 2010 {\em Micro- and Nanoscale Fluid Mechanics: Transport in
  Microfluidic Devices\/} (New York: Cambridge University Press)

\bibitem{KBA05}
Karniadakis G, Beskok A and Aluru N 2005 {\em Microflows and Nanoflows:
  Fundamentals and Simulation\/} ({\em Interdisciplinary Applied Mathematics\/}
  no~29) (New York, NY: Springer Science+Business Media)

\bibitem{T05}
Tabeling P 2005 {\em Introduction to Microfluidics\/} (Oxford, UK: Oxford
  University Press)

\bibitem{ADJRC19}
Anand V, Rathinaraj J~D~J and Christov I~C 2019 {\em J. Non-Newtonian Fluid
  Mech.\/} \href{http://dx.doi.org/10.1016/j.jnnfm.2018.12.008}{{\bf 264}
  62--72}

\bibitem{RS16}
Raj A and Sen A~K 2016 {\em Microfluid Nanofluid\/}
  \href{http://dx.doi.org/10.1007/s10404-016-1702-9}{{\bf 20} 31}

\bibitem{RCDC18}
Raj~M K, Chakraborty J, DasGupta S and Chakraborty S 2018 {\em
  Biomicrofluidics\/} \href{http://dx.doi.org/10.1063/1.5036632}{{\bf 12}
  034116}

\bibitem{DGNM16}
Del~Giudice F, Greco F, Netti P~A and Maffettone P~L 2016 {\em
  Biomicrofluidics\/} \href{http://dx.doi.org/10.1063/1.4945603}{{\bf 10}
  043501}

\bibitem{NDW19}
Nahar S, Dubey B~N and Windhab E~J 2019 {\em Phys. Fluids\/}
  \href{http://dx.doi.org/10.1063/1.5123182}{{\bf 31} 101905}

\bibitem{KK03}
Koo J and Kleinstreuer C 2003 {\em J. Micromech. Microeng.\/}
  \href{http://dx.doi.org/10.1088/0960-1317/13/5/307}{{\bf 13} 568--579}

\bibitem{Oh12}
Oh K~W, Lee K, Ahn B and Furlani E~P 2012 {\em Lab Chip\/}
  \href{http://dx.doi.org/10.1039/C2LC20799K}{{\bf 12} 515--545}

\bibitem{LME11}
Lugt P~M and Morales-Espejel G~E 2011 {\em Tribol. Trans.\/}
  \href{http://dx.doi.org/10.1080/10402004.2010.551804}{{\bf 54} 470--496}

\bibitem{CS97}
Carvalho M~S and Scriven L~E 1997 {\em J. Fluid Mech.\/}
  \href{http://dx.doi.org/10.1017/S0022112097005090}{{\bf 339} 143--172}

\bibitem{YK05}
Yin X and Kumar S 2005 {\em Phys. Fluids\/}
  \href{http://dx.doi.org/10.1063/1.1914819}{{\bf 17} 063101}

\bibitem{Purcell2013}
Purcell E~M and Morin D~J 2013 {\em Electricity and Magnetism\/} 3rd ed
  Electricity and Magnetism (New York, NY: Cambridge University Press)

\bibitem{Wang2008}
Wang Y, Bedekar A~S, Krishnamoorthy S, Sundaram S and Singhal A~K 2008 {Model
  Order Reduction (MOR)} {\em Encyclopedia of Microfluidics and Nanofluidics\/}
  ed Li D (Boston, MA: Springer)
  \href{http://dx.doi.org/10.1007/978-0-387-48998-8_1047}{pp 1382--1391}

\bibitem{SS93}
Sutera S~P and Skalak R 1993 {\em Annu. Rev. Fluid Mech.\/}
  \href{http://dx.doi.org/10.1146/annurev.fl.25.010193.000245}{{\bf 25} 1--19}

\bibitem{HB83}
Happel J~R and Brenner H 1983 {\em Low {Reynolds} Number Hydrodynamics with
  Special Applications to Particulate Media\/} 2nd ed (The Hague: Martinus
  Nijhoff Publishers)

\bibitem{LSS04}
Lauga E, Stroock A~D and Stone H~A 2004 {\em Phys. Fluids\/}
  \href{http://dx.doi.org/10.1063/1.1760105}{{\bf 16} 3051--3062}

\bibitem{W06_book}
White F~M 2006 {\em Viscous Fluid Flow\/} 3rd ed (New York, NY: McGraw-Hill
  Higher Education)

\bibitem{MOB05}
Mortensen N~A, Okkels F and Bruus H 2005 {\em Phys. Rev. E\/}
  \href{http://dx.doi.org/10.1103/PhysRevE.71.057301}{{\bf 71} 057301}

\bibitem{Case19}
Case D~J, Liu Y, Kiss I~Z, Angilella J~R and Motter A~E 2019 {\em Nature\/}
  \href{http://dx.doi.org/10.1038/s41586-019-1701-6}{{\bf 574} 647--652}

\bibitem{A04}
Ajdari A 2004 {\em C. R. Phys.\/}
  \href{http://dx.doi.org/10.1016/j.crhy.2004.02.012}{{\bf 5} 539--546}

\bibitem{HUZK09}
Hardy B~S, Uechi K, Zhen J and Kavehpour H~P 2009 {\em Lab Chip\/}
  \href{http://dx.doi.org/10.1039/b813061b}{{\bf 9} 935--938}

\bibitem{LESKUBL09}
Leslie D~C, Easley C~J, Seker E, Karlinsey J~M, Utz M, Begley M~R and Landers
  J~P 2009 {\em Nat. Phys.\/} \href{http://dx.doi.org/10.1038/NPHYS1196}{{\bf
  5} 231--235}

\bibitem{BULHLB09}
Begley M~R, Utz M, Leslie D~C, Haj-Hariri H, Landers J and Bart-Smith H 2009
  {\em Appl. Phys. Lett.\/} \href{http://dx.doi.org/10.1063/1.3266064}{{\bf 95}
  203501}

\bibitem{SLHLUB09}
Seker E, Leslie D~C, Haj-Hariri H, Landers J~P, Utz M and Begley M~R 2009 {\em
  Lab Chip\/} \href{http://dx.doi.org/10.1039/B903960K}{{\bf 9} 2691--2697}

\bibitem{ZM94}
Zengerle R and Richter M 1994 {\em J. Micromech. Microeng.\/}
  \href{http://dx.doi.org/10.1088/0960-1317/4/4/004}{{\bf 4} 192--204}

\bibitem{BAH87}
Bird R~B, Armstrong R~C and Hassager O 1987 {\em Dynamics of Polymeric
  Liquids\/} 2nd ed vol~1 (New York: John Wiley)

\bibitem{L98}
Larson R~G 1998 {\em The Structure and Rheology of Complex Fluids\/} Topics in
  Chemical Engineering (Oxford, UK: Oxford University Press)

\bibitem{CR08}
Chhabra R~P and Richardson J~F 2008 {\em Non-{Newtonian} Fluid Behaviour\/} 2nd
  ed (Oxford: Butterworth-Heinemann)

\bibitem{OP02}
Owens R~G and Phillips T~N 2002 {\em Computational Rheology\/} (London:
  Imperial College Press)

\bibitem{Netal07}
Nijenhuis K, McKinley G~H, Spiegelberg S, Barnes H~A, Aksel N, Heymann L and
  Odell J~A 2007 Non-{Newtonian} flows {\em Springer Handbook of Experimental
  Fluid Mechanics\/} ed Tropea C, Yarin A~L and Foss J~F (Berlin/Heidelberg:
  Springer-Verlag) \href{http://dx.doi.org/10.1007/978-3-540-30299-5_9}{chap~9,
  pp 619--743}

\bibitem{A08}
Anna S~L 2008 Non-{Newtonian} fluids in microfluidics {\em Encyclopedia of
  Microfluidics and Nanofluidics\/} ed Li D (Boston, MA: Springer)
  \href{http://dx.doi.org/10.1007/978-0-387-48998-8_1129}{pp 1480--1488}

\bibitem{C08}
Chakraborty S 2008 Non-{Newtonian} fluids in microchannel {\em Encyclopedia of
  Microfluidics and Nanofluidics\/} ed Li D (Boston, MA: Springer)
  \href{http://dx.doi.org/10.1007/978-0-387-48998-8_1128}{pp 1471--1480}

\bibitem{B77}
Boger D~V 1977 {\em Nature\/} \href{http://dx.doi.org/10.1038/265126a0}{{\bf
  265} 126--128}

\bibitem{P11}
Pritchard P~J 2011 {\em Fox {\&} McDonald's Introduction to Fluid Mechanics\/}
  8th ed (Hoboken, NJ: John Wiley {\&} Sons)

\bibitem{L07}
Leal L~G 2007 {\em Advanced Transport Phenomena: Fluid Mechanics and Convective
  Transport Processes\/} ({\em Cambridge Series in Chemical Engineering\/}
  vol~7) (New York, NY: Cambridge University Press)

\bibitem{Lee06}
Lee L~J 2006 {BioMEMS} {\em Encyclopedia of Chemical Processing\/} vol~1 ed Lee
  S (New York: Taylor {\&} Francis)
  \href{http://dx.doi.org/10.1081/E-ECHP-120030622}{pp 161--169}

\bibitem{BGB13}
Basaran O~A, Gao H and Bhat P~P 2013 {\em Annu. Rev. Fluid Mech.\/}
  \href{http://dx.doi.org/10.1146/annurev-fluid-120710-101148}{{\bf 45}
  85--113}

\bibitem{DC06}
Das S and Chakraborty S 2006 {\em Anal. Chim. Acta\/}
  \href{http://dx.doi.org/10.1016/j.aca.2005.11.046}{{\bf 559} 15--24}

\bibitem{S01}
Steller R~T 2001 {\em Polym. Eng. Sci.\/}
  \href{http://dx.doi.org/10.1002/pen.10883}{{\bf 41} 1859--1870}

\bibitem{CLLDF21}
Ciriello V, Lenci A, Longo S and {Di Federico} V 2021 {\em Adv. Water. Res.\/}
  \href{http://dx.doi.org/10.1016/j.advwatres.2021.103914}{{\bf 152} 103914}

\bibitem{BFO14}
Balmforth N~J, Frigaard I~A and Ovarlez G 2014 {\em Annu. Rev. Fluid Mech.\/}
  \href{http://dx.doi.org/10.1146/annurev-fluid-010313-141424}{{\bf 46}
  121--146}

\bibitem{F93}
Fung Y~C 1993 {\em Biomechanics: Mechanical Properties of Living Tissues\/}
  (New York, NY: Springer)

\bibitem{A16}
Ayyaswamy P~S 2016 Introduction to biofluid mechanics {\em Fluid Mechanics\/}
  ed Kundu P~K, Cohen I~M and Dowling D~R (San Diego, CA: Academic Press)
  \href{http://dx.doi.org/10.1016/B978-0-12-405935-1.00016-2}{chap~16, pp
  e1--e73} 6th ed

\bibitem{Ch05}
Chakraborty S 2005 {\em Lab Chip\/}
  \href{http://dx.doi.org/10.1039/B414566F}{{\bf 5} 421--430}

\bibitem{CC16}
Christov I~C and Christov C~I 2016 {\em Mech. Res. Commun.\/}
  \href{http://dx.doi.org/10.1016/j.mechrescom.2016.01.005}{{\bf 72} 59--63}

\bibitem{OAP11}
Oliveira M~S~N, Alves M~A and Pinho F~T 2011 Microfluidic flows of viscoelastic
  fluids {\em Transport and Mixing in Laminar Flows: From Microfluidics to
  Oceanic Currents\/} ed Grigoriev R (Weinheim, Germany: Wiley‐VCH Verlag)
  \href{http://dx.doi.org/10.1002/9783527639748.ch6}{chap~6, pp 131--174}

\bibitem{RM01}
Rothstein J~P and McKinley G~H 2001 {\em J. Non-Newtonian Fluid Mech.\/}
  \href{http://dx.doi.org/10.1016/S0377-0257(01)00094-5}{{\bf 98} 33--63}

\bibitem{KSAK09}
Koppol A~P, Sureshkumar R, Abedijaberi A and Khomami B 2009 {\em J. Fluid
  Mech.\/} \href{http://dx.doi.org/10.1017/S0022112009006922}{{\bf 631}
  231---253}

\bibitem{RAB21}
Ramos-Arzola L and Bautista O 2021 {\em J. Non-Newtonian Fluid Mech.\/}
  \href{http://dx.doi.org/10.1016/j.jnnfm.2021.104634}{{\bf 296} 104634}

\bibitem{CCSS18}
Christov I~C, Cognet V, Shidhore T~C and Stone H~A 2018 {\em J. Fluid Mech.\/}
  \href{http://dx.doi.org/10.1017/jfm.2018.30}{{\bf 841} 267--286}

\bibitem{R86}
Reynolds O 1886 {\em Phil. Trans. R. Soc. Lond.\/}
  \href{http://dx.doi.org/10.1098/rstl.1886.0005}{{\bf 177} 157--234}

\bibitem{R83}
Reynolds O 1883 {\em Phil. Trans. R. Soc. Lond. A\/}
  \href{http://dx.doi.org/10.1098/rstl.1883.0029}{{\bf 174} 935--982}

\bibitem{JL07}
Jackson D and Launder B 2007 {\em Annu. Rev. Fluid Mech.\/}
  \href{http://dx.doi.org/10.1146/annurev.fluid.39.050905.110241}{{\bf 39}
  19--35}

\bibitem{Stone2017}
Stone H~A 2017 Fundamentals of fluid dynamics with an introduction to the
  importance of interfaces {\em Soft Interfaces\/} ({\em Lecture Notes of the
  Les Houches Summer School\/} vol~98) ed Bocquet L, Qu{\'{e}}r{\'{e}} D,
  Witten T~A and Cugliandolo L~F (New York, NY: Oxford University Press)
  \href{http://dx.doi.org/10.1093/oso/9780198789352.003.0001}{chap~1, pp 3--76}

\bibitem{BBG17}
Boyko E, Bercovici M and Gat A~D 2017 {\em Phys. Rev. Fluids\/}
  \href{http://dx.doi.org/10.1103/PhysRevFluids.2.073301}{{\bf 2} 073301}

\bibitem{AC19c}
Anand V and Christov I~C 2021 {\em Z. Angew. Math. Mech. (ZAMM)\/}
  \href{http://dx.doi.org/10.1002/zamm.201900309}{{\bf 101} e201900309}

\bibitem{VD87}
{Van Dyke} M 1987 {\em Adv. Appl. Mech.\/}
  \href{http://dx.doi.org/10.1016/S0065-2156(08)70276-X}{{\bf 25} 1--45}

\bibitem{WC19}
Wang X and Christov I~C 2019 {\em Proc. R. Soc. A\/}
  \href{http://dx.doi.org/10.1098/rspa.2019.0513}{{\bf 475} 20190513}

\bibitem{EG14}
Elbaz S~B and Gat A~D 2014 {\em J. Fluid Mech.\/}
  \href{http://dx.doi.org/10.1017/jfm.2014.527}{{\bf 758} 221--237}

\bibitem{RK72}
Rubinow S~I and Keller J~B 1972 {\em J. Theor. Biol.\/}
  \href{http://dx.doi.org/10.1016/0022-5193(72)90041-0}{{\bf 34} 299--313}

\bibitem{IMP08}
Iliev O, Mikeli{\'c} A and Popov P 2008 {\em Multiscale Model. Simul.\/}
  \href{http://dx.doi.org/10.1137/06067732X}{{\bf 7} 93--123}

\bibitem{PRBM20}
Rosti M~E, Pramanik S, Brandt L and Mitra D 2020 {\em Soft Matter\/}
  \href{http://dx.doi.org/10.1039/C9SM01678C}{{\bf 16} 939--944}

\bibitem{UCTSQ00}
Unger M~A, Chou H~P, Thorsen T, Scherer A and Quake S~R 2000 {\em Science\/}
  \href{http://dx.doi.org/10.1126/science.288.5463.113}{{\bf 288} 113--116}

\bibitem{SC18}
Shidhore T~C and Christov I~C 2018 {\em J. Phys.: Condens. Matter\/}
  \href{http://dx.doi.org/10.1088/1361-648X/aaa226}{{\bf 30} 054002}

\bibitem{AC19b}
Anand V and Christov I~C 2019 On the deformation of a hyperelastic tube due to
  steady viscous flow within {\em Dynamical Processes in Generalized Continua
  and Structures\/} ({\em Advanced Structured Materials\/} vol 103) ed
  Altenbach H, Belyaev A, Eremeyev V, Krivtsov A and Porubov A (Cham,
  Switzerland: Springer Nature)
  \href{http://dx.doi.org/10.1007/978-3-030-11665-1_2}{pp 17--35}

\bibitem{AMC20}
Anand V, Muchandimath S~C and Christov I~C 2020 {\em ASME J. Appl. Mech.\/}
  \href{http://dx.doi.org/10.1115/1.4046297}{{\bf 87} 051012}

\bibitem{RDC17}
Raj M~K, DasGupta S and Chakraborty S 2017 {\em Microfluid. Nanofluid.\/}
  \href{http://dx.doi.org/10.1007/s10404-017-1908-5}{{\bf 21} 70}

\bibitem{Hunter20}
Hunter L, {Gala de Pablo} J, Stammers A~C, Thomson N~H, Evans S~D and Shim J~u
  2020 {\em SN Appl. Sci.\/}
  \href{http://dx.doi.org/10.1007/s42452-020-03288-8}{{\bf 2} 1501}

\bibitem{HKO09}
Howell P, Kozyreff G and Ockendon J 2009 {\em {Applied Solid Mechanics}\/}
  (Cambridge, UK: Cambridge University Press)

\bibitem{SM04}
Skotheim J~M and Mahadevan L 2004 {\em Phys. Rev. Lett.\/}
  \href{http://dx.doi.org/10.1103/PhysRevLett.92.245509}{{\bf 92} 245509}

\bibitem{SM05}
Skotheim J~M and Mahadevan L 2005 {\em Phys. Fluids\/}
  \href{http://dx.doi.org/10.1063/1.1985467}{{\bf 17} 092101}

\bibitem{CC11}
Chakraborty J and Chakraborty S 2011 {\em Phys. Fluids\/}
  \href{http://dx.doi.org/10.1063/1.3624615}{{\bf 23} 082004}

\bibitem{RDC19}
Kiran~Raj M, Dasgupta S and Chakraborty S 2019 {\em Biomicrofluidics\/}
  \href{http://dx.doi.org/10.1063/1.5065901}{{\bf 13} 14103}

\bibitem{RM18}
Ramachandran V and Majidi C 2018 {\em ASME J. Appl. Mech.\/}
  \href{http://dx.doi.org/10.1115/1.4040477}{{\bf 85} 101004}

\bibitem{WC21}
Wang X and Christov I~C 2021 {\em Phys. Fluids\/}
  \href{http://dx.doi.org/10.1063/5.0062252}{{\bf 33} 102004}

\bibitem{Dillard2018}
Dillard D~A, Mukherjee B, Karnal P, Batra R~C and Frechette J 2018 {\em Soft
  Matter\/} \href{http://dx.doi.org/10.1039/c7sm02062g}{{\bf 14} 3669--3683}

\bibitem{CV20}
Chandler T~G~J and Vella D 2020 {\em Proc. R. Soc. A\/}
  \href{http://dx.doi.org/10.1098/rspa.2020.0551}{{\bf 476} 20200551}

\bibitem{SGOG20}
Salem L, Gamus B, Or Y and Gat A~D 2020 {\em Soft Robotics\/}
  \href{http://dx.doi.org/10.1089/soro.2019.0005}{{\bf 7} 76--84}

\bibitem{LLMH21}
Lee G, Luner A, Marzuola J and Harris D~M 2021 {\em Microfluid. Nanofluid.\/}
  \href{http://dx.doi.org/10.1007/s10404-021-02436-9}{{\bf 25} 34}

\bibitem{S15}
Sochi T 2015 {\em Rheol. Acta\/}
  \href{http://dx.doi.org/10.1007/s00397-015-0863-x}{{\bf 54} 745--756}

\bibitem{PC15}
Pritchard D and Corson L~T 2015 {\em Rheol. Acta\/}
  \href{http://dx.doi.org/10.1007/s00397-015-0860-0}{{\bf 54} 657--659}

\bibitem{BS21}
Boyko E and Stone H~A 2021 {\em J. Fluid Mech.\/}
  \href{http://dx.doi.org/10.1017/jfm.2021.621}{{\bf 923} R5}

\bibitem{MY19}
Mehboudi A and Yeom J 2019 {\em Microfluid. Nanofluid.\/}
  \href{http://dx.doi.org/10.1007/s10404-019-2235-9}{{\bf 23} 66}

\bibitem{CYDHM10}
Christopher G~F, Yoo J~M, Dagalakis N, Hudson S~D and Migler K~B 2010 {\em Lab
  Chip\/} \href{http://dx.doi.org/10.1039/C005065B}{{\bf 10} 2749--2757}

\bibitem{MR_NIST}
Baum M 2010 Micro rheometer is latest lab on a chip device {NIST} News
  \urlprefix\url{https://www.nist.gov/news-events/news/2010/08/micro-rheometer-latest-lab-chip-device}

\bibitem{PM09}
Pipe C~J and McKinley G~H 2009 {\em Mech. Res. Commun.\/}
  \href{http://dx.doi.org/10.1016/j.mechrescom.2008.08.009}{{\bf 36} 110--120}

\bibitem{GWV16}
Gupta S, Wang W~S and Vanapalli S~A 2016 {\em Biomicrofluidics\/}
  \href{http://dx.doi.org/10.1063/1.4955123}{{\bf 10} 043402}

\bibitem{DDGM15}
{Del Giudice} F, {D'Avino} G, Greco F, De~Santo I, Netti P~A and Maffettone P~L
  2015 {\em Lab Chip\/} \href{http://dx.doi.org/10.1039/C4LC01157K}{{\bf 15}
  783--792}

\bibitem{SB06}
Srivastava N and Burns M~A 2006 {\em Anal. Chem.\/}
  \href{http://dx.doi.org/10.1021/ac0518046}{{\bf 78} 1690--1696}

\bibitem{GAO13}
Galindo-Rosales F~J, Alves M~A and Oliveira M~S~N 2013 {\em Microfluid.
  Nanofluid.\/} \href{http://dx.doi.org/10.1007/s10404-012-1028-1}{{\bf 14}
  1--19}

\bibitem{HSPM15}
Hudson S~D, Sarangapani P, Pathak J~A and Migler K~B 2015 {\em J. Pharm.
  Sci.\/} \href{http://dx.doi.org/10.1002/jps.24201}{{\bf 104} 678--685}

\bibitem{VJ20}
Vishwanathan G and Juarez G 2020 {\em Microfluid. Nanofluid.\/}
  \href{http://dx.doi.org/10.1007/s10404-020-02373-z}{{\bf 24} 69}

\bibitem{R01}
Riley N 2001 {\em Annu. Rev. Fluid Mech.\/}
  \href{http://dx.doi.org/10.1146/annurev.fluid.33.1.43}{{\bf 33} 43--65}

\bibitem{S12}
Sadhal S~S 2012 {\em Lab Chip\/}
  \href{http://dx.doi.org/10.1039/c2lc40202e}{{\bf 12} 2292--2300}

\bibitem{VJ19a}
Vishwanathan G and Juarez G 2019 {\em Phys. Fluids\/}
  \href{http://dx.doi.org/10.1063/1.5092634}{{\bf 31} 041701}

\bibitem{VJ19b}
Vishwanathan G and Juarez G 2019 {\em J. Non-Newtonian Fluid Mech.\/}
  \href{http://dx.doi.org/10.1016/j.jnnfm.2019.07.007}{{\bf 271} 104143}

\bibitem{ME08}
Muzychka Y~S and Edge J 2008 {\em ASME J. Fluids Eng.\/}
  \href{http://dx.doi.org/10.1115/1.2979005}{{\bf 130} 111201}

\bibitem{SASM01}
Sharp K~V, Adrian R~J, Santiago J~G and Molho J~I 2001 Liquid flows in
  microchannels {\em The {MEMS} Handbook\/} ed Gad-el Hak M (Boca Raton, FL:
  CRC Press) \href{http://dx.doi.org/10.1201/9781420050905.ch6}{chap~6}

\bibitem{YWHFJZM19}
Yang X, Weldetsadik N~T, Hayat Z, Fu T, Jiang S, Zhu C and Ma Y 2019 {\em
  Microfluid. Nanofluid.\/}
  \href{http://dx.doi.org/10.1007/s10404-019-2241-y}{{\bf 23} 75}

\bibitem{KLK05}
Kang K, Lee L~J and Koelling K~W 2005 {\em Exp. Fluids\/}
  \href{http://dx.doi.org/10.1007/s00348-004-0901-4}{{\bf 38} 222--232}

\bibitem{PMM08}
Pipe C~J, Majmudar T~S and McKinley G~H 2008 {\em Rheol. Acta\/}
  \href{http://dx.doi.org/10.1007/s00397-008-0268-1}{{\bf 47} 621--642}

\bibitem{SLVYW21}
Shiba K, Li G, Virot E, Yoshikawa G and Weitz D~A 2021 {\em Lab Chip\/}
  \href{http://dx.doi.org/10.1039/D1LC00202C}{{\bf 21} 2805--2811}

\bibitem{sdamsw02}
Stroock A~D, Dertinger S~K~W, Ajdari A, Mezi\'c I, Stone H~A and Whitesides F~M
  2002 {\em Science\/} \href{http://dx.doi.org/10.1126/science.1066238}{{\bf
  295} 647--651}

\bibitem{VK13}
Verma M~K~S and Kumaran V 2013 {\em J. Fluid Mech.\/}
  \href{http://dx.doi.org/10.1017/jfm.2013.264}{{\bf 727} 407--455}

\bibitem{SK17}
Srinivas S~S and Kumaran V 2017 {\em J. Fluid Mech.\/}
  \href{http://dx.doi.org/10.1017/jfm.2016.839}{{\bf 812} 1076--1118}

\bibitem{CSD19}
Chandra B, Shankar V and Das D 2019 {\em Phys. Fluids\/}
  \href{http://dx.doi.org/10.1063/1.5122867}{{\bf 31} 114103}

\bibitem{RDIAM02}
Reyes D~R, Iossifidis D, Auroux P~A and Manz A 2002 {\em Anal. Chem.\/}
  \href{http://dx.doi.org/10.1021/ac0202435}{{\bf 74} 2623--2636}

\bibitem{AIRM02}
Auroux P~A, Iossifidis D, Reyes D~R and Manz A 2002 {\em Anal. Chem.\/}
  \href{http://dx.doi.org/10.1021/ac020239t}{{\bf 74} 2637--2652}

\bibitem{Yetisen13}
Yetisen A~K, Akram M~S and Lowe C~R 2013 {\em Lab Chip\/}
  \href{http://dx.doi.org/10.1039/C3LC50169H}{{\bf 13} 2210--2251}

\bibitem{SFB14}
Sackmann E~K, Fulton A~L and Beebe D~J 2014 {\em Nature\/}
  \href{http://dx.doi.org/10.1038/nature13118}{{\bf 507} 181--189}

\bibitem{Yager06}
Yager P, Edwards T, Fu E, Helton K, Nelson K, Tam M~R and Weigl B~H 2006 {\em
  Nature\/} \href{http://dx.doi.org/10.1038/nature05064}{{\bf 442} 412--418}

\bibitem{K13}
Karnik R 2013 Microfluidic mixing {\em Encyclopedia of Microfluidics and
  Nanofluidics\/} ed Li D (Boston, MA: Springer)
  \href{http://dx.doi.org/10.1007/978-3-642-27758-0_939-2}{pp 1--13}

\bibitem{Taylor_NCFMF}
{National Committee for Fluid Mechanics Films} 1967 Low {Reynolds} number flow
  Education Development Center, Inc.
  \urlprefix\url{http://web.mit.edu/hml/ncfmf.html}

\bibitem{OW04}
Ottino J~M and Wiggins S 2004 {\em Phil. Trans. R. Soc. A\/}
  \href{http://dx.doi.org/10.1098/rsta.2003.1355}{{\bf 362} 923--935}

\bibitem{L92}
Larson R~G 1992 {\em Rheol. Acta\/}
  \href{http://dx.doi.org/10.1007/BF00366504}{{\bf 31} 213--263}

\bibitem{Shaq96}
Shaqfeh E~S~G 1996 {\em Annu. Rev. Fluid Mech.\/}
  \href{http://dx.doi.org/10.1146/annurev.fl.28.010196.001021}{{\bf 28}
  129--185}

\bibitem{S21}
Steinberg V 2021 {\em Annu. Rev. Fluid Mech.\/}
  \href{http://dx.doi.org/10.1146/annurev-fluid-010719-060129}{{\bf 53} 27--58}

\bibitem{Li12}
Li X~B, Li F~C, Cai W~H, Zhang H~N and Yang J~C 2012 {\em Exp. Thermal Fluid
  Sci.\/} \href{http://dx.doi.org/10.1016/j.expthermflusci.2011.12.014}{{\bf
  39} 1--16}

\bibitem{GR14}
Galindo-Rosales F~J, Campo-Dea{\~{n}}o L, Sousa P~C, Ribeiro V~M, Oliveira
  M~S~N, Alves M~A and Pinho F~T 2014 {\em Exp. Thermal Fluid Sci.\/}
  \href{http://dx.doi.org/10.1016/j.expthermflusci.2014.03.004}{{\bf 59}
  128--139}

\bibitem{a84}
Aref H 1984 {\em J. Fluid Mech.\/}
  \href{http://dx.doi.org/10.1017/S0022112084001233}{{\bf 143} 1--21}

\bibitem{o89}
Ottino J~M 1989 {\em The Kinematics of Mixing: Stretching, Chaos, and
  Transport\/} ({\em Cambridge Texts in Applied Mathematics\/} vol~3)
  (Cambridge, UK: Cambridge University Press)

\bibitem{Thakur03}
Thakur R~K, Vial C, Nigam K~D~P, Nauman E~B and Djelveh G 2003 {\em Chem. Eng.
  Res. Des.\/} \href{http://dx.doi.org/10.1205/026387603322302968}{{\bf 81}
  787--826}

\bibitem{Ghanem14}
Ghanem A, Lemenand T, Valle D~D and Peerhossaini H 2014 {\em Chem. Eng. Res.
  Des.\/} \href{http://dx.doi.org/10.1016/j.cherd.2013.07.013}{{\bf 92}
  205--228}

\bibitem{KS79}
Krindel P and Silberberg A 1979 {\em J. Colloid Interface Sci.\/}
  \href{http://dx.doi.org/10.1016/0021-9797(79)90219-4}{{\bf 71} 39--50}

\bibitem{VK12}
Verma M~K~S and Kumaran V 2013 {\em J. Fluid Mech.\/}
  \href{http://dx.doi.org/10.1017/jfm.2011.55}{{\bf 705} 322--347}

\bibitem{KB16}
Kumaran V and Bandaru P 2016 {\em Chem. Eng. Sci.\/}
  \href{http://dx.doi.org/10.1016/j.ces.2016.04.001}{{\bf 149} 156--168}

\bibitem{ESHW07}
Eckhardt B, Schneider T~M, Hof B and Westerweel J 2007 {\em Annu. Rev. Fluid
  Mech.\/} \href{http://dx.doi.org/10.1146/annurev.fluid.39.050905.110308}{{\bf
  39} 447--468}

\bibitem{SA04}
Sharp K~V and Adrian R~J 2004 {\em Experiments in Fluids\/}
  \href{http://dx.doi.org/10.1007/s00348-003-0753-3}{{\bf 36} 741--747}

\bibitem{LCWF11}
Lee C~Y, Chang C~L, Wang Y~N and Fu L~M 2011 {\em Int. J. Mol. Sci.\/}
  \href{http://dx.doi.org/10.3390/ijms12053263}{{\bf 12} 3263--3287}

\bibitem{FY11}
Friend J and Yeo L~Y 2011 {\em Rev. Mod. Phys.\/}
  \href{http://dx.doi.org/10.1103/RevModPhys.83.647}{{\bf 83} 647--704}

\bibitem{RWH17}
Rallabandi B, Wang C and Hilgenfeldt S 2017 {\em Phys. Rev. Fluids\/}
  \href{http://dx.doi.org/10.1103/PhysRevFluids.2.064501}{{\bf 2} 064501}

\bibitem{CK07}
Chokshi P and Kumaran V 2007 {\em Phys. Fluids\/}
  \href{http://dx.doi.org/10.1063/1.2798069}{{\bf 19} 104103}

\bibitem{CBK15}
Chokshi P, Bhade P and Kumaran V 2015 {\em Phys. Rev. E\/}
  \href{http://dx.doi.org/10.1103/PhysRevE.91.023007}{{\bf 91} 023007}

\bibitem{Toner2018}
Toner M 2018 ``{E}xtreme'' microfluidics: {L}arge-volumes and complex fluids
  Nobel Symposium 162 - Microfluidics, Nobel162/2017/09 Nobel Foundation
  \urlprefix\url{https://arxiv.org/abs/1802.05600}

\bibitem{BKP08}
Bhagat A~A~S, Kuntaegowdanahalli S~S and Papautsky I 2008 {\em Lab Chip\/}
  \href{http://dx.doi.org/10.1039/B807107A}{{\bf 8} 1906--1914}

\bibitem{DC09}
{Di Carlo} D 2009 {\em Lab Chip\/}
  \href{http://dx.doi.org/10.1039/b912547g}{{\bf 9} 3038--3046}

\bibitem{DiCarlo2007}
Di~Carlo D, Irimia D, Tompkins R~G and Toner M 2007 {\em Proc. Natl Acad. Sci.
  USA\/} \href{http://dx.doi.org/10.1073/pnas.0704958104}{{\bf 104}
  18892--18897}

\bibitem{Lim2014}
Lim E~J, Ober T~J, Edd J~F, Desai S~P, Neal D, Bong K~W, Doyle P~S, McKinley
  G~H and Toner M 2014 {\em Nat. Commun.\/}
  \href{http://dx.doi.org/10.1038/ncomms5120}{{\bf 5} 4120}

\bibitem{RGHM88}
Riley J~J, {Gad-el-Hak} M and Metcalfe R~W 1988 {\em Annu. Rev. Fluid Mech.\/}
  \href{http://dx.doi.org/10.1146/annurev.fl.20.010188.002141}{{\bf 20}
  393--420}

\bibitem{G96}
Gad-el Hak M 1996 {\em Appl. Mech. Rev.\/}
  \href{http://dx.doi.org/10.1115/1.3101966}{{\bf 49} S147--S157}

\bibitem{G02}
Gad-el Hak M 2002 {\em Prog. Aerospace Sci.\/}
  \href{http://dx.doi.org/10.1016/S0376-0421(01)00020-3}{{\bf 38} 77--99}

\bibitem{K61}
Kramer M~O 1961 {\em J. Am. Soc. Naval Eng.\/}
  \href{http://dx.doi.org/10.1111/j.1559-3584.1961.tb02422.x}{{\bf 73}
  103--108}

\bibitem{B60}
Benjamin T~B 1960 {\em J. Fluid Mech.\/}
  \href{http://dx.doi.org/10.1017/S0022112060001286}{{\bf 9} 513--532}

\bibitem{NS15}
Neelamegam R and Shankar V 2015 {\em Phys. Fluids\/}
  \href{http://dx.doi.org/10.1063/1.4907246}{{\bf 27} 024102}

\bibitem{PS19}
Patne R and Shankar V 2019 {\em J. Fluid Mech.\/}
  \href{http://dx.doi.org/10.1017/jfm.2018.908}{{\bf 860} 837--885}

\bibitem{TPS18}
Tanmay V~S, Patne R and Shankar V 2018 {\em Phys. Fluids\/}
  \href{http://dx.doi.org/10.1063/1.5041771}{{\bf 30} 074103}

\bibitem{K21}
Kumaran V 2021 {\em J. Fluid Mech.\/}
  \href{http://dx.doi.org/10.1017/jfm.2021.602}{{\bf 924} P1}

\bibitem{Karan2020b}
Karan P, Das S~S, Mukherjee R, Chakraborty J and Chakraborty S 2020 {\em Soft
  Matter\/} \href{http://dx.doi.org/10.1039/D0SM00333F}{{\bf 16} 5777--5786}

\bibitem{DGPHD07}
Dendukuri D, Gu S~S, Pregibon D~C, Hatton T~A and Doyle P~S 2007 {\em Lab
  Chip\/} \href{http://dx.doi.org/10.1039/b703457a}{{\bf 7} 818--828}

\bibitem{MCC13}
Mukherjee U, Chakraborty J and Chakraborty S 2013 {\em Soft Matter\/}
  \href{http://dx.doi.org/10.1039/c2sm27247d}{{\bf 9} 1562--1569}

\bibitem{MCSPS19}
Mart{\'{i}}nez-Calvo A, Sevilla A, Peng G~G and Stone H~A 2020 {\em J. Fluid
  Mech.\/} \href{http://dx.doi.org/10.1017/jfm.2019.994}{{\bf 885} A25}

\bibitem{EJG18}
Elbaz S~B, Jacob H and Gat A~D 2018 {\em J. Fluid Mech.\/}
  \href{http://dx.doi.org/10.1017/jfm.2018.287}{{\bf 846} 460--481}

\bibitem{AC20}
Anand V and Christov I~C 2020 {\em Phys. Fluids\/}
  \href{http://dx.doi.org/10.1063/5.0022406}{{\bf 32} 112014}

\bibitem{H13}
Holmes M~H 2013 {\em Introduction to Perturbation Methods\/} 2nd ed ({\em Texts
  in Applied Mathematics\/} vol~20) (New York, NY: Springer)

\bibitem{B96}
Barenblatt G~I 1996 {\em Similarity, Self-Similarity, and Intermediate
  Asymptotics\/} ({\em Cambridge Texts in Applied Mathematics\/} vol~14) (New
  York, NY: Cambridge University Press)

\bibitem{IWC20}
Inamdar T~C, Wang X and Christov I~C 2020 {\em Phys. Rev. Fluids\/}
  \href{http://dx.doi.org/10.1103/PhysRevFluids.5.064101}{{\bf 5} 064101}

\bibitem{GKATB17}
Garg V, Kamat P~M, Anthony C~R, Thete S~S and Basaran O~A 2017 {\em J. Fluid
  Mech.\/} \href{http://dx.doi.org/10.1017/jfm.2017.446}{{\bf 826} 455--483}

\bibitem{VEG19}
Vurgaft A, Elbaz S~B and Gat A~D 2019 {\em J. Fluid Mech.\/}
  \href{http://dx.doi.org/10.1017/jfm.2019.789}{{\bf 881} 1048--1072}

\bibitem{PSMG20}
Peretz O, Mishra A~K, Shepherd R~F and Gat A~D 2020 {\em Proc. Natl Acad. Sci.
  USA\/} \href{http://dx.doi.org/10.1073/pnas.1919738117}{{\bf 117} 5217--5221}

\bibitem{EG16}
Elbaz S~B and Gat A~D 2016 {\em J. Fluid Mech.\/}
  \href{http://dx.doi.org/10.1017/jfm.2016.587}{{\bf 806} 580--602}

\bibitem{Poly17}
Polygerinos P, Correll N, Morin S~A, Mosadegh B, Onal C~D, Petersen K,
  Cianchetti M, Tolley M~T and Shepherd R~F 2017 {\em Adv. Eng. Mater.\/}
  \href{http://dx.doi.org/10.1002/adem.201700016}{{\bf 19} 1700016}

\bibitem{CCLDF19}
Chiapponi L, Ciriello V, Longo S and Di~Federico V 2019 {\em Water Res. Res.\/}
  \href{http://dx.doi.org/10.1029/2019WR026071}{{\bf 55} 10144--10158}

\bibitem{BDRM16}
Barbati A~C, Desroches J, Robisson A and McKinley G~H 2016 {\em Annu. Rev.
  Chem. Biomol. Eng.\/}
  \href{http://dx.doi.org/10.1146/annurev-chembioeng-080615-033630}{{\bf 7}
  415--453}

\bibitem{Dana18}
Dana A, Zheng Z, Peng G~G, Stone H~A, Huppert H~E and Ramon G~Z 2018 {\em J.
  Fluid Mech.\/} \href{http://dx.doi.org/10.1017/jfm.2017.778}{{\bf 836}
  828--849}

\bibitem{P94}
Probstein R~F 1994 {\em Physicochemical Hydrodynamics: An Introduction\/} 2nd
  ed (New York, NY: John Wiley \& Sons)

\bibitem{L04}
Li D 2004 {\em Electrokinetics in Microfluidics\/} ({\em Interface Science and
  Technology\/} vol~2) (San Diego, CA: Elsevier)

\bibitem{C10_book_Ch1}
Chakraborty D and Chakraborty S 2010 Microfluidic transport and micro-scale
  flow physics: An overview {\em Microfluidics and Microfabrication\/} ed
  Chakraborty S (New York: Springer Science+Business Media)
  \href{http://dx.doi.org/10.1007/978-1-4419-1543-6_1}{chap~1, pp 1--85}

\bibitem{C10_book_Ch2}
Ghosal S 2010 Mathematical modeling of electrokinetic effects in micro and nano
  fluidics {\em Microfluidics and Microfabrication\/} ed Chakraborty S (New
  York: Springer Science+Business Media)
  \href{http://dx.doi.org/10.1007/978-1-4419-1543-6_2}{chap~2, pp 87--112}

\bibitem{AWWD21}
Alizadeh A, Hsu W~L, Wang M and Daiguji H 2021 {\em Electrophoresis\/}
  \href{http://dx.doi.org/10.1002/elps.202000313}{{\bf 42} 834--868}

\bibitem{CC10}
Chakraborty J and Chakraborty S 2010 {\em Phys. Fluids\/}
  \href{http://dx.doi.org/10.1063/1.3524530}{{\bf 22} 122002}

\bibitem{SCKC08}
Steinberger A, Cottin-Bizonne C, Kleimann P and Charlaix E 2008 {\em Phys. Rev.
  Lett.\/} \href{http://dx.doi.org/10.1103/PhysRevLett.100.134501}{{\bf 100}
  134501}

\bibitem{C13_book_Ch3}
Das S, Chakraborty J and Chakraborty S 2013 Electrokinetics in narrow
  confinements {\em Microfluidics and Microscale Transport Processes\/} IIT
  Kharagpur Research Monograph Series ed Chakraborty S (Boca Raton, FL: CRC
  Press) \href{http://dx.doi.org/10.1201/b12976-9}{chap~2, pp 49--110}

\bibitem{RTGB17}
Rubin S, Tulchinsky A, Gat A~D and Bercovici M 2017 {\em J. Fluid Mech.\/}
  \href{http://dx.doi.org/10.1017/jfm.2016.830}{{\bf 812} 841--865}

\bibitem{BEGGB19}
Boyko E, Eshel R, Gommed K, Gat A~D and Bercovici M 2019 {\em J. Fluid Mech.\/}
  \href{http://dx.doi.org/10.1017/jfm.2018.967}{{\bf 862} 732--752}

\bibitem{Rutte16}
de~Rutte J~M, Janssen K~G~H, Tas N~R, Eijkel J~C~T and Pennathur S 2016 {\em
  Microfluid. Nanofluid.\/}
  \href{http://dx.doi.org/10.1007/s10404-016-1815-1}{{\bf 20} 150}

\bibitem{BEGB20}
Boyko E, Eshel R, Gat A~D and Bercovici M 2020 {\em Phys. Rev. Lett.\/}
  \href{http://dx.doi.org/10.1103/PhysRevLett.124.024501}{{\bf 124} 024501}

\bibitem{BIBG20}
Boyko E, Ilssar D, Bercovici M and Gat A~D 2020 {\em Phys. Rev. Fluids\/}
  \href{http://dx.doi.org/10.1103/PhysRevFluids.5.104201}{{\bf 5} 104201}

\bibitem{HM04}
Hosoi A~E and Mahadevan L 2004 {\em Phys. Rev. Lett.\/}
  \href{http://dx.doi.org/10.1103/PhysRevLett.93.137802}{{\bf 93} 137802}

\bibitem{ZY13}
Zhao C and Yang C 2013 {\em Adv. Colloid Interface Sci.\/}
  \href{http://dx.doi.org/10.1016/j.cis.2013.09.001}{{\bf 201-202} 94--108}

\bibitem{ZZMY08}
Zhao C, Zholkovskij E, Masliyah J~H and Yang C 2008 {\em J. Colloid Interface
  Sci.\/} \href{http://dx.doi.org/10.1016/j.jcis.2008.06.028}{{\bf 326}
  503--510}

\bibitem{ZY11}
Zhao C and Yang C 2011 {\em J. Non-Newtonian Fluid Mech.\/}
  \href{http://dx.doi.org/10.1016/j.jnnfm.2011.05.006}{{\bf 166} 1076--1079}

\bibitem{ZRC06}
Zimmerman W~B, Rees J~M and Craven T~J 2006 {\em Microfluid. Nanofluid.\/}
  \href{http://dx.doi.org/10.1007/s10404-006-0089-4}{{\bf 2} 481--492}

\bibitem{S19}
Chakraborty S 2019 {\em Electrophoresis\/}
  \href{http://dx.doi.org/10.1002/elps.201800353}{{\bf 40} 180--189}

\bibitem{KCC20}
Karan P, Chakraborty J and Chakraborty S 2020 {\em Phys. Fluids\/}
  \href{http://dx.doi.org/10.1063/1.5134149}{{\bf 32} 022002}

\bibitem{KCC21}
Karan P, Chakraborty J and Chakraborty S 2021 {\em J. Fluid Mech.\/}
  \href{http://dx.doi.org/10.1017/jfm.2021.595}{{\bf 923} A32}

\bibitem{NCC17}
Naik K~G, Chakraborty S and Chakraborty J 2017 {\em Soft Matter\/}
  \href{http://dx.doi.org/10.1039/C7SM01423F}{{\bf 13} 6422--6429}

\bibitem{Dhong18}
Dhong C, Edmunds S~J, Ram{\'{i}}rez J, Kayser L~V, Chen F, Jokerst J~V and
  Lipomi D~J 2018 {\em Nano Lett.\/}
  \href{http://dx.doi.org/10.1021/acs.nanolett.8b02292}{{\bf 18} 5306--5311}

\bibitem{OYE13}
Ozsun O, Yakhot V and Ekinci K~L 2013 {\em J. Fluid Mech.\/}
  \href{http://dx.doi.org/10.1017/jfm.2013.474}{{\bf 734} R1}

\bibitem{NNBR17}
Niu P, Nablo B~J, Bhadriraju K and Reyes D~R 2017 {\em Anal. Chem.\/}
  \href{http://dx.doi.org/10.1021/acs.analchem.7b02287}{{\bf 89} 11372--11377}

\bibitem{Chen19}
Chen X, Falzon L, Zhang J, Zhang X, Wang R~Q and Du K 2019 {\em
  Nanotechnology\/} \href{http://dx.doi.org/10.1088/1361-6528/ab2279}{{\bf 31}
  05LT01}

\bibitem{EPKVS21}
Essink M~H, Pandey A, Karpitschka S, Venner C~H and Snoeijer J~H 2021 {\em J.
  Fluid Mech.\/} \href{http://dx.doi.org/10.1017/jfm.2021.96}{{\bf 915} A49}

\bibitem{DGM17}
D'Avino G, Greco F and Maffettone P~L 2017 {\em Annu. Rev. Fluid Mech.\/}
  \href{http://dx.doi.org/10.1146/annurev-fluid-010816-060150}{{\bf 49}
  341--360}

\bibitem{ZP20}
Zhou J and Papautsky I 2020 {\em Microsyst. Nanoeng.\/}
  \href{http://dx.doi.org/10.1038/s41378-020-00218-x}{{\bf 6} 113}

\bibitem{Wexler13}
Wexler J~S, Trinh P~H, Berthet H, Quennouz N, du~Roure O, Huppert H~E, Lindner
  A and Stone H~A 2013 {\em J. Fluid Mech.\/}
  \href{http://dx.doi.org/10.1017/jfm.2013.49}{{\bf 720} 517--544}

\bibitem{DMSLR20}
Dey A~A, Modarres-Sadeghi Y, Lindner A and Rothstein J~P 2020 {\em Soft
  Matter\/} \href{http://dx.doi.org/10.1039/C9SM01794A}{{\bf 16} 1227--1235}

\bibitem{TH06}
Turek S and Hron J 2006 Proposal for numerical benchmarking of fluid-structure
  interaction between an elastic object and laminar incompressible flow {\em
  Fluid-Structure Interaction\/} ({\em Lecture Notes in Computational Science
  and Engineering\/} vol~53) ed Bungartz H~J and Sch{\"a}fer M
  (Berlin/Heidelberg: Springer)
  \href{http://dx.doi.org/10.1007/3-540-34596-5_15}{pp 371--385}

\bibitem{ACLH17}
Alvarado J, Comtet J, de~Langre E and Hosoi A~E 2017 {\em Nat. Phys.\/}
  \href{http://dx.doi.org/10.1038/nphys4225}{{\bf 13} 1014--1019}

\bibitem{Poole12}
Poole R~J 2012 {\em Rheology Bulletin\/} {\bf 53(2)} 32--39
  \urlprefix\url{http://pcwww.liv.ac.uk/~robpoole/PAPERS/POOLE_45.pdf}

\bibitem{DMSR20b}
Dey A~A, Modarres-Sadeghi Y and Rothstein J~P 2020 {\em J. Non-Newtonian Fluid
  Mech.\/} \href{http://dx.doi.org/10.1016/j.jnnfm.2020.104433}{{\bf 286}
  104433}

\bibitem{ML18}
Moukhtari F~E and Lecampion B 2018 {\em J. Fluid Mech.\/}
  \href{http://dx.doi.org/10.1017/jfm.2017.900}{{\bf 838} 573--605}

\bibitem{BS21b}
Boyko E and Stone H~A 2021 {\em Phys. Rev. Fluids\/}
  \href{http://dx.doi.org/10.1103/PhysRevFluids.6.L081301}{{\bf 6} L081301}

\bibitem{GQ04}
Groisman A and Quake S~R 2004 {\em Phys. Rev. Lett.\/}
  \href{http://dx.doi.org/10.1103/PhysRevLett.92.094501}{{\bf 92} 094501}

\bibitem{GEQ03}
Groisman A, Enzelberger M and Quake S~R 2003 {\em Science\/}
  \href{http://dx.doi.org/10.1126/science.1083694}{{\bf 300} 955--958}

\bibitem{GZL03}
Granick S, Zhu Y and Lee H 2003 {\em Nature Materials\/}
  \href{http://dx.doi.org/10.1038/nmat854}{{\bf 2} 221--227}

\bibitem{FNP12}
Ferr\'{a}s L~L, N\'{o}brega J~M and Pinho F~T 2012 {\em J. Non-Newtonian Fluid
  Mech.\/} \href{http://dx.doi.org/10.1016/j.jnnfm.2012.03.004}{{\bf 175-176}
  76--88}

\bibitem{KCCWC21}
Karan P, Chakraborty J, Chakraborty S, Wereley S~T and Christov I~C 2021 {\em
  Phys. Rev. E\/} \href{http://dx.doi.org/10.1103/PhysRevE.104.015108}{{\bf
  104} 015108}

\bibitem{GMV17}
Gomez M, Moulton D~E and Vella D 2017 {\em Phys. Rev. Lett.\/}
  \href{http://dx.doi.org/10.1103/PhysRevLett.119.144502}{{\bf 119} 144502}

\bibitem{S11}
Schomburg W~K 2011 {\em Introduction to Microsystem Design\/}
  (Berlin/Heidelberg: Springer-Verlag)

\bibitem{WHJW10}
Whittaker R~J, Heil M, Jensen O~E and Waters S~L 2010 {\em Q. J. Mech. Appl.
  Math.\/} \href{http://dx.doi.org/10.1093/qjmam/hbq020}{{\bf 63} 465--496}

\bibitem{N11}
Nunes L~C~S 2011 {\em Mat. Sci. Eng. A\/}
  \href{http://dx.doi.org/10.1016/j.msea.2010.11.025}{{\bf 528} 1799--1804}

\bibitem{HS12}
Hu Y and Suo Z 2012 {\em Acta Mech. Solida Sin.\/}
  \href{http://dx.doi.org/10.1016/S0894-9166(12)60039-1}{{\bf 25} 441--458}

\bibitem{LOLLCZ09}
Lin I~K, Ou K~S, Liao Y~M, Liu Y, Chen K~S and Zhang X 2009 {\em J.
  Microelectromech. Syst.\/}
  \href{http://dx.doi.org/10.1109/JMEMS.2009.2029166}{{\bf 18} 1087--1099}

\bibitem{Huang11}
Huang G~Y, Zhou L~H, Zhang Q~C, Chen Y~M, Sun W, Xu F and Lu T~J 2011 {\em
  Biofabrication\/} \href{http://dx.doi.org/10.1088/1758-5082/3/1/012001}{{\bf
  3} 012001}

\bibitem{AM17}
Auton L~C and MacMinn C~W 2017 {\em Proc. R. Soc. A\/}
  \href{http://dx.doi.org/10.1098/rspa.2016.0753}{{\bf 473} 20160753}

\bibitem{AM18}
Auton L~C and MacMinn C~W 2018 {\em Proc. R. Soc. A\/}
  \href{http://dx.doi.org/10.1098/rspa.2018.0284}{{\bf 474} 20180284}

\bibitem{AENY19}
Ambartsumyan I, Ervin V~J, Nguyen T and Yotov I 2019 {\em ESAIM: M2AN\/}
  \href{http://dx.doi.org/10.1051/m2an/2019061}{{\bf 53} 1915--1955}

\end{thebibliography}
\balance

\end{document}